\documentclass[a4paper,12pt]{article}
\usepackage{jheppub}
\usepackage{graphicx} 
\usepackage{amsmath}
\usepackage{amssymb}

\newcommand{\ptj}{p_T^{\rm jet}}
\newcommand{\yj}{y^{\rm jet}}
\newcommand{\hthat}{\hat{H}_T}
\newcommand{\mur}{\mu_{\rm R}}
\newcommand{\muf}{\mu_{\rm F}}
\newcommand{\muj}{\mu_{\rm J}}

\newcommand{\nn}{\nonumber}

\title{Les Houches study on inclusive jet production at NNLO+NNLL}

\author[a]{Terry Generet,}
\affiliation[a]{Cavendish Laboratory, University of Cambridge, Cambridge CB3 0HE, UK}
\emailAdd{generet@hep.phy.cam.ac.uk}

\author[b]{Joey Huston,}
\affiliation[b]{Department of Physics and Astronomy, Michigan State University, East Lansing, MI 48824, USA}
\emailAdd{huston@msu.edu}

\author[c]{Kyle Lee,}
\affiliation[c]{High Energy Physics Division, Argonne National Laboratory, Lemont, IL 60439, USA}
\emailAdd{kyle@anl.gov}

\author[d]{Ian Moult,}
\affiliation[d]{Department of Physics, Yale University, New Haven, CT 06511, USA}
\emailAdd{ian.moult@yale.edu}

\author[e]{Rene Poncelet,}
\affiliation[e]{Institute of Nuclear Physics, ul. Radzikowskiego 152, 31--342 Krakow, Poland}
\emailAdd{rene.poncelet@ifj.edu.pl}

\author[f]{Xiaoyuan Zhang}
\affiliation[f]{Center for Theoretical Physics - a Leinweber Institute, Massachusetts Institute of Technology, Cambridge, MA 02139, USA}
\emailAdd{xyz2@mit.edu}

\preprint{\vbox{%
    \hbox{MIT-CTP/6029}
    \hbox{Cavendish-HEP-26/02}
    \hbox{IFJPAN-IV-2026-12}
}}

\abstract{
Jet production at the LHC is a powerful probe of QCD, making it ideal for precision tests and determinations of QCD parameters such as parton distribution functions and the strong coupling constant.
To make the most of the abundant jet production data collected at the LHC, precise calculations are required.
While state-of-the-art calculations reach next-to-next-to-leading order (NNLO) QCD accuracy, a critical assessment of the remaining uncertainties arising from non-perturbative effects and missing higher orders remains crucial for correctly interpreting comparisons between theory and data.
Scale variation is nearly always used to determine effects from missing higher orders.
In this article, we reassess this method in the context of inclusive jet production by performing NNLO QCD calculations supplemented by small-jet-radius resummation through next-to-next-to-leading-logarithmic accuracy (NNLL). We find that NNLL resummation can have an appreciable impact on the scale uncertainty for inclusive jet cross sections, and, for some scale choices, can lead to sizeable shifts of the central cross section.
We conclude that scale variations in fixed-order and resummed calculations can drastically underestimate the impact of higher orders for commonly used jet radius parameters, and that missing higher-order estimates obtained via scale variations should be considered unreliable. Our findings add further evidence to the importance of going beyond scale variations in jet and jet substructure calculations.
}

\date{\today}

\begin{document}

\maketitle

\section{Introduction}

The LHC has become a precision machine for a variety of experimental measurements, with uncertainties at or approaching a few percent~\cite{Huss:2025nlt}.
Such a level of experimental precision demands a corresponding level of theoretical precision to interpret the data meaningfully.
This effort involves continuing improvement in parton distribution functions (PDFs), extending matrix element and cross section calculations to higher orders in QCD and electroweak theory, and, in some cases, to higher logarithmic precision in resummation.
The latter is especially true when the process of interest involves a hierarchy of scales that generates logarithmically enhanced contributions.
Such hierarchies can arise in QCD processes if observables are sensitive to additional parton emissions, such as the transverse momentum of a color-singlet, or if the phase space for such emissions is restricted by fiducial cuts, for example, through jet vetoes.
A more subtle hierarchy, which will be of particular interest in the present work, is introduced through the jet algorithm used to define infrared-safe jet cross sections.
Jet algorithms restrict the emission phase space ``within a jet" through the cone radius parameter $R$, which leads to logarithmically enhanced terms $\sim \ln^n R$ that spoil the perturbative convergence in the small $R$ limit.

Differential measurements of inclusive jet or di-jet production are powerful probes of QCD at hadron colliders and have been used to determine the strong coupling constant $\alpha_S$ and have a significant impact on the determination of the gluon distribution in recent PDF fits \cite{Nocera:2017zge, Harland-Lang:2017ytb, Hou:2019efy, Ablat:2024uvg, AbdulKhalek:2020jut, Bailey:2020ooq, NNPDF:2021njg, H1:2021xxi}.
On the theoretical side, jet cross sections, for up to three jets at the Born level and possibly involving additional electroweak bosons, have been computed through NNLO QCD~\cite{Boughezal:2015dva, Boughezal:2015dra, Boughezal:2015aha, Chen:2014gva, Gehrmann-DeRidder:2015wbt, Boughezal:2015ded, Campbell:2016lzl, Chawdhry:2021hkp, Badger:2023mgf, Currie:2016bfm, Currie:2017eqf, Currie:2018xkj, Czakon:2019tmo, Chen:2022tpk, Czakon:2021mjy, Alvarez:2023fhi} at fixed-order in perturbation theory.

Parton showers provide a generic method to resum logarithmically enhanced contributions, but their accuracy is so far limited to the leading logarithm for multi-jet processes.
The state of the art is the NLO+PS-matched computation \cite{Alioli:2010xa} for up to three jets, or corresponding multi-jet-merged NLO QCD samples~\cite{Gleisberg:2008ta, Reyer:2019obz}.
Further, dedicated computations have achieved the inclusion of specific logarithmically enhanced contributions. For example, high-energy resummation of invariant masses has been implemented in HEJ \cite{Andersen:2009he, Andersen:2009nu, Andersen:2011hs, Andersen:2023kuj}, and analytic resummation of small-$R$ logarithms has been performed at NLO+LL \cite{Dasgupta:2014yra, Dasgupta:2016bnd}, NLO+NLL \cite{Kang:2016mcy, Dai:2016hzf}, and NNLO+NNLL \cite{Generet:2025vth}. Threshold logarithms \cite{Kidonakis:1998bk, Kidonakis:2000gi} have been included through NLL in Refs.~\cite{Liu:2017pbb, Liu:2018ktv,Moch:2018hgy}.

Beyond the nominal perturbative precision, it is crucial to estimate the theory uncertainty.
In particular, the uncertainty arising from missing higher-order terms in the perturbative expansion, dubbed missing higher-order uncertainty (MHOU), is challenging to estimate but essential for assessing data-theory agreement.
The MHOU for a fixed-order prediction is often evaluated by varying the renormalization and factorization scales around a central value characteristic of the process being investigated by a factor of 2, avoiding cases where the renormalization and factorization scales differ by a factor of 4.
This procedure has well-known shortcomings: the range of scale variation is arbitrary, and the correlations of uncertainties across phase space are unclear.
Another ambiguity arises from the choice of the central scale and the question of which kinematic variable constitutes the best characteristic scale.
Alternative approaches to estimate MHOU, such as those based on Bayesian estimates \cite{Cacciari:2011ze, Bonvini:2020xeo, Duhr:2021mfd, Ghosh:2022lrf} or theory-nuisance parameters \cite{Tackmann:2024kci, Lim:2024nsk}, are challenging to implement for complicated processes such as jet production at hadron colliders.

For inclusive jet production, two commonly used scale choices are (1) $\ptj$, the transverse momentum of each jet in the final state being considered, and (2) $\hthat$, the scalar sum of the transverse momenta of all partons in the final state. The two scales yield similar NNLO predictions for inclusive jet production at the LHC, although arguments have been made to reduce the scale to $\hthat/2$, which is also commonly used, including for central di-jet production.\footnote{Another suggested scale, the transverse momentum of the leading jet, has also been proposed, but has proven to be problematic~\cite{Currie:2018xkj}.}
Global PDF fits typically use the scale of $\ptj$, while many collider comparisons to data use a scale of $\hthat$.

As mentioned above, calculations for jet production involve applying a jet algorithm, most commonly the anti-$k_T$ algorithm, with a finite jet size.
Typical jet sizes are $R=0.4$, $0.6$, and $0.7$, depending on the physics process being targeted.
It is uncommon that more than one jet size will be used in a typical LHC measurement.
One disadvantage of using only one jet size is that it makes it harder to visualize the overall picture in the theoretical calculation.

For example, a jet radius scan from $0.3$ to $1.0$ for NNLO inclusive jet production (or for Z + jet production), reveals that there are accidental cancellations in the scale-dependent terms that result in artificially low MHOU estimates from scale variation near the commonly-used value of $0.4$, thus ascribing a greater precision to the NNLO prediction than warranted~\cite{Currie:2018xkj, Bellm:2019yyh}.
To make the situation more complex, the behavior is different for a scale of $\ptj$ than for $\hthat$. Studies of jet cross sections for $R$ values outside those typically measured by LHC experiments can provide a perspective on the behavior of perturbative QCD that can be lacking in predictions based on a single jet radius.

This peculiar behavior motivates the reinvestigation of these observables, including small-$R$ resummation.
The approach in Ref.~\cite{Dasgupta:2014yra, Kang:2016mcy, Dai:2016hzf, Dasgupta:2016bnd, Generet:2025vth} factorizes the cross section into a hard function, encoding the hard scattering of partons, and a semi-inclusive jet function, describing the fragmentation of the parton into a jet of a given size $R$.
This approach allows the resummation of large $\ln R$ contributions in the small-$R$ limit using renormalization-group equations.

The present work aims to extend the study of Ref.~\cite{Generet:2025vth} by including the $\hthat$ scale choice in the analysis and perform a detailed comparison of the MHOU estimates based on scale variations using fixed-order and resummed $\ptj$ spectra.
We will introduce the computational setup in Section~\ref{sec:comp} and follow with the phenomenological results in Section~\ref{sec:pheno}.
The discussion and conclusions follow in Section~\ref{sec:discussion}.

\section{Computational setup}\label{sec:comp}

We focus on the process of inclusive jet production,
\begin{equation}
 pp \to \text{jet} + X\;.
\end{equation}
In particular, we will consider the perturbative expansion of the (rapidity-dependent) transverse momentum spectrum $\ptj$:
\begin{equation}
\frac{\text{d}^2\sigma }{\text{d}\ptj \text{d} y^{\rm jet}}  = \frac{\text{d}^2\sigma^{(0)} (\mur,\muf)}{\text{d}\ptj \text{d} \yj} + \frac{\text{d}^2\sigma^{(1)} (\mur,\muf)}{\text{d}\ptj \text{d} \yj} + \frac{\text{d}^2\sigma^{(2)} (\mur,\muf)}{\text{d}\ptj \text{d} \yj} + \mathcal{O}(\alpha_S^5)\;.
\end{equation}
Each jet is binned individually, implying that a single event can contribute 1-4 entries at NNLO QCD. The LO, NLO, and NNLO fixed-order predictions are defined as:
\begin{align}
\frac{\text{d}^2\sigma^{\rm LO} (\mur,\muf)}{\text{d}\ptj \text{d} \yj} &= \frac{\text{d}^2\sigma^{(0)} (\mur,\muf)}{\text{d}\ptj \text{d} \yj}\,,\nn\\
\frac{\text{d}^2\sigma^{\rm NLO} (\mur,\muf)}{\text{d}\ptj \text{d} \yj}&= \frac{\text{d}^2\sigma^{\rm LO} (\mur,\muf)}{\text{d}\ptj \text{d} \yj}+ \frac{\text{d}^2\sigma^{(1)} (\mur,\muf)}{\text{d}\ptj \text{d} \yj}\,,\nn \\
\frac{\text{d}^2\sigma^{\rm NNLO} (\mur,\muf)}{\text{d}\ptj \text{d} \yj} &= \frac{\text{d}^2\sigma^{\rm NLO} (\mur,\muf)}{\text{d}\ptj \text{d} \yj}+ \frac{\text{d}^2\sigma^{(2)} (\mur,\muf)}{\text{d}\ptj \text{d} \yj}\,.
\end{align}
We employ the sector-improved residue subtraction scheme, a fully automated NNLO QCD subtraction scheme used for a broad range of LHC processes \cite{Czakon:2010td, Czakon:2014oma, Czakon:2019tmo}.
The extension of the scheme to support parton-to-hadron fragmentation has been performed in Ref.~\cite{Czakon:2021ohs} and has been used to study light and heavy hadron production at the LHC \cite{Czakon:2024tjr, Czakon:2025yti}.
The implementation has been further extended to allow more general convolution structures and applied to the resummation of small-$R$ logarithms, where, to leading power in $R$, the cross section factorizes as
\begin{align}\label{eq:factorization}
    \frac{d\sigma_{\rm LP}}{\text{d}\ptj \text{d} \yj}=&\sum_{i,j,k}\int_{x_{i,\rm min}}^1 \frac{dx_i}{x_i}f_{i/P}(x_i,\muf)\int_{x_{j,\rm min}}^1 \frac{dx_j}{x_j}f_{j/P}(x_j,\muf) \int_{z_{\text{min}}}^1 \frac{dz}{z}\nn\\
    & \;\mathcal{H}_{ij}^k(x_i,x_j,\ptj/z,\yj,\mur,\muf,\muj)\times J_k\left(z, \ln\frac{(\ptj)^2 R^2}{z^2 \muj^2},\muj\right)\,.
\end{align}
Here $\mathcal{H}_{ij}^k$ is the partonic hard function, and $J_k$ the semi-inclusive jet function \cite{Kang:2016mcy, Lee:2024icn}.
The computation of the necessary terms of $J_k$ to achieve NNLL accuracy, further details on the computation of the hard function, and the convolution with the jet function are discussed in detail in Ref.~\cite{Generet:2025vth}.

We define the matched resummed predictions $\text{d}\sigma^{\rm LO+LL}$, $\text{d}\sigma^{\rm NLO+NLL}$, as well as $\text{d}\sigma^{\rm NNLO+NNLL}$ by the minimal necessary orders to achieve the claimed formal accuracy.
This means that we convolute for the NNLO+NNLL cross section, the LO hard function with the NNLL jet function, the NLO coefficient of the hard function with the NLL jet function, and the NNLO coefficient of the hard function with the LL jet function.

The factorization introduces the ``jet" scale, $\muj$, which is chosen similarly to the renormalization and factorization scales.
The scale $\muj$ is to be interpreted as a `hard' scale to which we evolve the jet function from its canonical scale $R\,\ptj$ and thereby resum logarithms of $R$. In practice, we choose $\muj$ to be the same as the renormalization and factorization scale. The central scale choice is therefore:
\begin{equation}
\mur = \muf = \muj = \mu\,.
\end{equation}
For $\mu$ we use either $\ptj$ (implying that each jet is binned with a different scale) or
\begin{align}
    \hthat = \sum_{i \in {\rm partons}} p_{T,i}\;,\quad p_{T,i}: \text{transverse momentum of parton}~i\;.
\end{align}
For the MHOU estimation, we use a 7-point variation for fixed-order calculations:
\begin{align}
&\text{7-point variation:}\quad \{\sigma(f_{\rm R} \mur,f_{\rm F} \muf)\}_{(f_{\rm R},f_{\rm F})}\nn\\
&\;\;\text{where}\quad (f_{\rm R},f_{\rm F}) \in \left\{ \left(1,1\right),\left(2,2\right),\left(\frac{1}{2},\frac{1}{2}\right), \left(1,2\right), \left(2,1\right),  \left(1,\frac{1}{2}\right), \left(\frac{1}{2},1\right)\right\},  
\end{align}
and a 15-point variation for resummed calculations:
\begin{align}
&\text{15-point variation:}\quad \{\sigma(f_{\rm R} \mur,f_{\rm F} \muf,f_{\rm J}\muj)\}_{(f_{\rm R},f_{\rm F},f_{\rm J})}\nn\\
&\;\;\text{where}\quad (f_{\rm R},f_{\rm F},f_{\rm J}) \in \bigg\{ \left(1,1,1\right),\left(2,2,2\right),\left(\frac{1}{2},\frac{1}{2},\frac{1}{2}\right), \left(1,1,2\right), \left(1,2,1\right),\nn\\
&\quad\left(2,1,1\right),\left(1,1,\frac{1}{2}\right), \left(1,\frac{1}{2},1\right),\left(\frac{1}{2},1,1\right),\left(1,2,2\right),\left(2,1,2\right),\left(2,2,1\right),\nn\\
&\quad\left(1,\frac{1}{2},\frac{1}{2}\right),\left(\frac{1}{2},1,\frac{1}{2}\right),\left(\frac{1}{2},\frac{1}{2},1\right)\bigg\}.  
\end{align}
For the PDF, we use the \textit{NNPDF31\_nnlo\_as0118} set \cite{NNPDF:2017mvq} for all perturbative orders.

\section{Phenomenology}\label{sec:pheno}

We study inclusive jet production within the fiducial phase space defined by the experimental analysis in Ref.~\cite{CMS:2020caw} by the CMS collaboration.
This measurement extracted the differential cross sections in bins of $\ptj$ and $|\yj|$.
Specifically, they provide the ratio of the cross sections with respect to $R=0.4$ for jet radii $R=0.1,0.2, \dots 1.2$. This data provides an experimental laboratory needed for a better understanding of the perturbative QCD aspects of the $R$-dependence of inclusive jet cross sections.
After discussing the absolute cross-section spectra in  Section~\ref{sec:pheno-abs}, we compare NNLO+NNLL predictions for ratios between different jet radii to CMS data in Section~\ref{sec:pheno-ratios}.

\subsection{Absolute transverse momentum spectra}\label{sec:pheno-abs}

\begin{figure}
\includegraphics[width=0.49\textwidth]{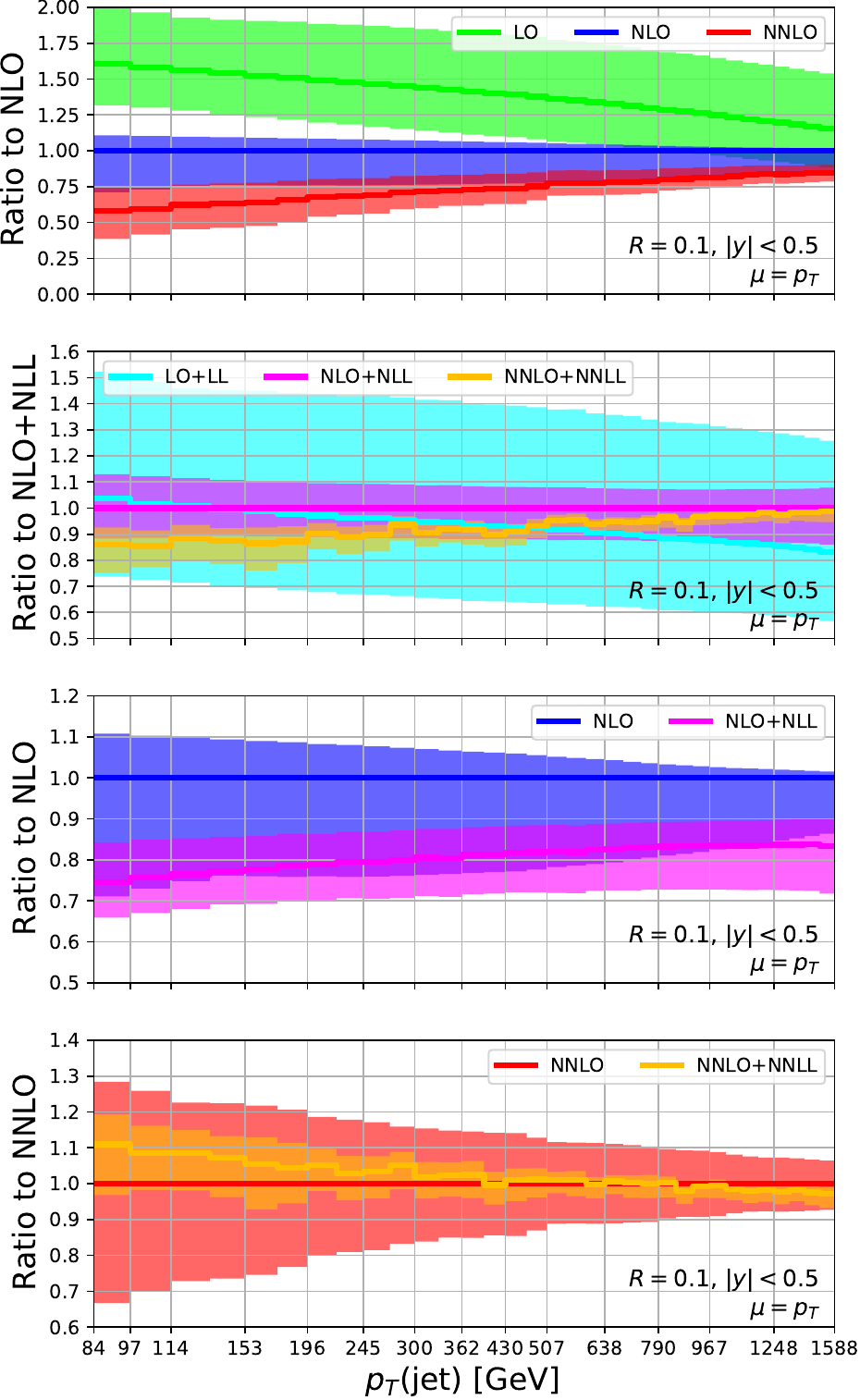}
\includegraphics[width=0.49\textwidth]{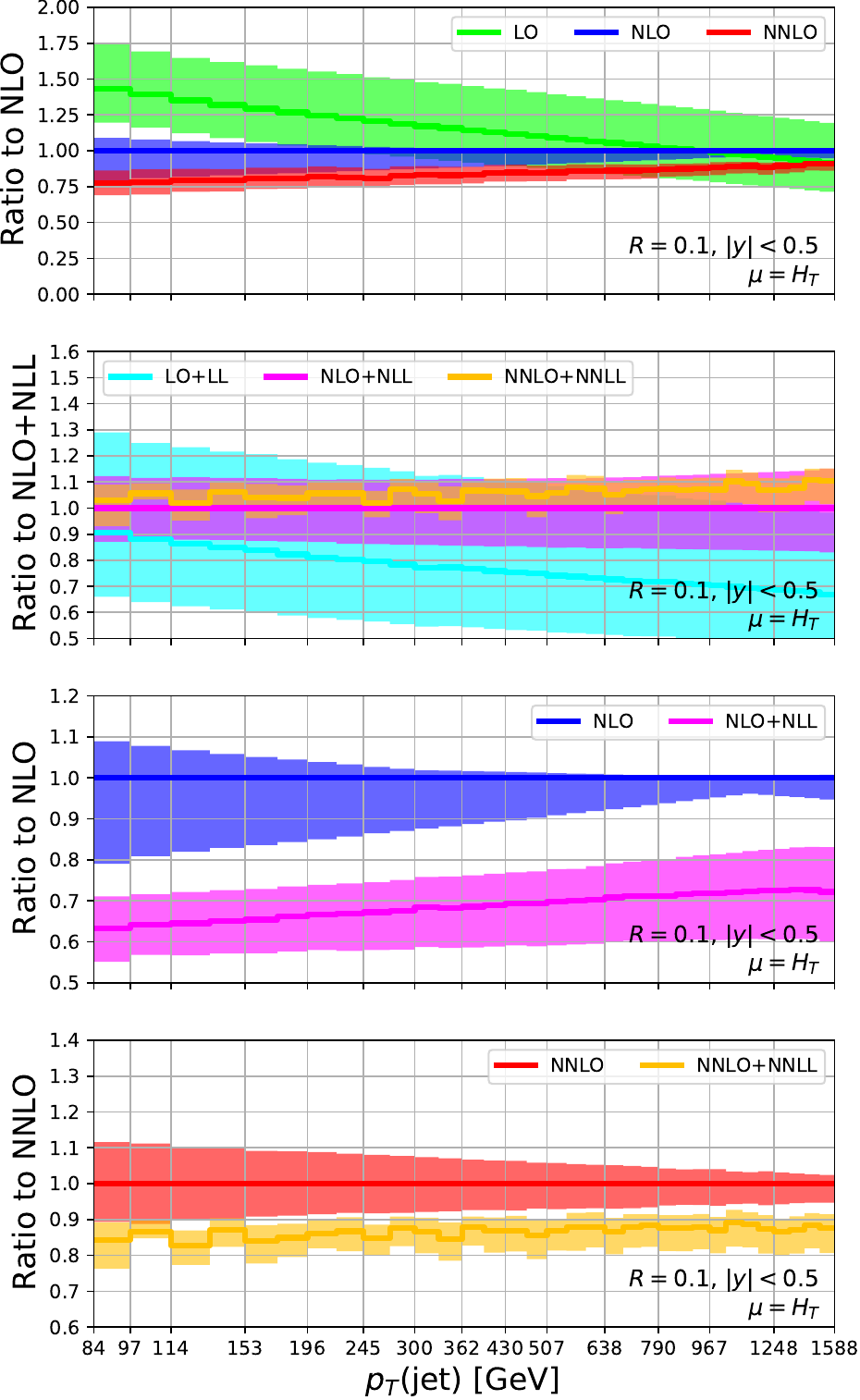}
\caption{The comparison of theory predictions for the absolute spectrum for $R=0.1$. Shown are predictions for $\mu=\ptj$ (left) and for $\mu=\hthat$ (right).}
\label{fig:AbsoluteR0.1}
\end{figure}

\begin{figure}
\includegraphics[width=0.49\textwidth]{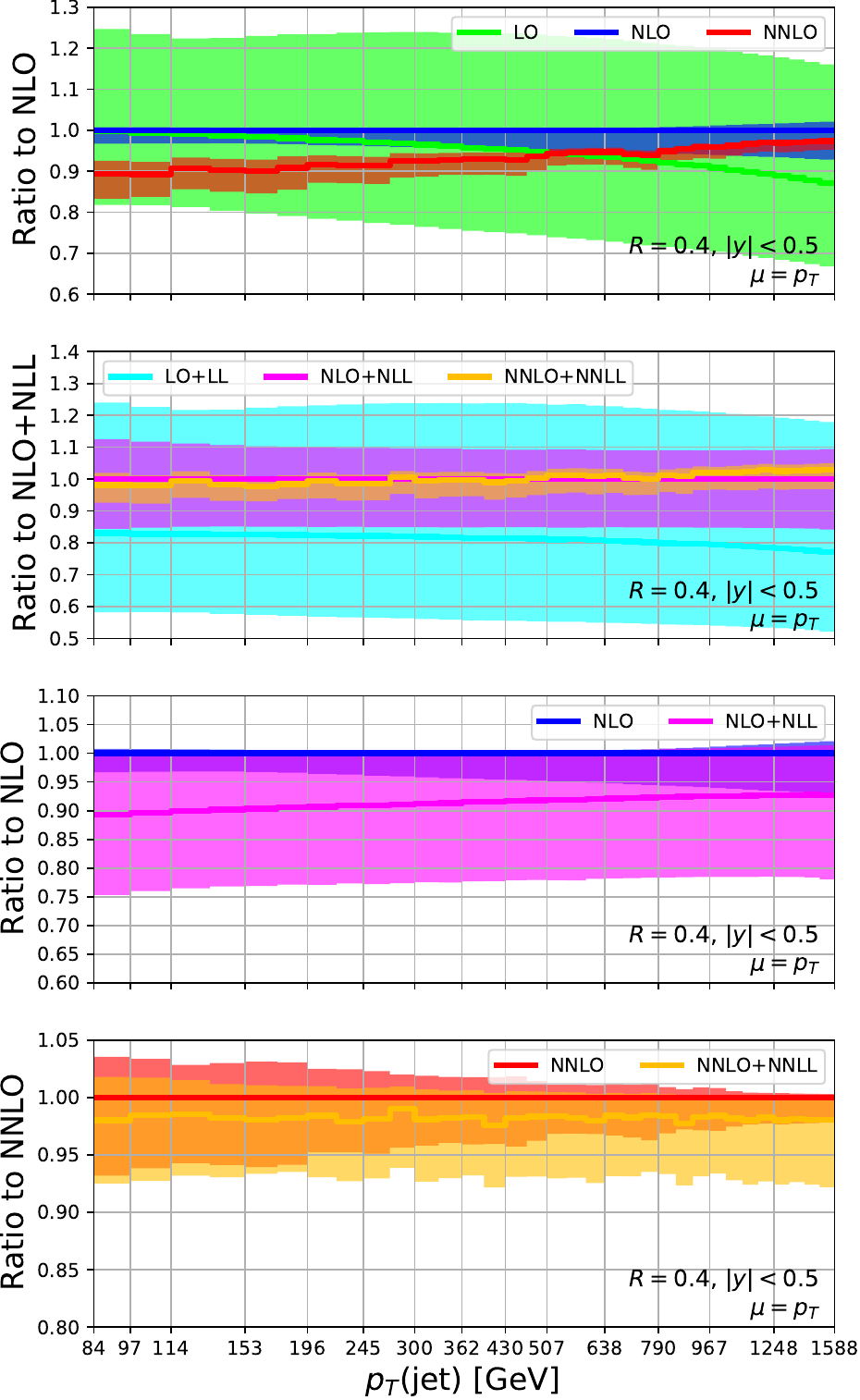}
\includegraphics[width=0.49\textwidth]{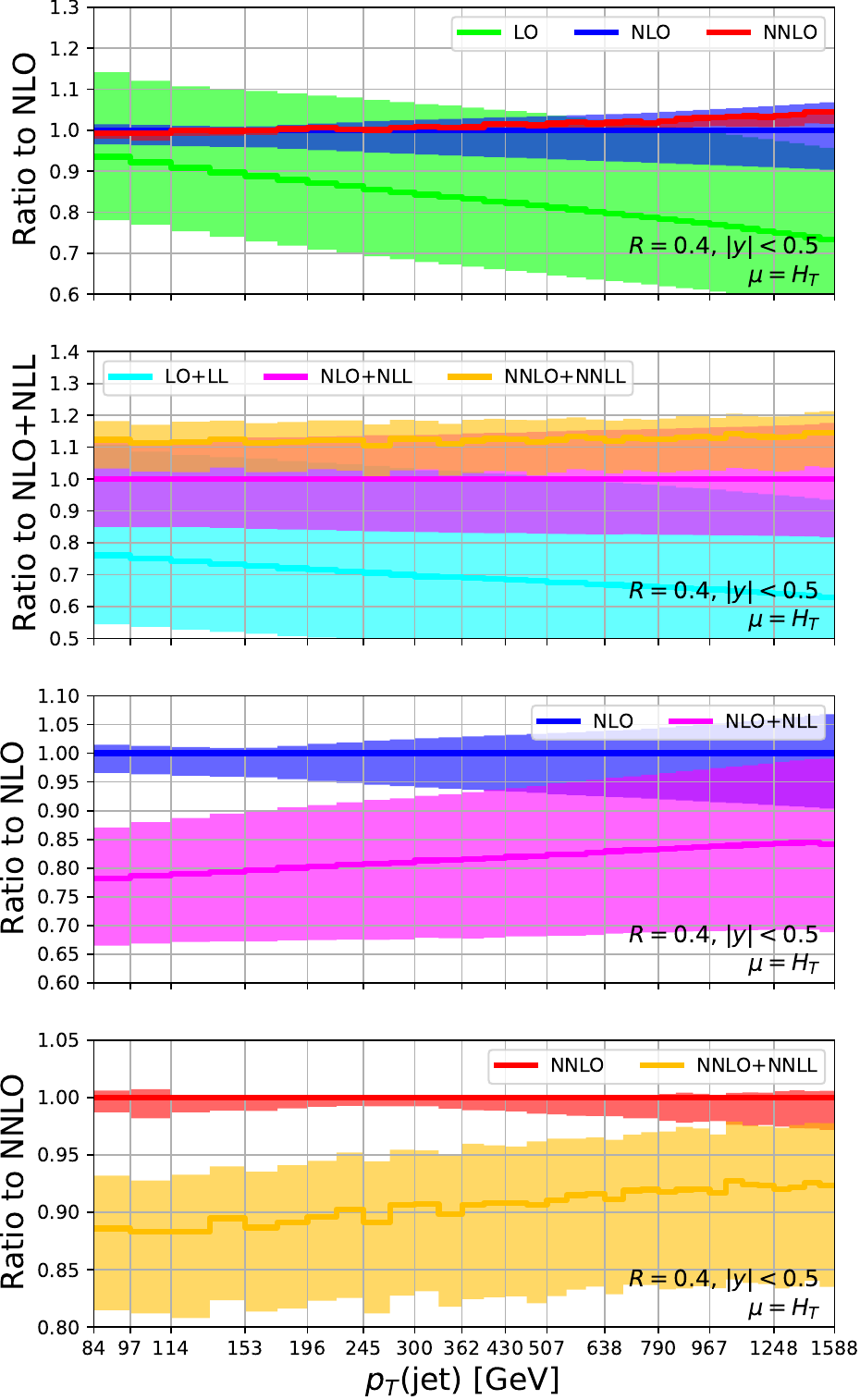}
\caption{As in Figure~\ref{fig:AbsoluteR0.1}, but for $R=0.4$.}
\label{fig:AbsoluteR0.4}
\end{figure}

\begin{figure}
\includegraphics[width=0.49\textwidth]{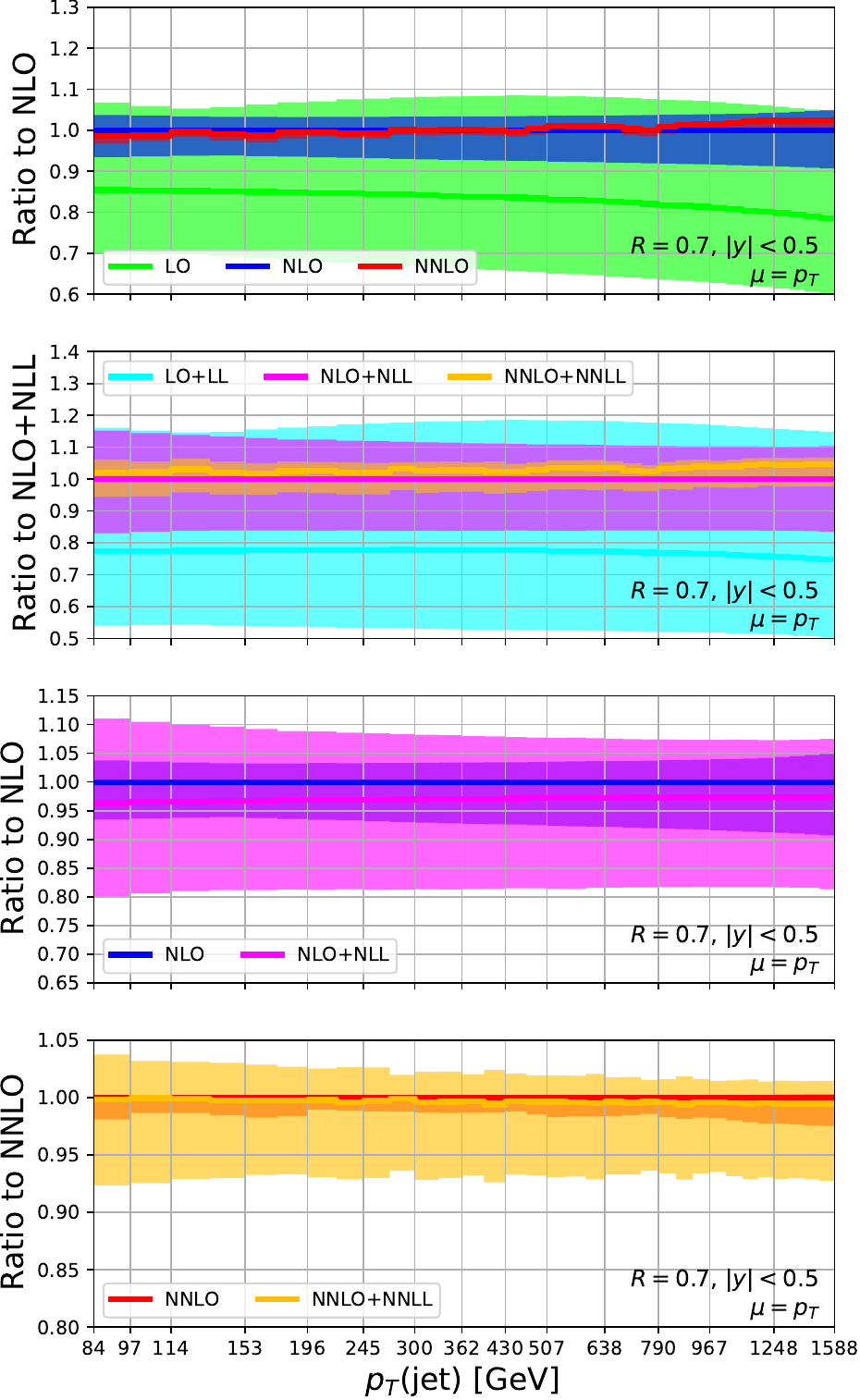}
\includegraphics[width=0.49\textwidth]{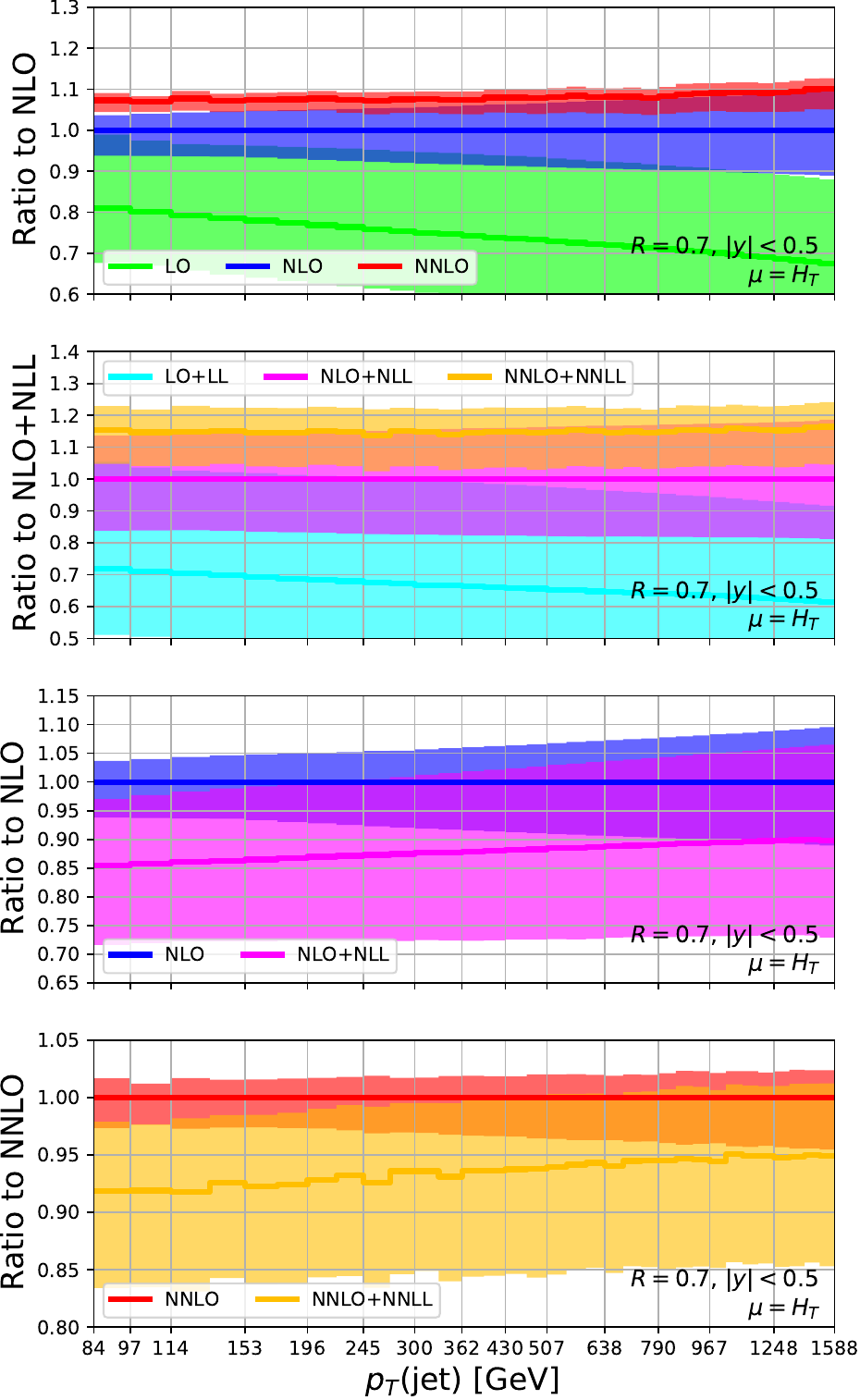}
\caption{As in Figure~\ref{fig:AbsoluteR0.1}, but for $R=0.7$.}
\label{fig:AbsoluteR0.7}
\end{figure}

\begin{figure}
\includegraphics[width=0.49\textwidth]{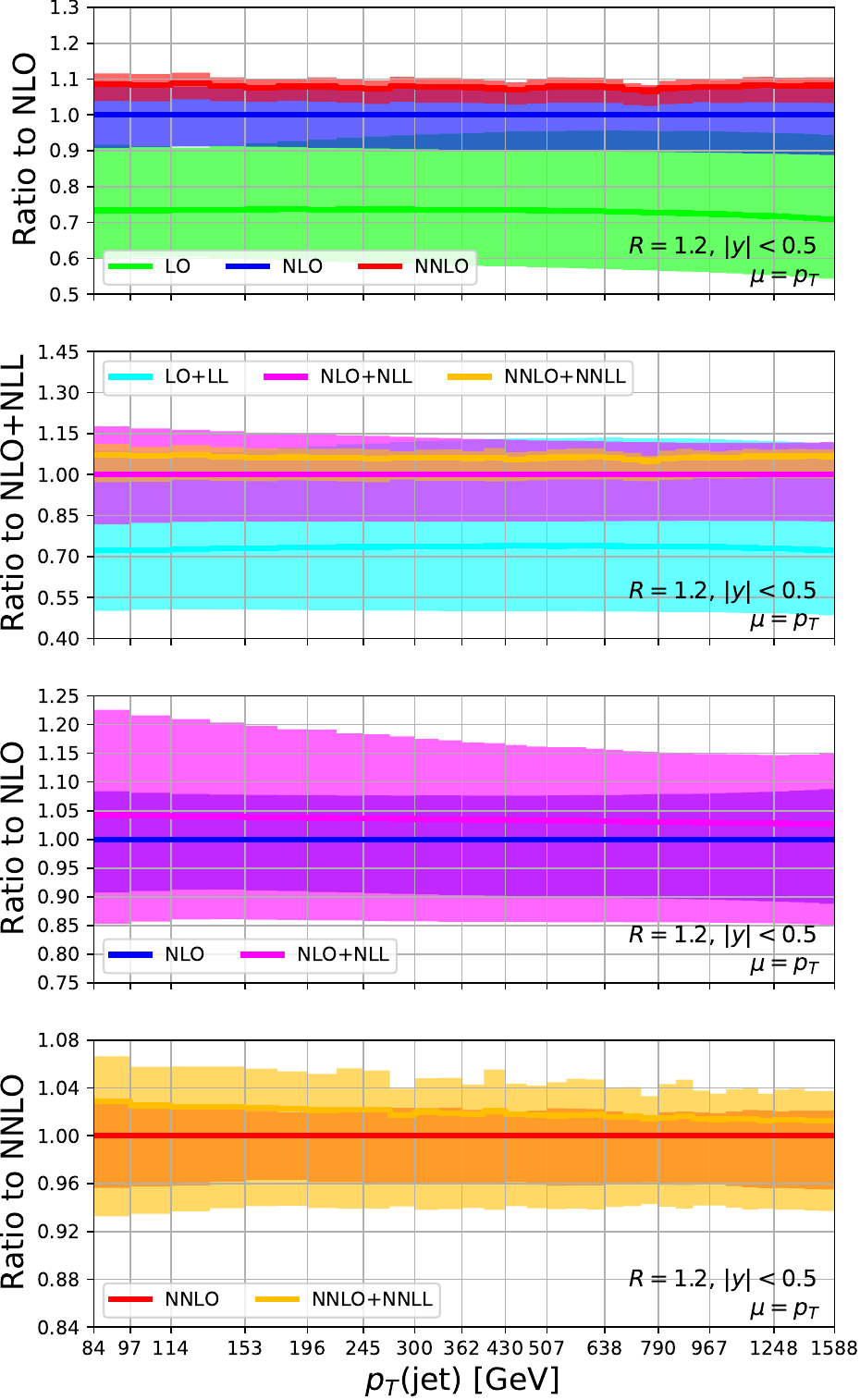}
\includegraphics[width=0.49\textwidth]{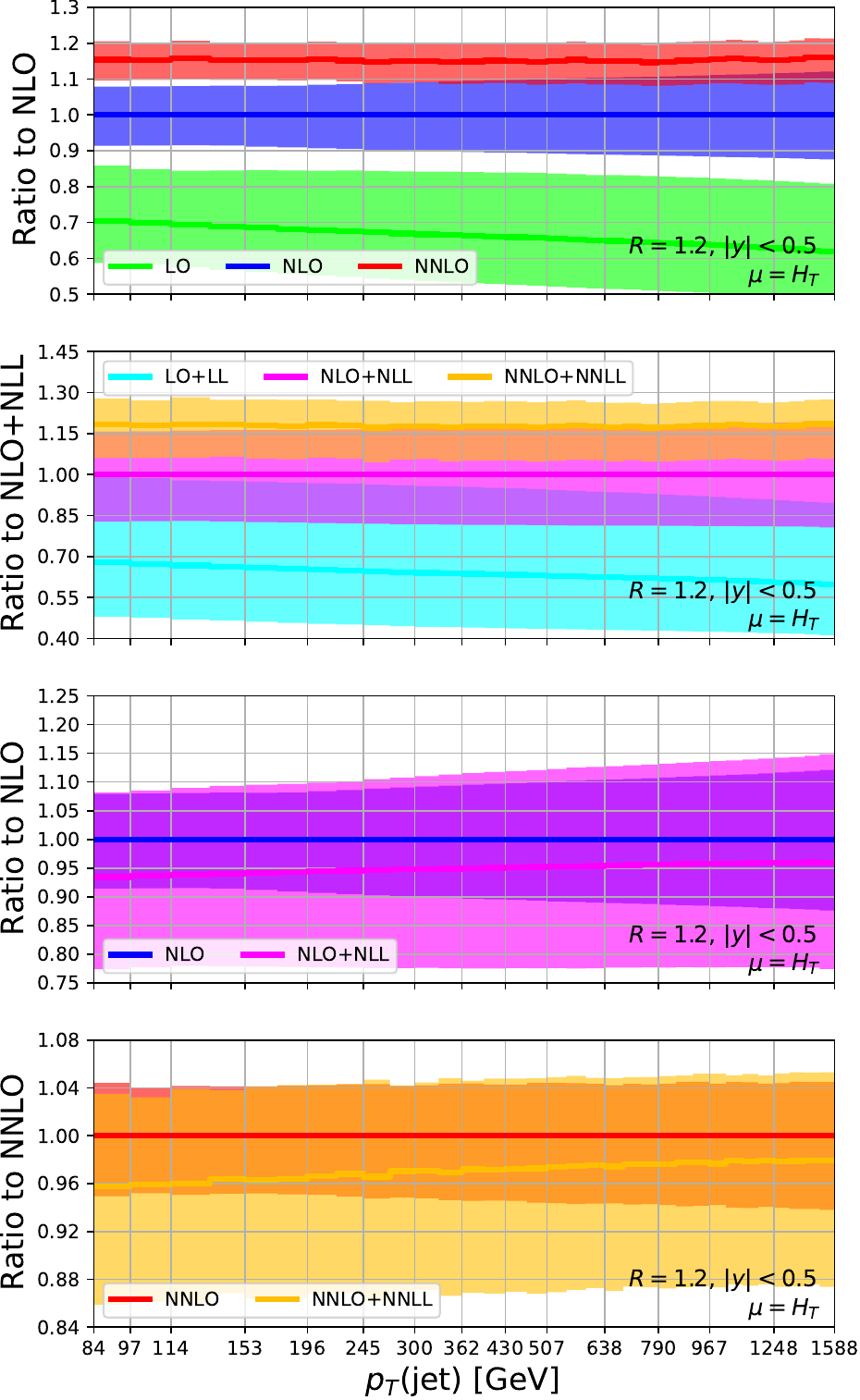}
\caption{As in Figure~\ref{fig:AbsoluteR0.1}, but for $R=1.2$.}
\label{fig:AbsoluteR1.2}
\end{figure}

In Figure~\ref{fig:AbsoluteR0.1} to Figure~\ref{fig:AbsoluteR1.2}, LO, NLO, NNLO, LO+LL, NLO+NLL and NNLO+NNLL predictions are shown for jet sizes of $0.1$, $0.4$, $0.7$, and $1.2$ in the central rapidity region $|\yj| < 0.5$, for central scale choices of $\ptj$ and $\hthat$.
Besides the nominal predictions, we show the envelopes arising from 7-point scale variations (15-point in the resummed cases).

At leading order, all jet algorithms are equivalent, as each jet consists of just a single parton, and by definition, there is no $R$ dependence.
At LO, the scale uncertainty is typically very large because there are no partial scale-compensating terms.
Calculations at NLO and NNLO include scale-compensating terms, as well as additional contributions that may alter the shape and normalization of the cross sections.
In general, corrections and scale uncertainties, which we use as a proxy for the MHOU, are expected to decrease with increasing perturbative order, provided there are no large logarithms.

\begin{figure}
\includegraphics[width=0.49\textwidth]{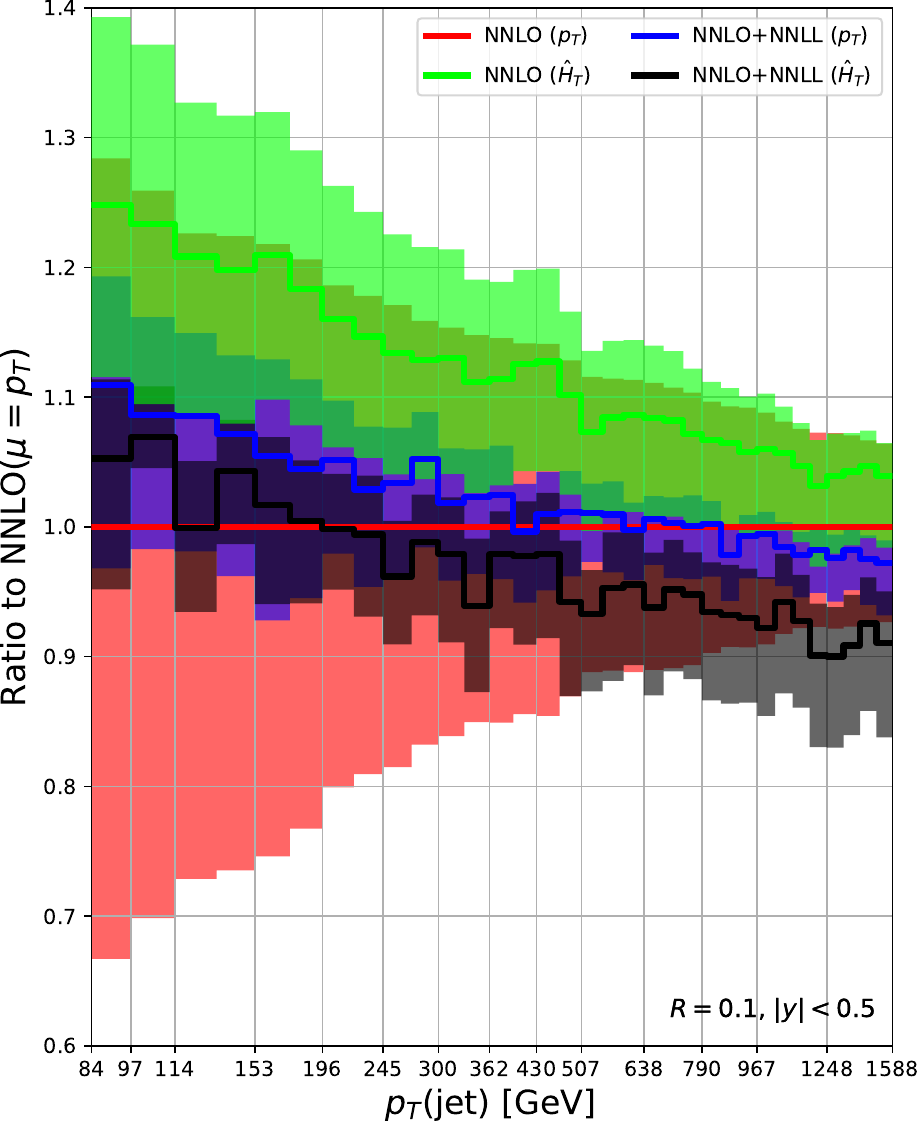}
\includegraphics[width=0.49\textwidth]{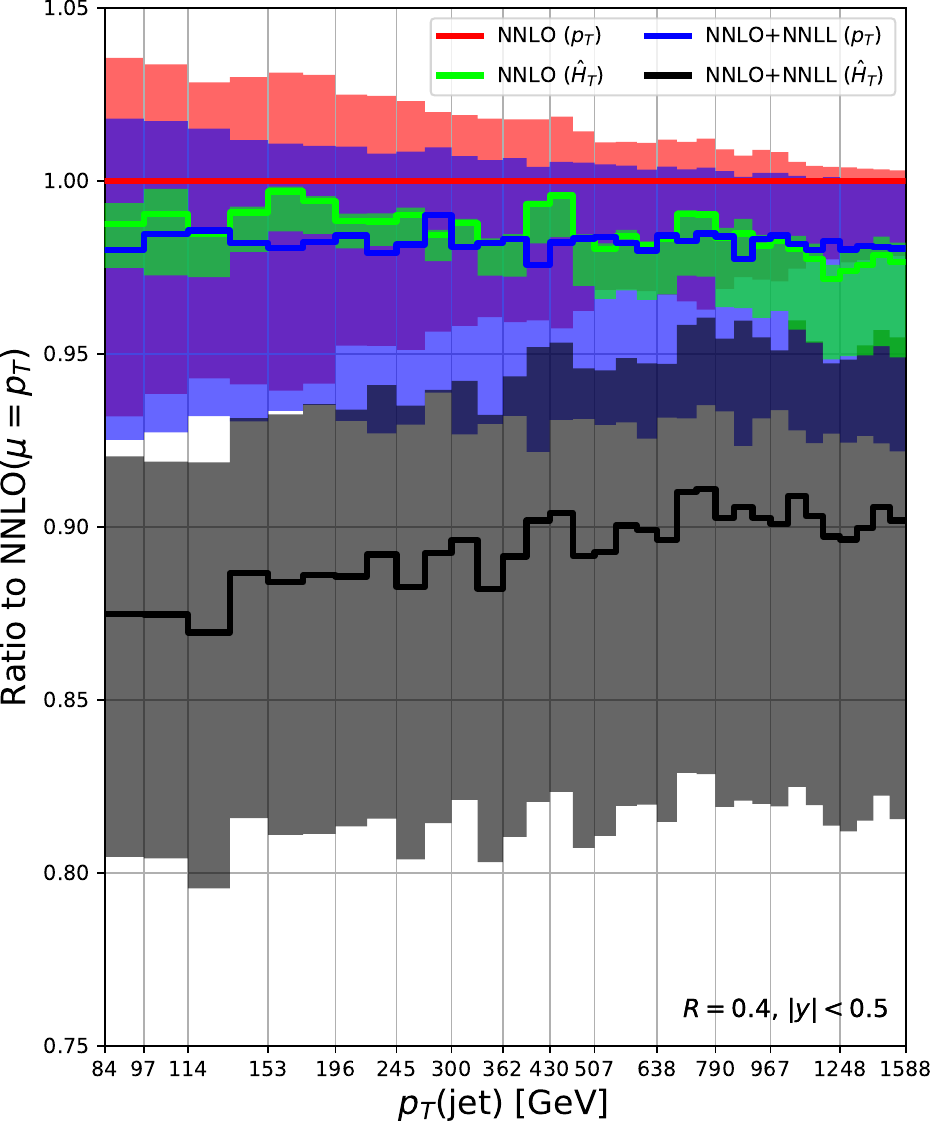}
\includegraphics[width=0.49\textwidth]{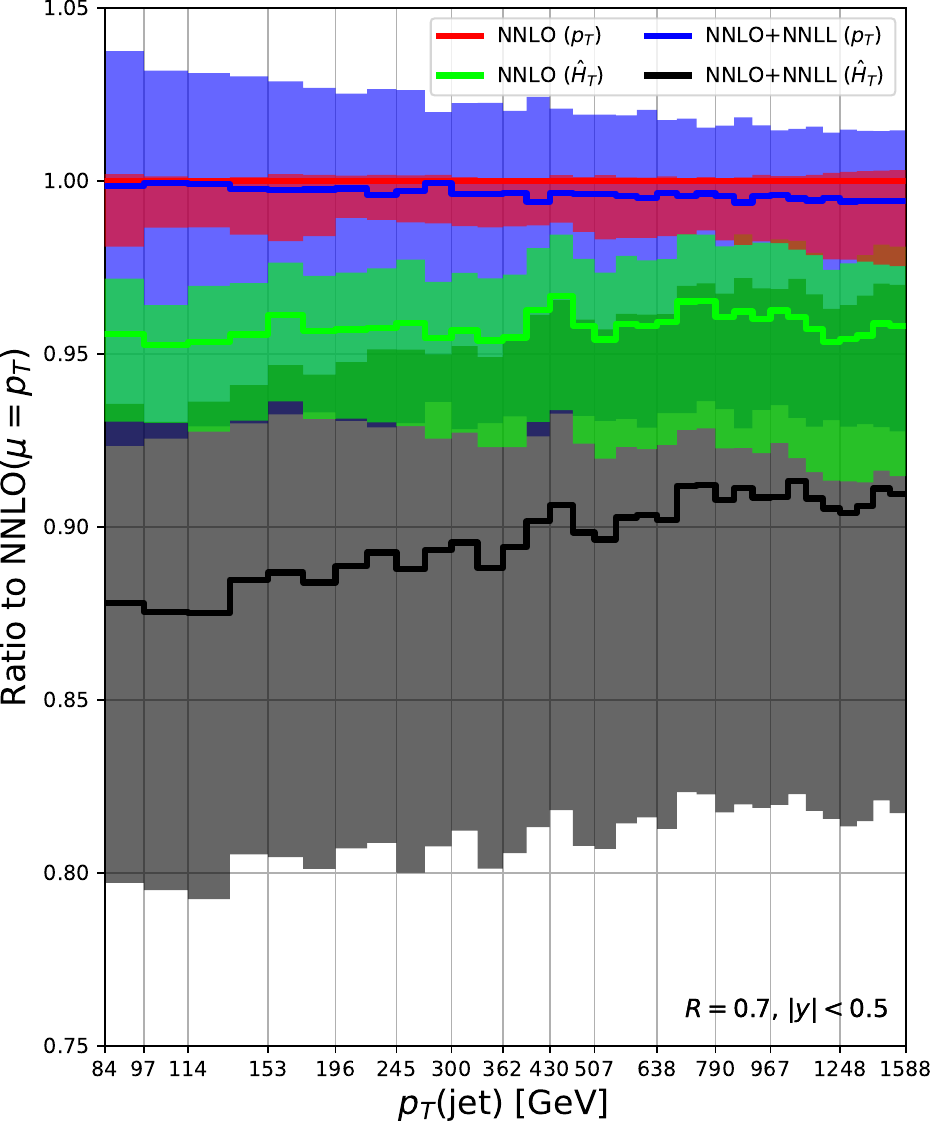}
\includegraphics[width=0.49\textwidth]{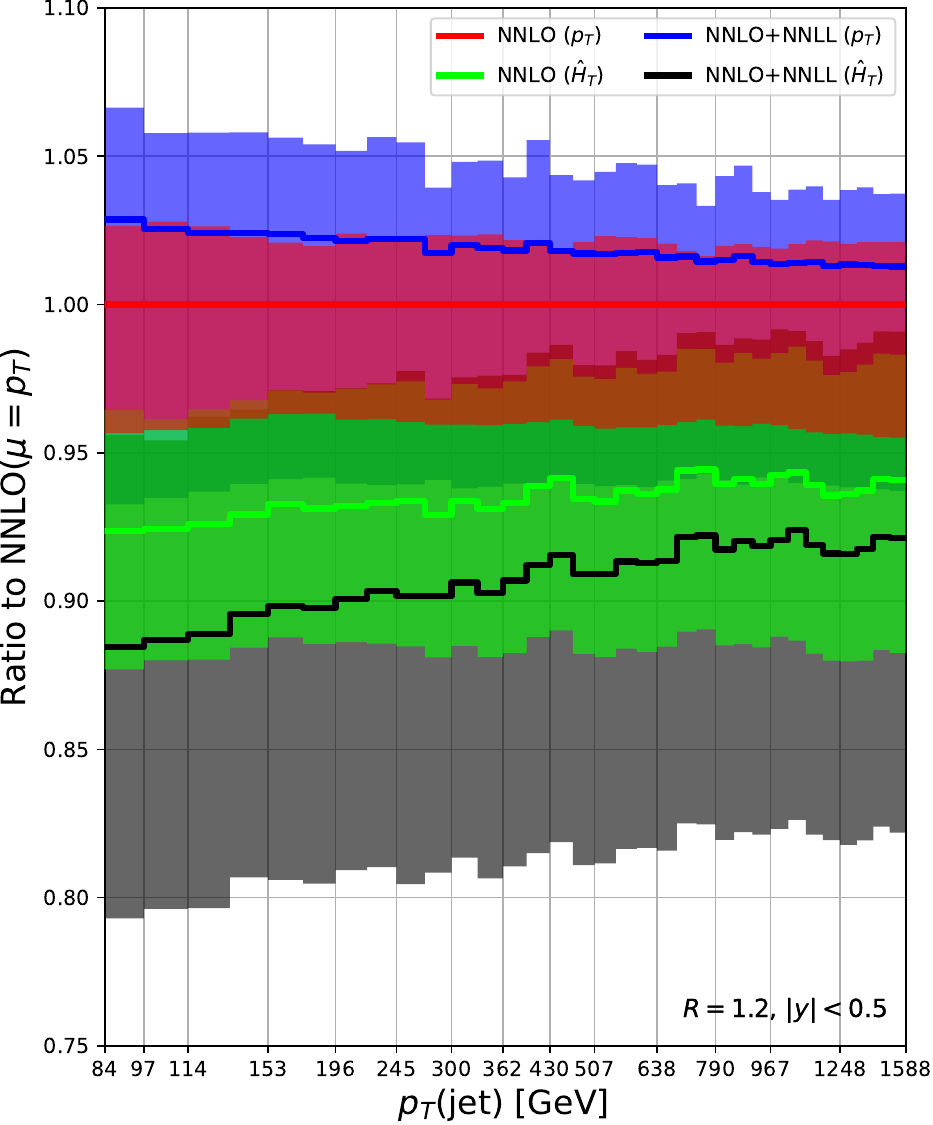}
\caption{A comparison of the NNLO and NNLO+NNLL predictions for the absolute spectrum for $R=0.1$ (top left), $R=0.4$ (top right), $R=0.7$ (bottom left), $R=1.2$ (bottom right) for both scale choices.}\label{fig:Comparison}
\end{figure}

For small $R$, where the corresponding logarithms are large, fixed-order perturbation theory is expected to fail.
This breakdown can indeed be seen in the top panels of Figure~\ref{fig:AbsoluteR0.1} for $ R=0.1$, for both central scale choices: the corrections are large and negative, and are typically outside the estimated uncertainty.
The reason for the large perturbative corrections lies in out-of-cone emissions or in jet broadening due to hard gluon emissions, with a sizeable fraction of the jet energy outside the jet radius.
Additionally, this effect leads to incomplete cancellation with the virtual contributions, which are independent of the jet radius.
This mis-cancellation continues even when going from NLO to NNLO, as the NNLO prediction provides a better description of the jet shape, and thus of the hard gluon emissions.
Including the small-$R$ resummation, we see that the perturbative series is stabilized (second panel from the top).
The NNLO scale dependence is very different between the two scale choices, specifically, the $\ptj$ envelope is much larger than the $\hthat$ one.
The differences between the scale choices also become apparent when comparing NLO and NLO+NLL, as well as NNLO and NNLO+NNLL, directly in the lowest panels of Figure~\ref{fig:AbsoluteR0.1}.
For $\ptj$, we see that the resummation corrections are small in general and only increase the scale-variation envelope.
For $\hthat$, the corrections from resummation are similar (albeit negative) but now lie outside the fixed order uncertainty band, which is much smaller.
It becomes even clearer in the top-left panel of Figure~\ref{fig:Comparison}, where the NNLO and NNLO+NNLL predictions using both scales are directly compared.
The differences between fixed-order and resummed predictions decrease with increasing perturbative order as expected.
Notably, we see that at fixed order, predictions with different scale choices lie within their respective MHOU estimates, but just so.
While the resummed predictions are much closer to each other, their MHOU estimates also barely capture the differences.
The smaller differences in the resummed case are consistent with the notion that, at small $R$, resummation effects dominate.

For larger radii, i.e., more inclusive phase spaces, we expect fixed-order perturbation theory to be more reliable; however, as pointed out in Ref.~\cite{Bellm:2019yyh}, accidental cancellations in the phenomenologically most relevant region $R \sim 0.4-0.7$ might lead to an artificially reduced sensitivity to the scale and thus to underestimated MHOU estimates.
As an example, we show in Figure~\ref{fig:AbsoluteR0.4} the predictions for a jet size of $R=0.4$.
Since $R=0.4$ is probably the most common jet size used for physics applications at the LHC, it is most important to understand the phenomenology of jet predictions in this case.
The cross-section corrections from LO to NLO to NNLO are less severe for this jet size, as expected.
The uncertainty estimates drop quickly from LO to NLO to NNLO for both scale choices.
At NNLO, the estimated uncertainty is very small, of order 1\%, specifically with the $\hthat$ scale, due to the aforementioned accidental cancellations.
The NNLL resummation results show a larger, perhaps more physically reasonable (and more constant as a function of jet $\ptj$), uncertainty for both scale choices.

At NLO for the $\ptj$ scale, we find that the fixed-order prediction is about $10\%$ higher than NLO+NLL but is within the estimated uncertainty of the NLO+NLL prediction, whose uncertainty is a factor $3-10$ larger than the fixed-order one.
For $\hthat$ the impact of the resummation is even larger, about $-20\%$, which leads to a significant difference between the two predictions measured by their respective uncertainty estimates.
The NNLO+NNLL prediction is within the NNLO uncertainty band (and the central values are very similar) for a scale of $\ptj$, but is outside that band for a scale of $\hthat$; however, they are generally closer than at NLO and NLO+NLL.
The uncertainty band at fixed order also shows a strong $\ptj$ dependence; in particular, for the scale of $\ptj$, we see a continuous decrease up to the highest investigated $\ptj$ bin.
For the $\hthat$ scale, we see a slight increase in the scale dependence as we approach higher $\ptj$ values.

The results for $R=0.7$ in Figure~\ref{fig:AbsoluteR0.7} are somewhat less alarming.
There is reasonable agreement among LO, NLO, and NNLO predictions, and a reasonable overlap in the uncertainty bands at fixed order, consistent with the expectation that a more inclusive phase space captures more of the emissions and leads to better cancellation between real and virtual contributions.
The NNLO corrections and MHOU uncertainty estimates are quite small, only about $\pm 2\%$. However, directly comparing the fixed-order predictions (see Figure~\ref{fig:Comparison}) reveals that the predictions for the two scale choices do not overlap within their respective uncertainties at NNLO.
For a scale of $\ptj$, there is again an increase in the scale uncertainty in going from NNLO to NNLO+NNLL to about $\pm (5-8)\%$.
This observation is true to a somewhat greater extent for the results with a scale of $\hthat$, where, again, there is a sizeable difference between the central values for NNLO and NNLO+NNLL, and the MHOU estimate at NNLO+NNLL is about 10\%.
Despite the large correction in the $\hthat$ case, the agreement between the two predictions at NNLO+NNLL is a bit better than at fixed order, as the uncertainty bands at least overlap.

The trends observed in Figure~\ref{fig:AbsoluteR0.4} and Figure~\ref{fig:AbsoluteR0.7} continue in Figure~\ref{fig:AbsoluteR1.2}.
With a jet radius of 1.2, essentially all radiation is captured within the jet cone.
At fixed order, the predicted cross-section increases with the perturbative order for both scale choices.
The impact of the resummation is an increase in the cross section from NNLO for $\ptj$ and a somewhat larger decrease for $\hthat$.
Also, the remaining scale dependence is larger for $\hthat$ than for $\ptj$ by a similar margin.

A breakdown of the scale bands of the resummed predictions into separate variations of the $\mur$, $\muf$, and $\muj$ scale is provided in the Appendix~\ref{app:scales}.
A detailed discussion is provided in the appendix as well; in summary, the insights from the breakdown are that the increase in scale dependence for $R=0.4$ to $0.7$ arises from the variation in $\muj$.
In the regions where the fixed-order scale variations are large, we see that the $\mur$ and $\muj$ scales contribute similarly to the band.

\begin{figure}
\centering
\includegraphics[width=0.49\textwidth]{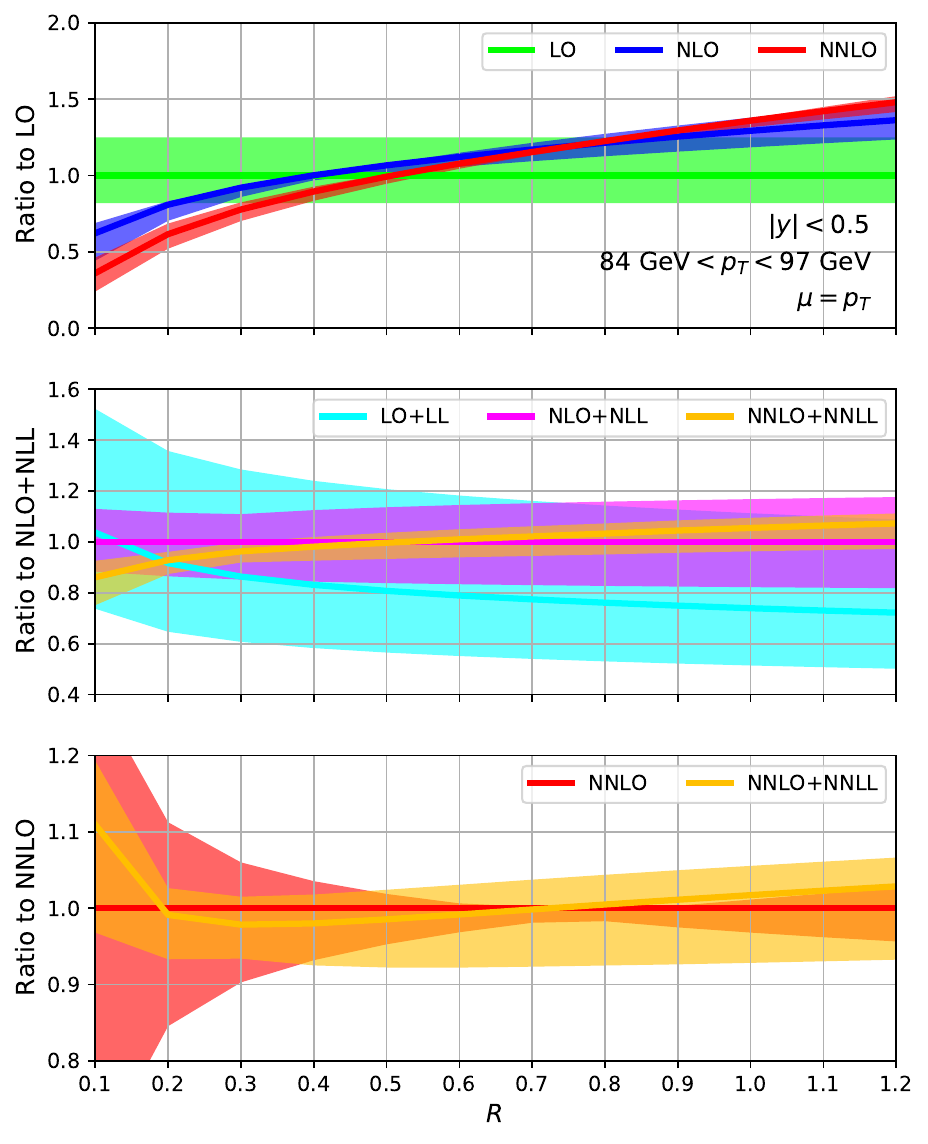}
\includegraphics[width=0.49\textwidth]{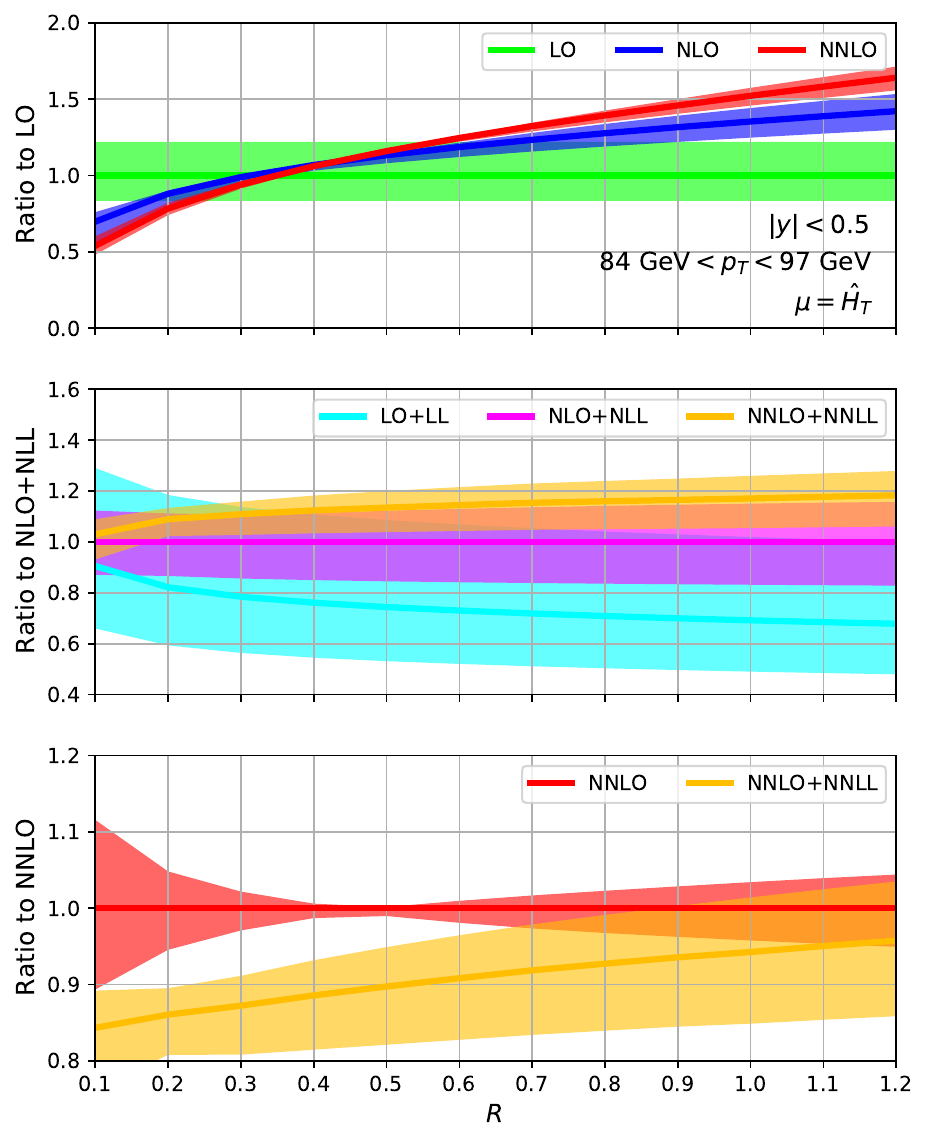}
\caption{The $R$-dependence of the absolute cross section for the lowest rapidity and lowest $\ptj$ bin for $\mu=\ptj$ (left) and $\mu=\hthat$ (right).}\label{fig:R-dependence1}
\end{figure}

\begin{figure}
\centering
\includegraphics[width=0.49\textwidth]{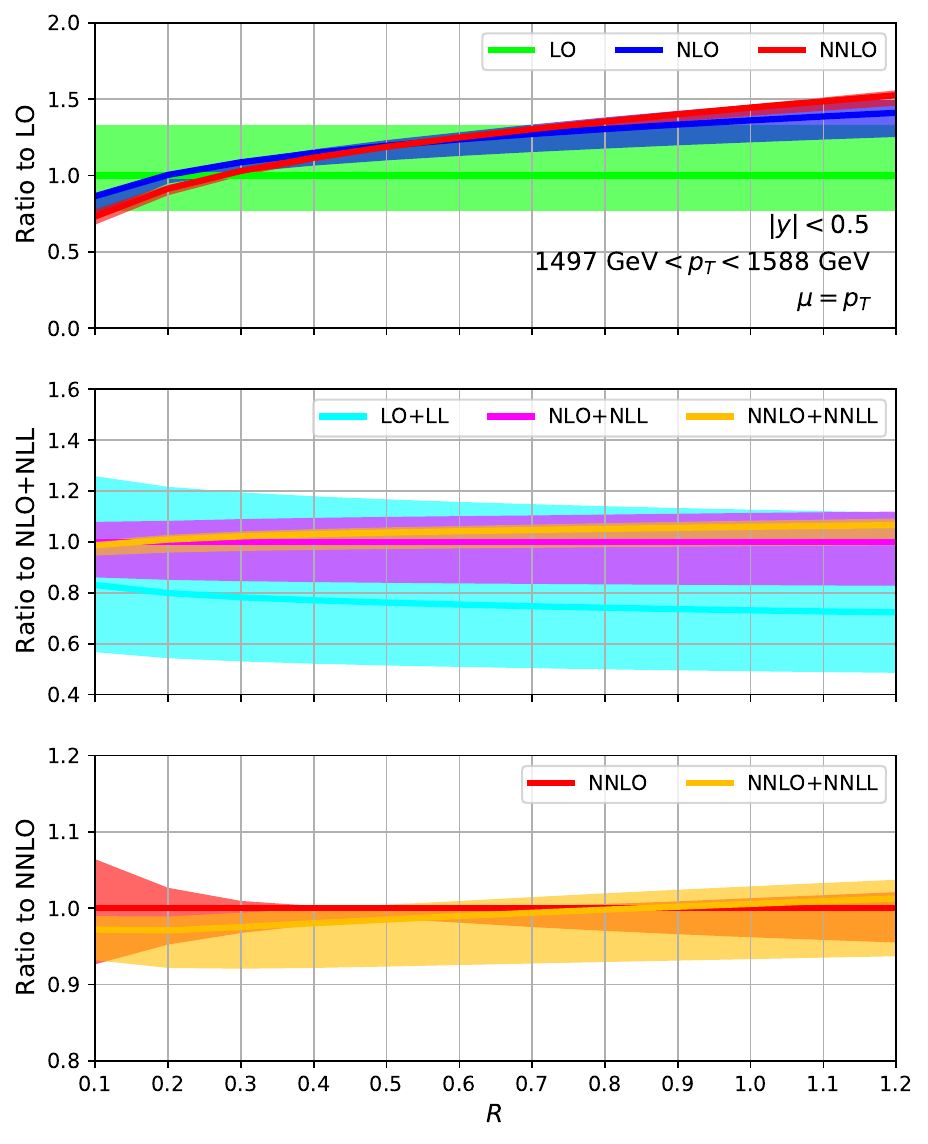}
\includegraphics[width=0.49\textwidth]{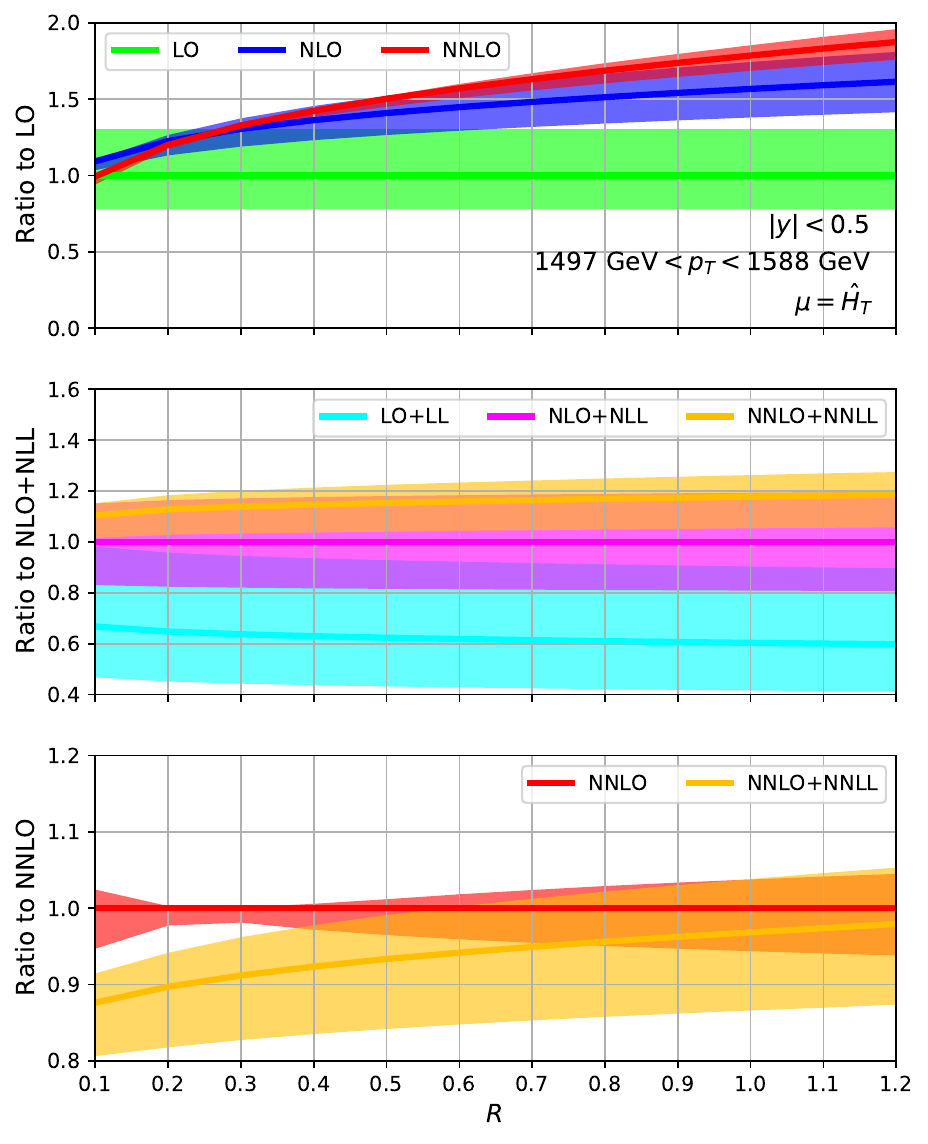}
\caption{The $R$-dependence of the absolute cross section for the lowest rapidity and highest $\ptj$ bin for $\mu=\ptj$ (left) and $\mu=\hthat$ (right).}\label{fig:R-dependence2}
\end{figure}

\begin{figure}
\centering
\includegraphics[width=0.49\textwidth]{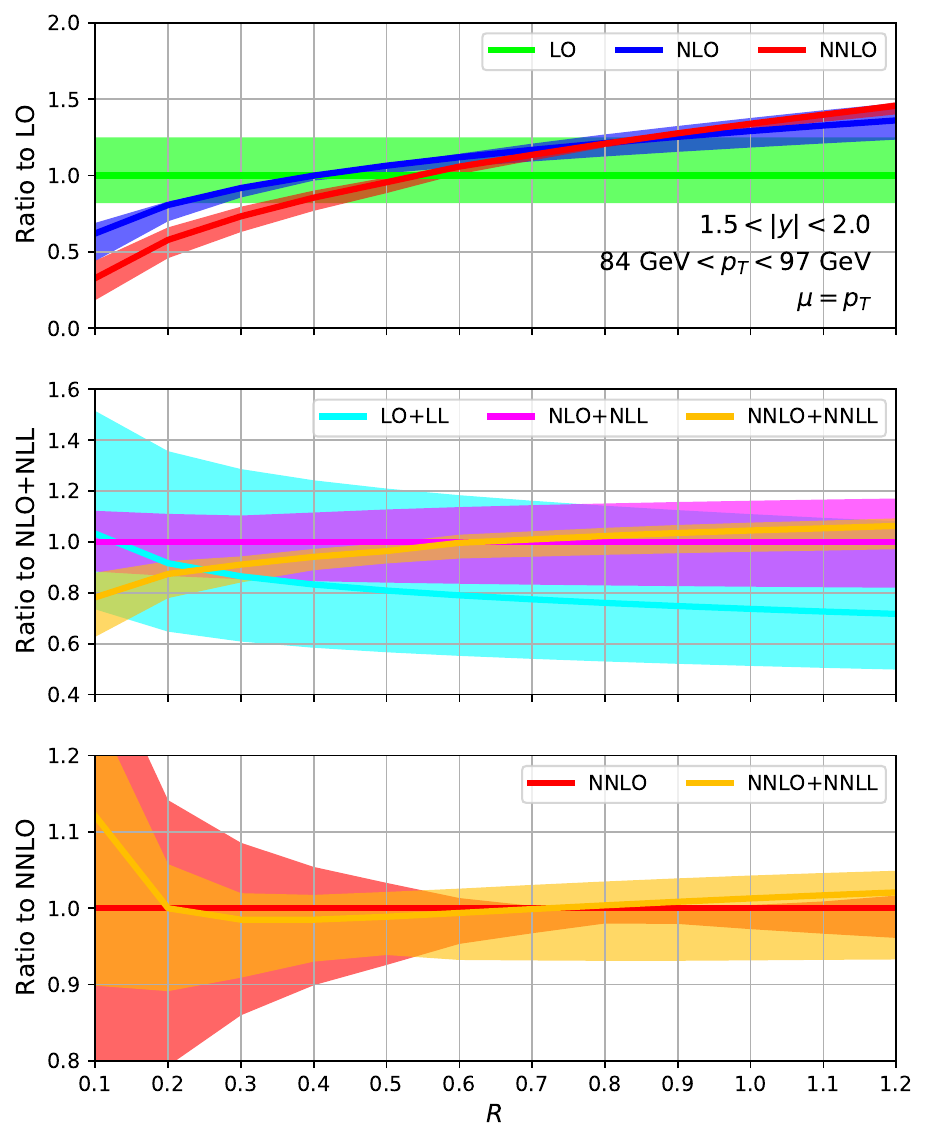}
\includegraphics[width=0.49\textwidth]{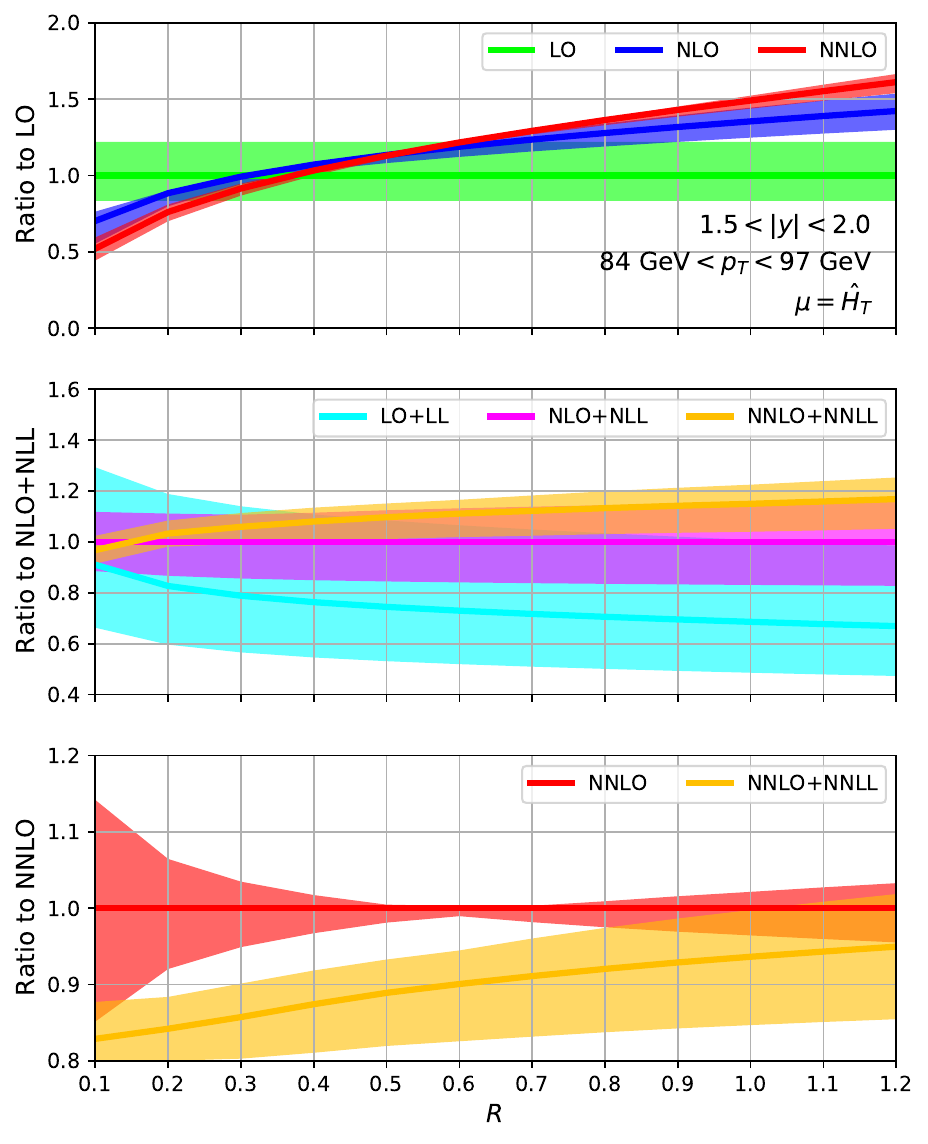}
\caption{The $R$-dependence of the absolute cross section for the highest rapidity and lowest $\ptj$ bin for $\mu=\ptj$ (left) and $\mu=\hthat$ (right).}\label{fig:R-dependence3}
\end{figure}

\begin{figure}
\centering
\includegraphics[width=0.49\textwidth]{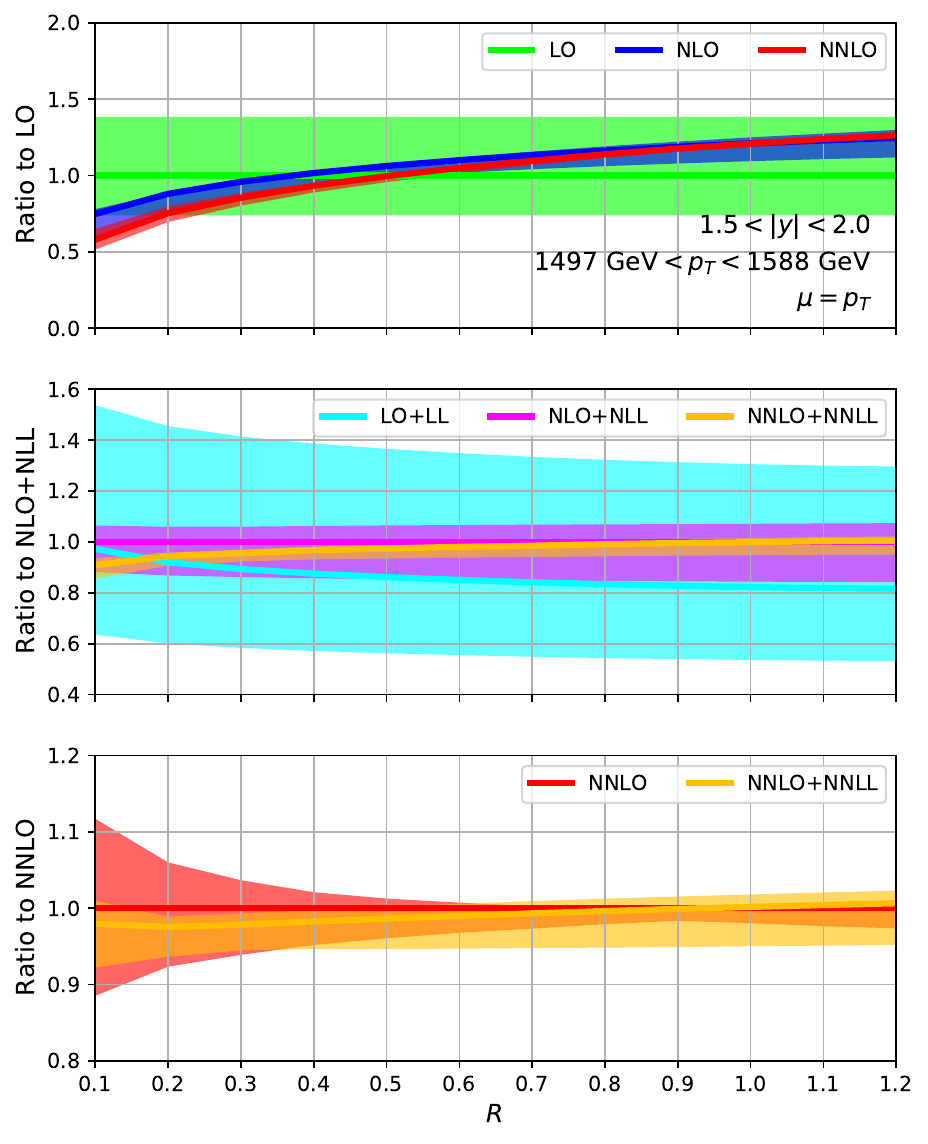}
\includegraphics[width=0.49\textwidth]{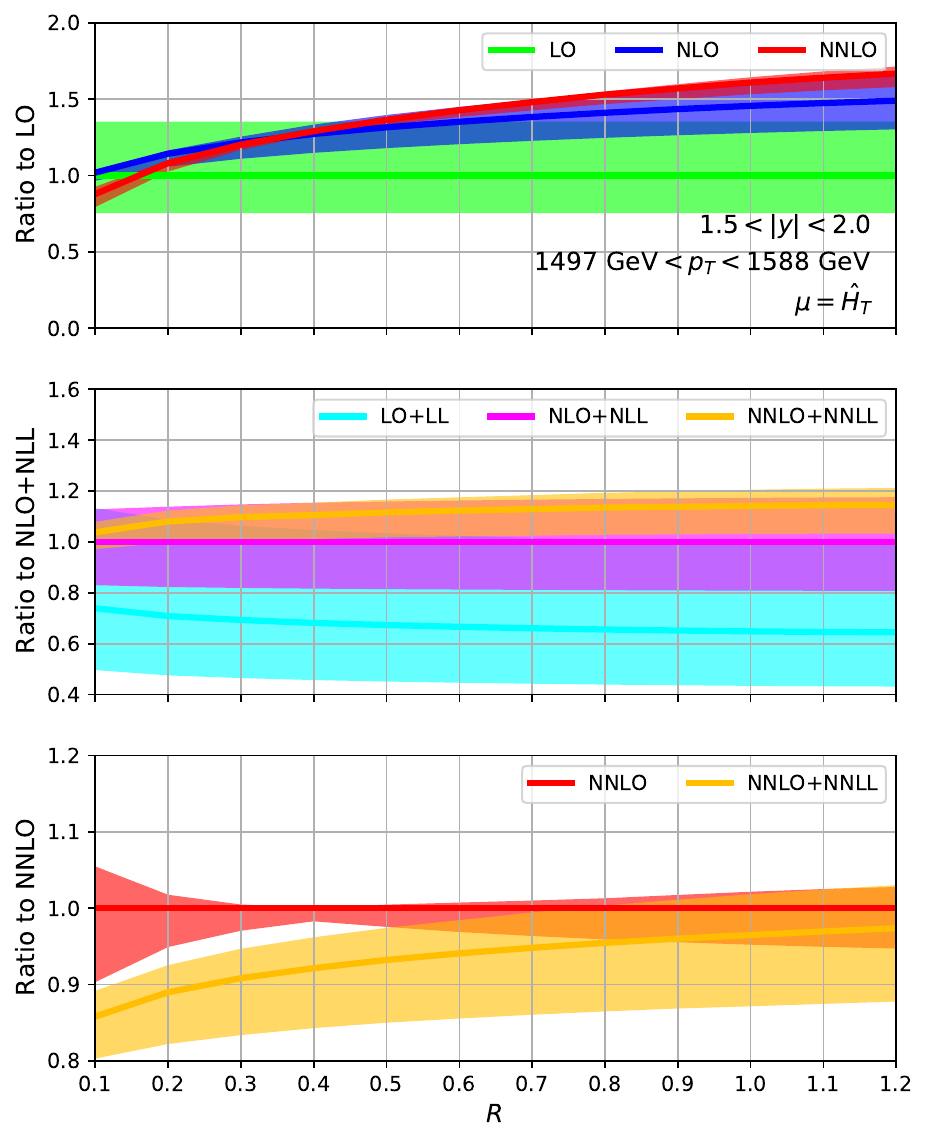}
\caption{The $R$-dependence of the absolute cross section for the highest rapidity and highest $\ptj$ bin for $\mu=\ptj$ (left) and $\mu=\hthat$ (right).}\label{fig:R-dependence4}
\end{figure}

Figures~\ref{fig:R-dependence1} to Figure~\ref{fig:R-dependence4} show the $R$-dependence of the absolute cross sections for the ``four corners'' of the phase space defined by the lowest and highest rapidity and transverse momentum bins.
The fixed-order predictions (shown in the top panels) exhibit a peculiar behavior, common to both scales, in which the NNLO predictions are typically well below the NLO prediction and well outside the estimated uncertainty at small $R$. In contrast, for large $R$, we observe the opposite: the NNLO corrections are typically positive and again typically outside the estimated NLO uncertainty.
Consequently, there is a region for medium $R$, where we find perturbative convergence at fixed order in the sense that the NNLO predictions lie within the NLO uncertainty bands.
The region where this happens depends on the scale choice, though; typically, it is at a higher $R$ for the $\hthat$ scale than for the $\ptj$ scale.
However, the phenomenologically most relevant $R$ range, $0.4$ to $0.7$, overlaps that area for both scales.
The resummed predictions demonstrate good perturbative convergence in all regions, shown in the panel second from the top of  Figure~\ref{fig:R-dependence1} to \ref{fig:R-dependence4}.
The $\hthat$ scale typically yields a larger remaining scale dependence, particularly for larger $R$.
The only region where one can see some tension is in the high-rapidity, low transverse momentum, small R limit, where, for the $\ptj$ scale, we see NNLO+NNLL corrections slightly outside the NLO+NLL uncertainty band.

A direct comparison between the fixed-order and resummed prediction in the lowest panels in Figure~\ref{fig:R-dependence1} to Figure~\ref{fig:R-dependence4} reveals another interesting observation.
For the $\ptj$ scale, we find overall good agreement between the resummed and fixed-order central values; the differences lie mostly in the much larger scale dependence for the resummed case.
For the $\hthat$ scale, on the other hand, we see again the large shift due to resummation, which appears independent of the $R$ value and phase space region.

\begin{figure}[t]
\includegraphics[width=0.49\textwidth]{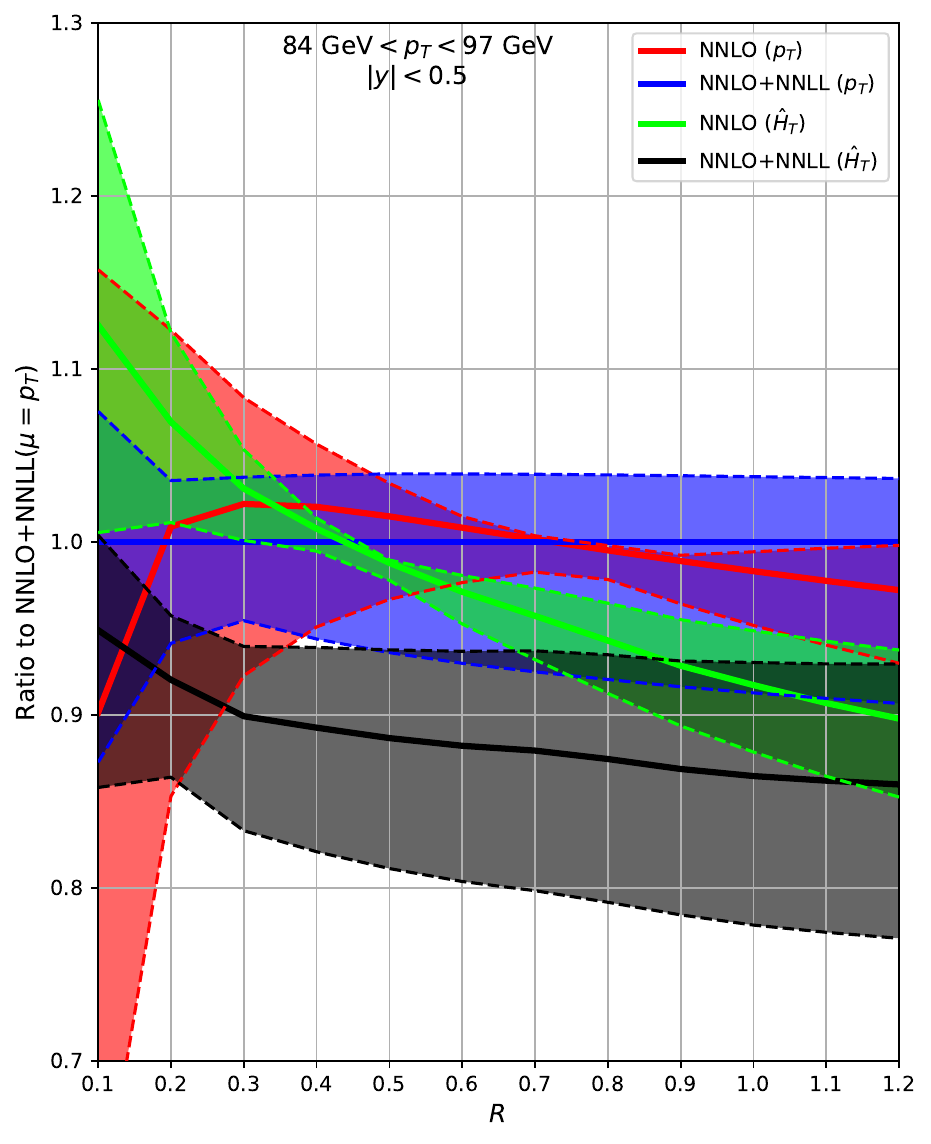}
\includegraphics[width=0.49\textwidth]{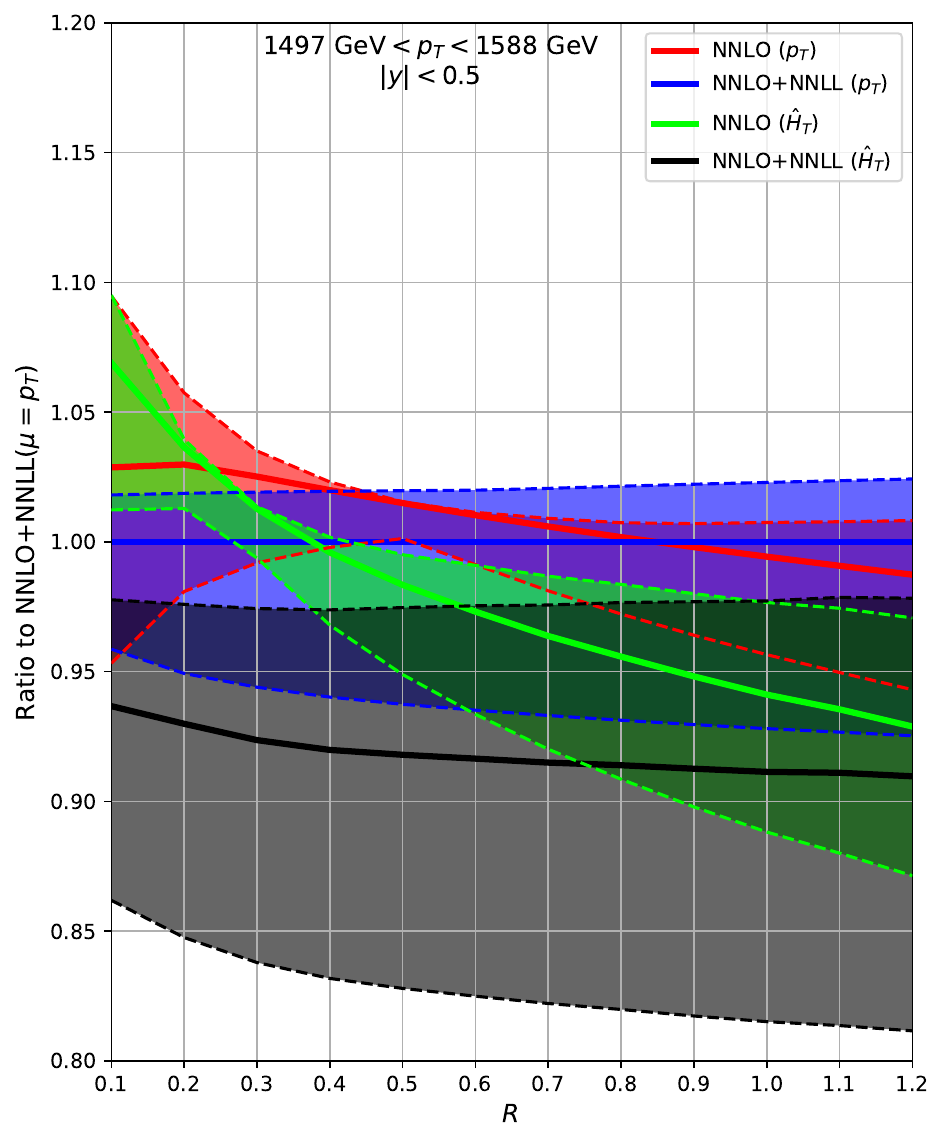}
\includegraphics[width=0.49\textwidth]{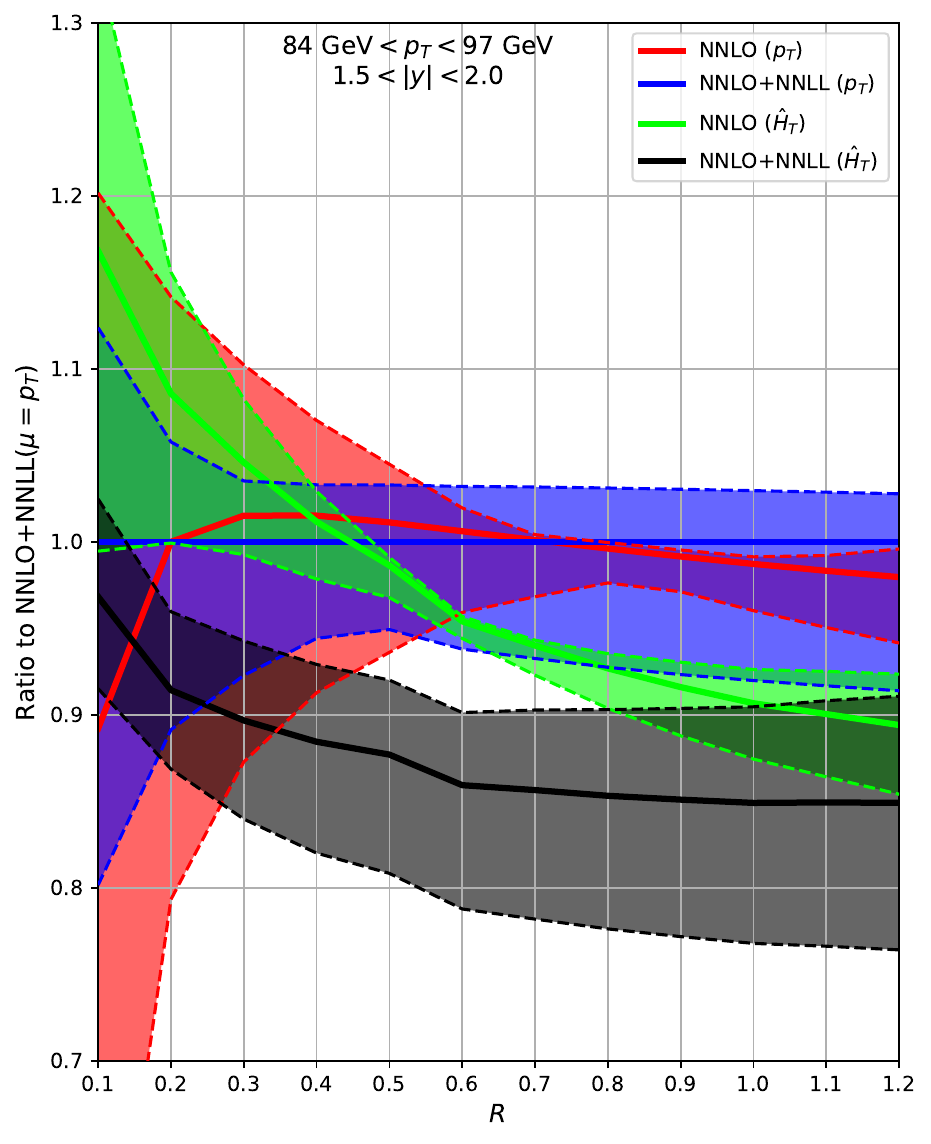}
\includegraphics[width=0.49\textwidth]{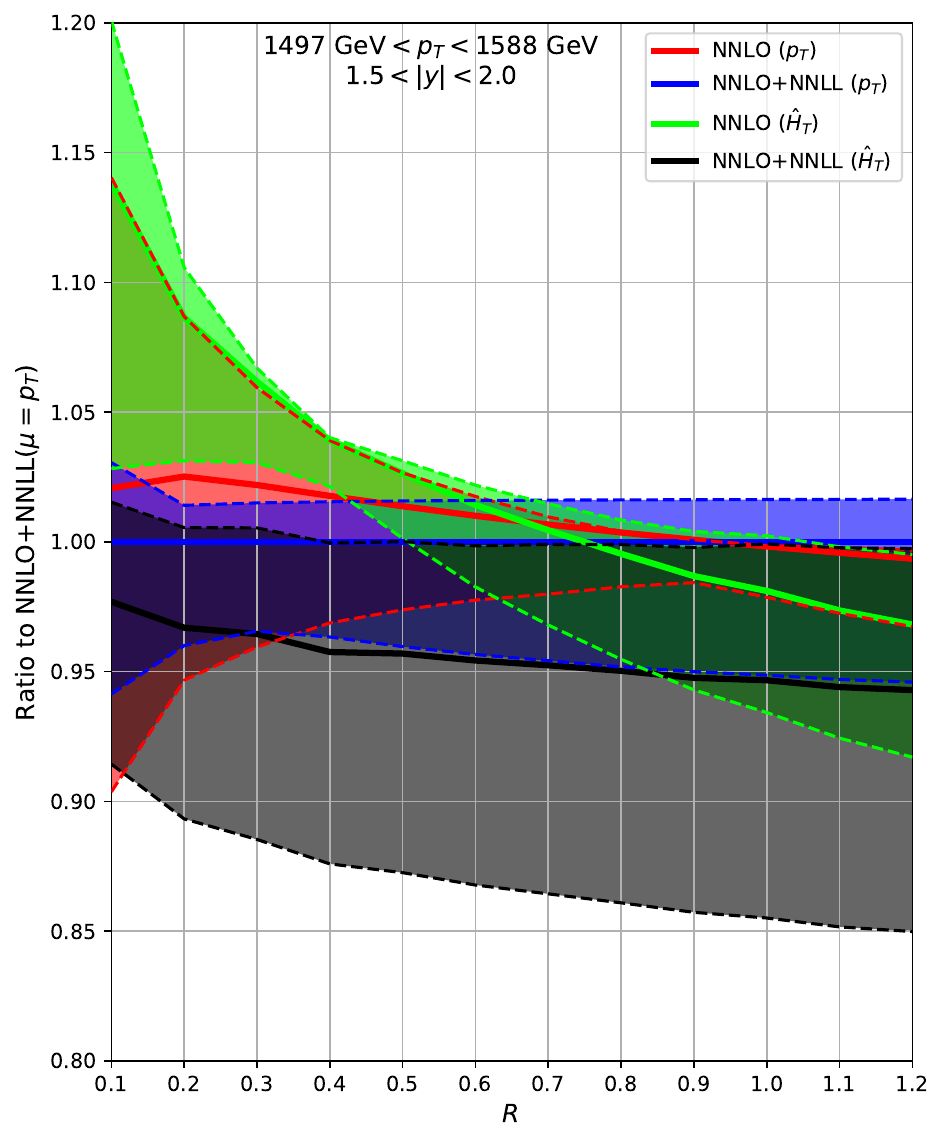}
\caption{A comparison of the $R$-dependence of the NNLO and NNLO+NNLL predictions for the absolute spectrum for the most extreme kinematics of our setup.}\label{fig:ComparisonR}
\end{figure}

A direct comparison between the NNLO and NNLO+NNLL predictions for different scale choices is provided in Figure~\ref{fig:ComparisonR}, again for the most extreme kinematic bins.
At fixed order, we can observe the strong $R$ dependence of the scale variation, which, depending on the scale choice and kinematic region, shows a minimum in the range $R\sim0.4-0.9$.
For $R$ values with very small scale dependence, we generally find that the predictions for the two scale choices do not agree within their respective uncertainties, except for the high transverse momentum and large rapidity region.
The difference between the two resummed predictions is, as observed before, about $10\%$, but given the larger uncertainty, they are largely in agreement.
The only exception is again the low $\ptj$, high rapidity region, where the differences exceed the uncertainty estimates substantially.

To summarize the observation from the absolute spectra: fixed-order scale variations tend to be unphysically small for phenomenologically important $R$ values; the resummed predictions partially rectify this. But even for the resummed predictions, we see substantial differences in the corners of the phase spaces not covered by the enlarged scale dependence.
The absolute predictions for $\hthat$ have a substantial downward resummation correction, which will cancel out in the ratio for cross sections with different $R$, see next section.
The downward correction is partially captured by the uncertainty band of the resummed $\ptj$ scale-based predictions, but, as before, we find this an insufficient uncertainty estimate.

\begin{figure}
\includegraphics[width=0.49\textwidth]{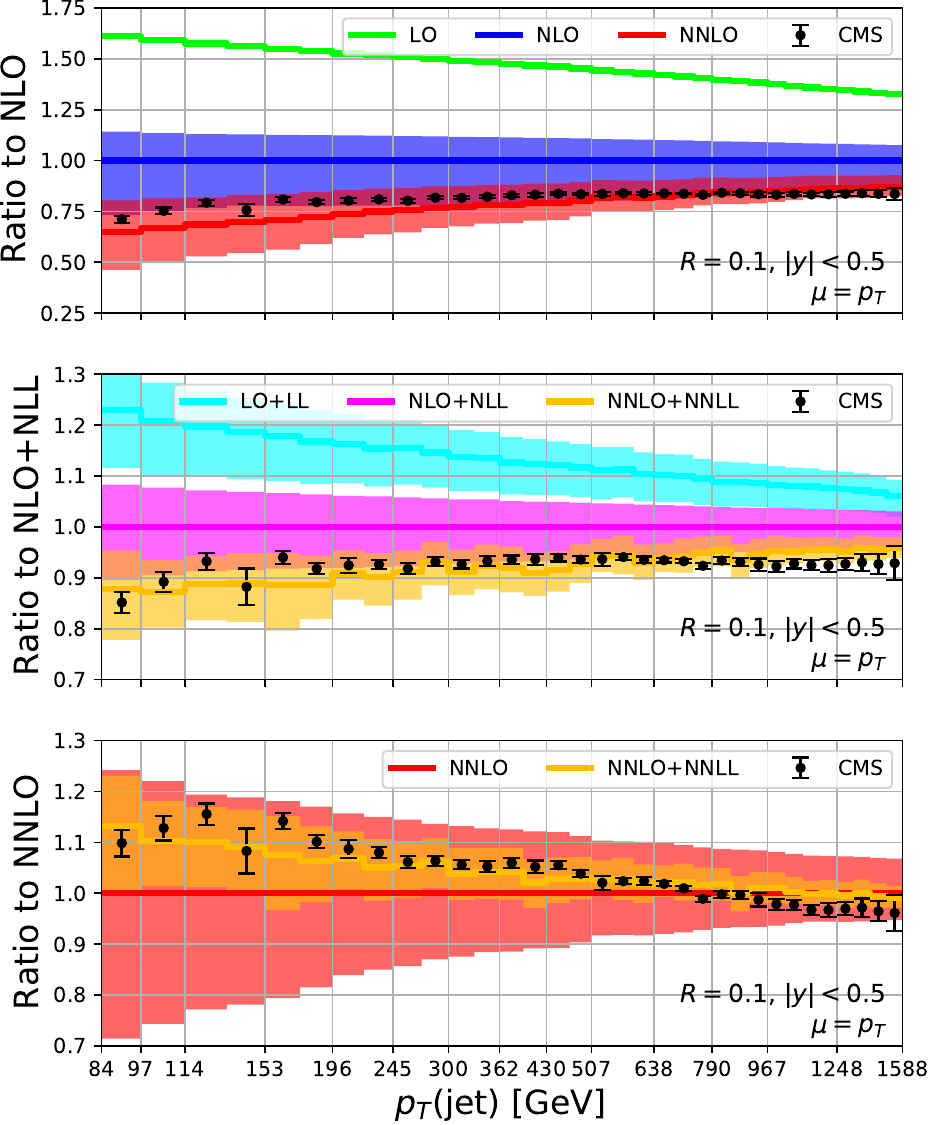}
\includegraphics[width=0.49\textwidth]{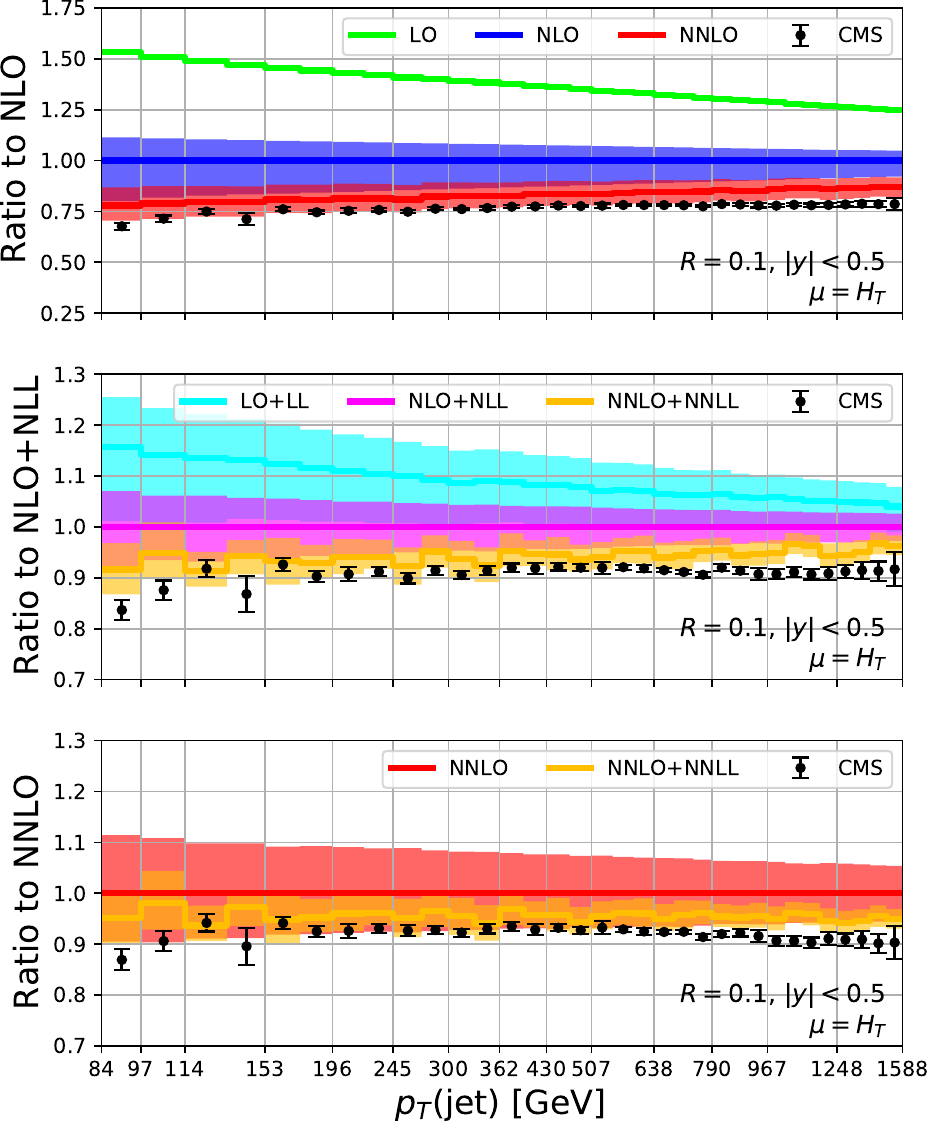}
\caption{Comparison of theory predictions for the ratio between $R=0.1$ and $R=0.4$. Shown are predictions for $\mu=\ptj$ (left) and for $\mu=\hthat$ (right).}
\label{fig:RatioR0.1}
\end{figure}

\subsection{Ratios and comparison to data}\label{sec:pheno-ratios}

\begin{figure}[t]
\includegraphics[width=0.49\textwidth]{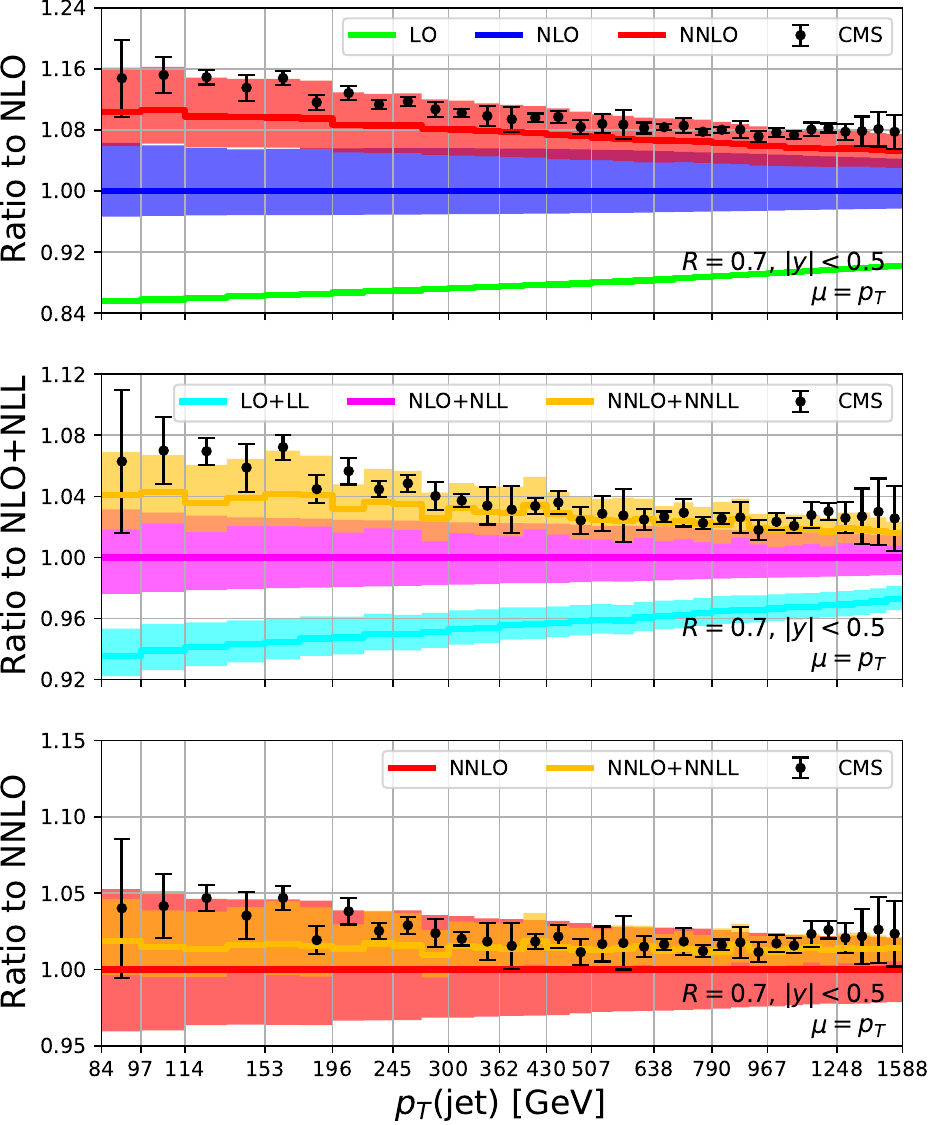}
\includegraphics[width=0.49\textwidth]{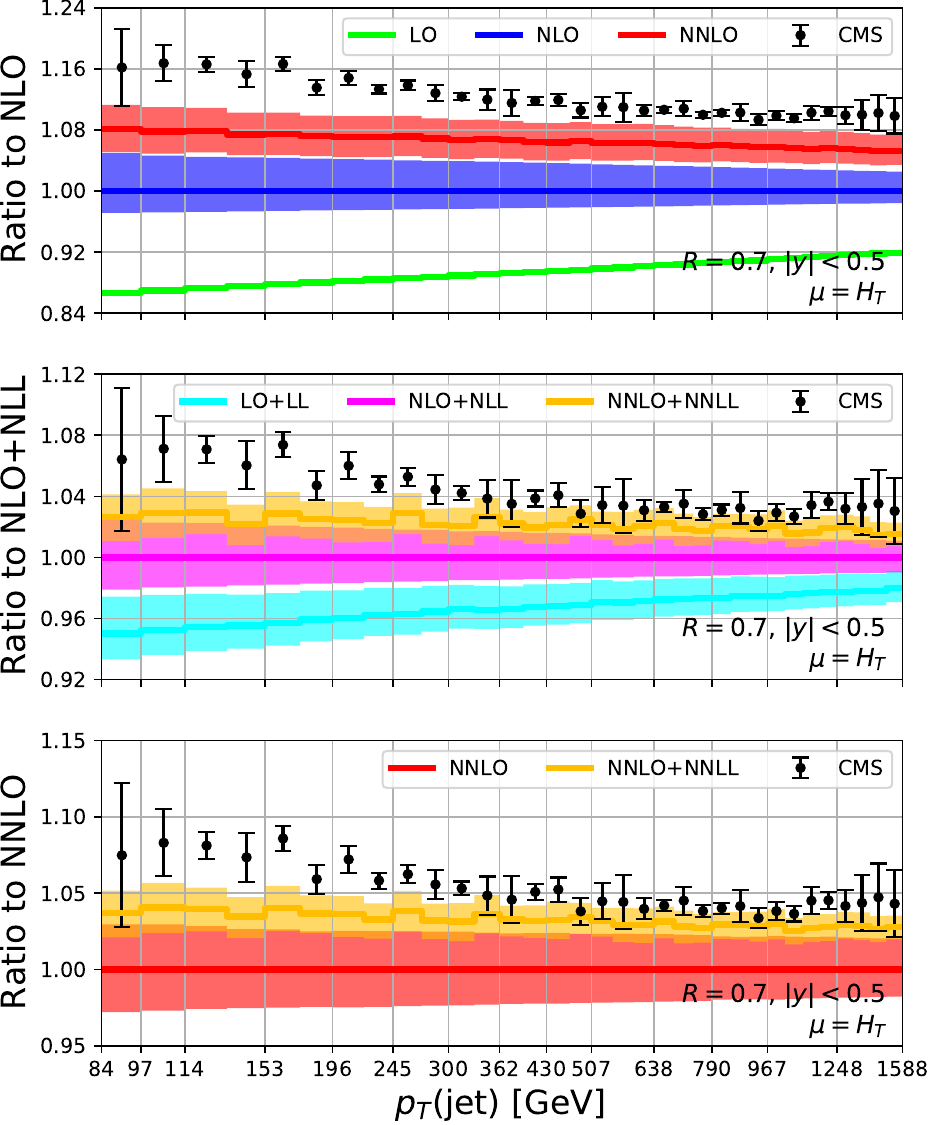}
\caption{As in Figure~\ref{fig:RatioR0.1}, but for the ratio between $R=0.7$ and $R=0.4$.}
\label{fig:RatioR0.7}
\end{figure}

\begin{figure}[t]
\includegraphics[width=0.49\textwidth]{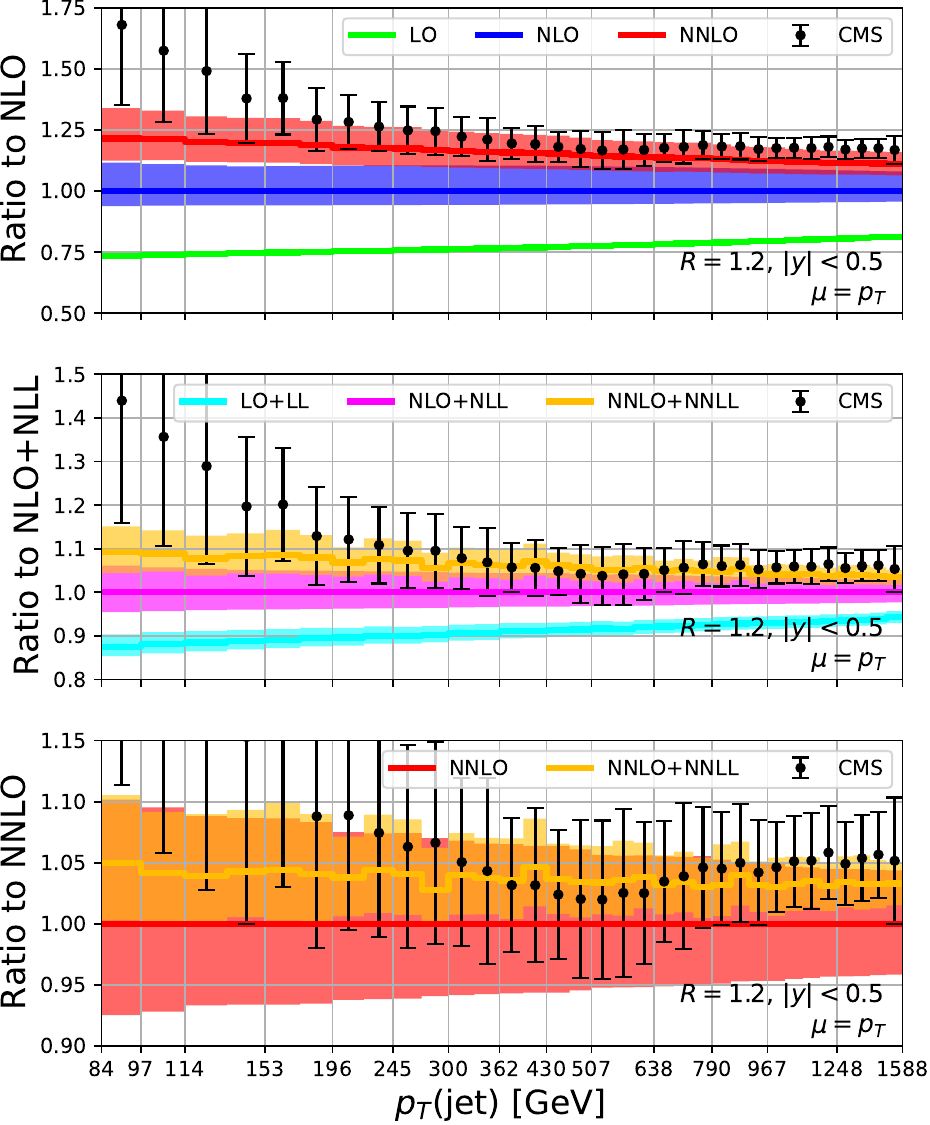}
\includegraphics[width=0.49\textwidth]{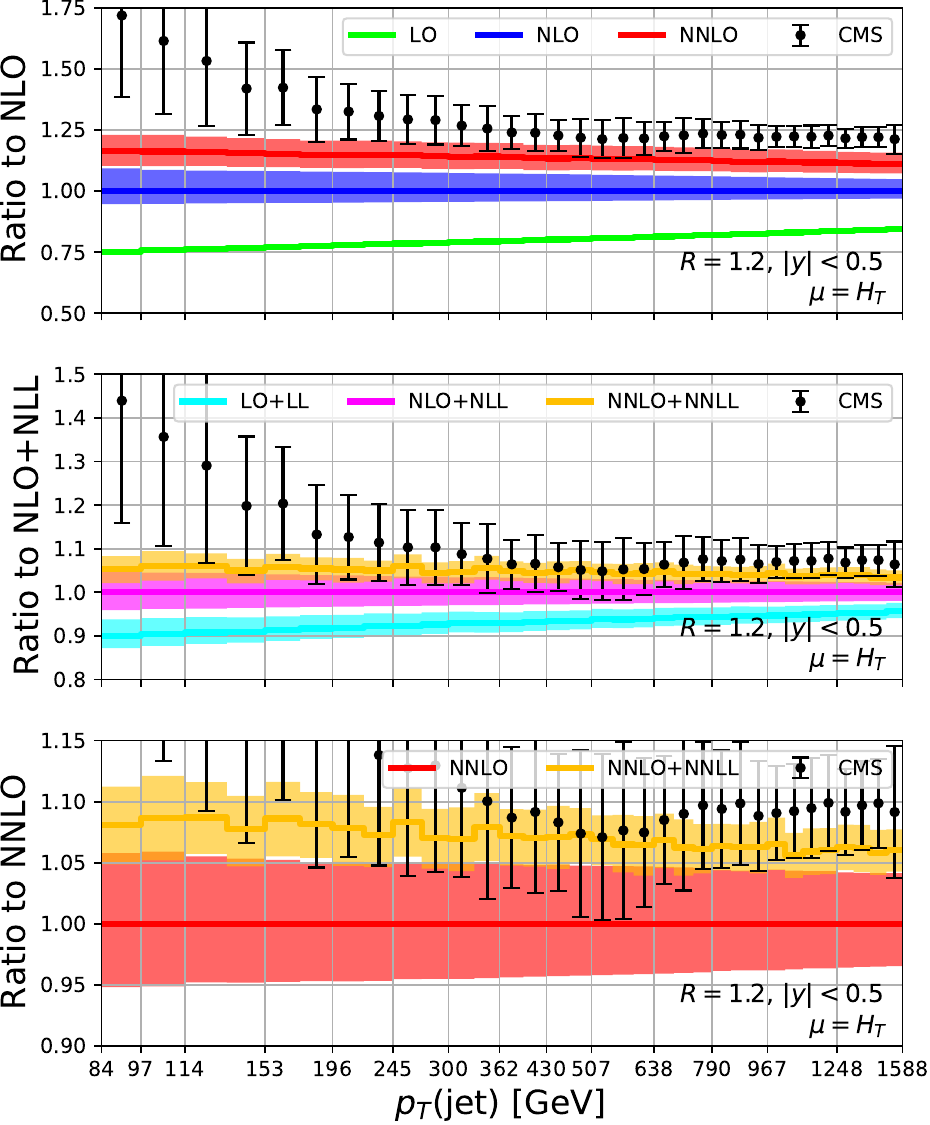}
\caption{As in Figure~\ref{fig:RatioR0.1}, but for the ratio between $R=1.2$ and $R=0.4$.}
\label{fig:RatioR1.2}
\end{figure}

In Figure~\ref{fig:RatioR0.1} to Figure~\ref{fig:RatioR1.2}, we show the ratio of the $R=0.1,0.7,1.2$ spectra with respect to $R=0.4$ with non-perturbative corrections\footnote{
with some modifications to the CMS factors as discussed in Appendix B of Ref.~\cite{Generet:2025vth}.
}
from CMS applied to all predictions.
Additionally, we present CMS data from the measurement in Ref.~\cite{CMS:2020caw}.

Overall, we rediscover the observations from Ref.~\cite{Generet:2025vth}, which can be summarized as an improved description of the data and a smaller MHOU when small-$R$ resummation is included.
The reduction of the uncertainty for the ratio in the resummed case indicates the strong correlation of the scale dependence between different values of $R$, which is not present in the fixed-order predictions.
The correlation is particularly present for the $\muj$ and $\mur$ scale variations, such that the uncertainties for fixed-order and resummed predictions are both dominated by $\mur$ variations, as can be observed explicitly in Appendix~\ref{app:scales}.
The impact of resummation is, as expected, most striking in the small-$R$ limit.

Focusing on the difference between the $\ptj$ and $\hthat$ scale choice, we see that for the $R=0.1$ ratio, the fixed order predictions do not describe the shape of the data very well, and that the predictions are quite different: NNLO fixed order results for $\hthat$ overshoot the data, $\ptj$ results typically undershoot it. For $\mu=\ptj$, the ratios involving larger $R$ remain below the data at NNLO, and are consistent with the data within the uncertainties, though only barely. For $\mu=\hthat$, the data is not described well by fixed-order NNLO predictions.

\begin{figure}[t]
\centering
\includegraphics[width=0.32\textwidth]{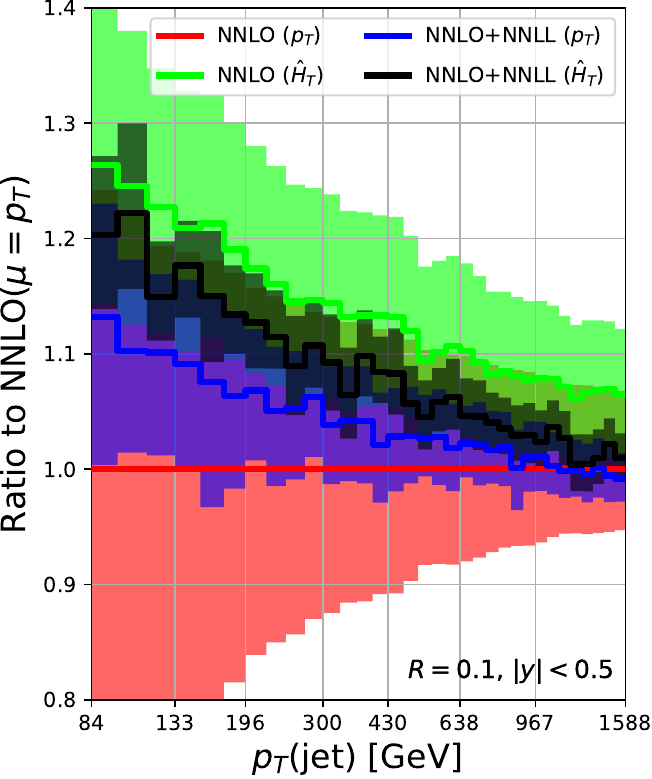}
\includegraphics[width=0.32\textwidth]{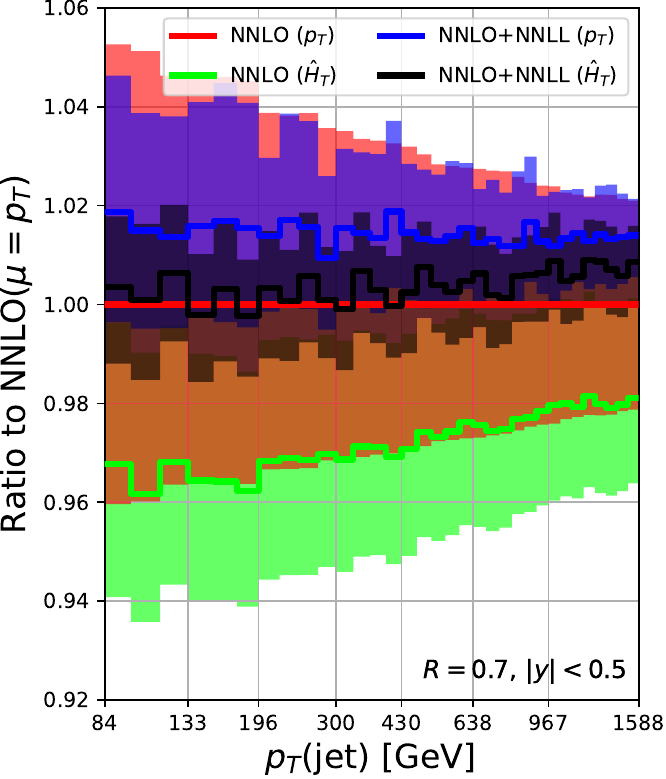}
\includegraphics[width=0.32\textwidth]{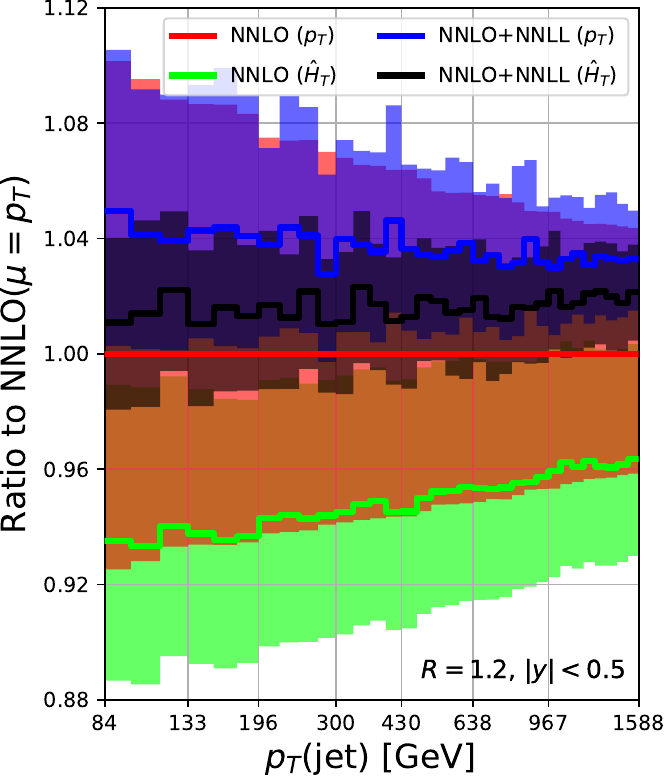}
\caption{A comparison of the NNLO and NNLO+NNLL predictions for the ratio between $R=0.1$ and $R=0.4$ (left), $R=0.7$ and $R=0.4$ (center) and  $R=1.2$ and $R=0.4$ (right) for both scale choices.}\label{fig:RatioComparison}
\end{figure}

Including the resummation brings the results for both scale choices quite close together, as shown in Figure~\ref{fig:RatioComparison}, where the best ratio predictions are directly compared.
For the absolute spectra, we observed a significant shift in the $\hthat$ scale when resummation was included; however, this shift is quite independent of $R$ and therefore cancels out to a large degree in these ratios.
The uncertainties are also drastically reduced by the correlation between numerator and denominator variations, as with the $\ptj$ scale.
For larger $R$ values, we see in the resummed case a slightly lower prediction for the $\hthat$ scale, which increases the overall tension with the data for $\hthat$ compared to $\ptj$.
For the ratio, we also see that the uncertainty estimates for $\hthat$ are roughly a factor of 2 smaller than those for $\ptj$.
Since the $\hthat$-based absolute predictions showed a larger scale dependence, this observation implies a much stronger correlation of the scale variations across different values of $R$ than in the $\ptj$ case.
Despite the improved behavior in the resummed case, for medium $R$ values, we see that the disagreement between the $\hthat$-based predictions and the data is often larger than the estimated uncertainties.

In addition to the data aligning very well with the $\ptj$-based predictions, we have to conclude that the discrepancies are rooted in a remaining underestimation of the MHOU in the $\hthat$ case, even with resummation included.

For the large $R=1.2$ jets, the data has large uncertainties at small transverse momenta; there, the agreement with the NNLO+NNLL prediction is worse in terms of the distribution's shape.
In the same region, non-perturbative corrections are also large and come with substantial uncertainties \cite{CMS:2020caw, Generet:2025vth}, making it difficult to reach a conclusion.

\section{Discussion}\label{sec:discussion}

We presented a thorough analysis of the scale dependence of fixed-order and small-$R$ resummed cross sections for inclusive jet production at the LHC, and compared absolute rapidity and transverse-momentum spectra, as well as ratios of different jet radii.
We studied in particular the reliability of the commonly used scale variation as an estimator of missing higher-order corrections in both cases.

Not only in regions where fixed-order predictions are expected to struggle but across the range of studied $R$ values and phase space, we find an improved perturbative behavior when small-$R$ resummation is included.
Specifically, pathological variation patterns at fixed orders are resolved, leading to an increase in the MHOU estimate for the phenomenologically most relevant $R$ values between $0.4$ and $0.7$.
We highlighted the phenomenological differences between the $\ptj$ and $\hthat$ scale choices in terms of perturbative behavior and resummation.
Crucially, we found that in the $\hthat$ case resummation corrections are substantially larger than in the $\ptj$ case, and that the differences cannot be captured by conventional scale variations, indicating an underestimation of MHOU. The reasons for this are not currently understood.

Additionally, we compared ratios between jet spectra using different jet radii to LHC data.
Overall, we find a better description of LHC data when small-$R$ resummation is included, not only in small-$R$ regions but also at large $R$.
We observe that the differences in the absolute spectra across different central scale choices largely cancel in ratios.
However, when using scale variation as the MHOU estimator, we still find discrepancies between the data and the predictions.

In conclusion, we deem scale variation in the case of fixed-order predictions unreliable to predict higher-order corrections, particularly for phenomenologically relevant $R=0.4$ to $R=0.7$ jets.
The inclusion of small-$R$ resummation improves the situation to some degree and stabilizes the phenomenology of inclusive jet spectra.
However, even in this case, we observe inconsistencies, in particular when directly comparing the $\ptj$ and $\hthat$ scale choices. This analysis has focused on inclusive jet production, but similar effects may well be present in other LHC processes involving jets. A better theoretical understanding is needed to exploit the experimental precision provided by these measurements fully.

Looking ahead, exploring alternative methods to estimate MHOU is desirable and necessary to re-establish confidence in MHOU estimates for inclusive jet production. More broadly, it will also be important to extend the study presented here to more differential jet substructure observables, where the presence of additional scales further complicates the estimation of MHOU using scale variations.

\section*{Acknowledgments}

We acknowledge the hospitality of the \'Ecole de Physique des Houches, where this work was initiated during the 2025 ``Physics at TeV Colliders and Beyond the Standard Model'' program.
This work was performed in part using the Cambridge Service for Data Driven Discovery (CSD3), part of which is operated by the University of Cambridge Research Computing on behalf of the STFC DiRAC HPC Facility (www.dirac.ac.uk). The DiRAC component of CSD3 was supported by STFC grants ST/P002307/1, ST/R002452/1 and ST/R00689X/1. T.G. has been supported by STFC consolidated HEP theory grants ST/T000694/1 and ST/X000664/1. K.L is supported in part by the U.S. Department of Energy under contracts DE-AC02-06CH11357. I.M. is supported by the DOE Early Career Award
DE-SC0025581, and the Sloan Foundation. 
X.Z. is supported by the U.S. Department of Energy under contract DE-SC0013607 and the MIT Pappalardo Fellowship.

\newpage
\appendix

\section{Additional breakdowns of scale dependence}\label{app:scales}

\begin{figure}
\includegraphics[width=0.49\textwidth]{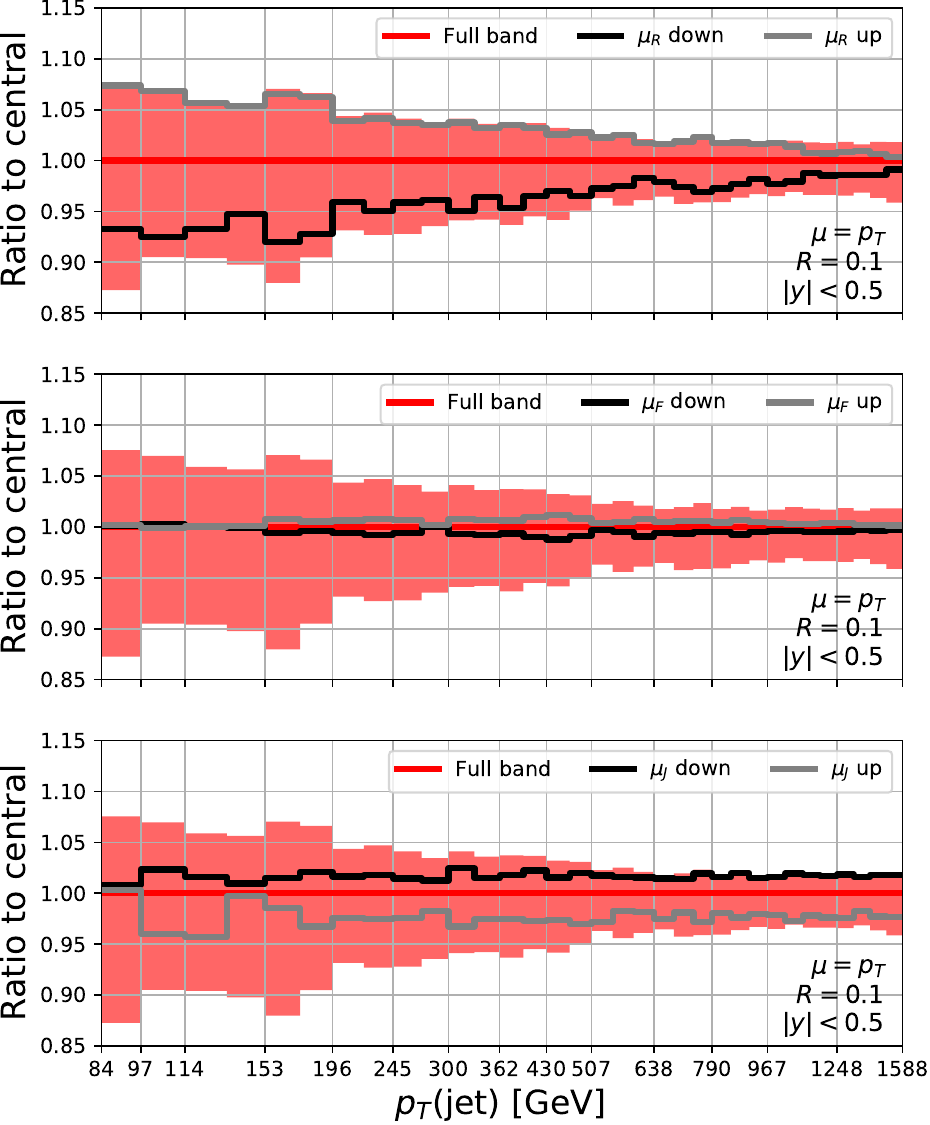}
\includegraphics[width=0.49\textwidth]{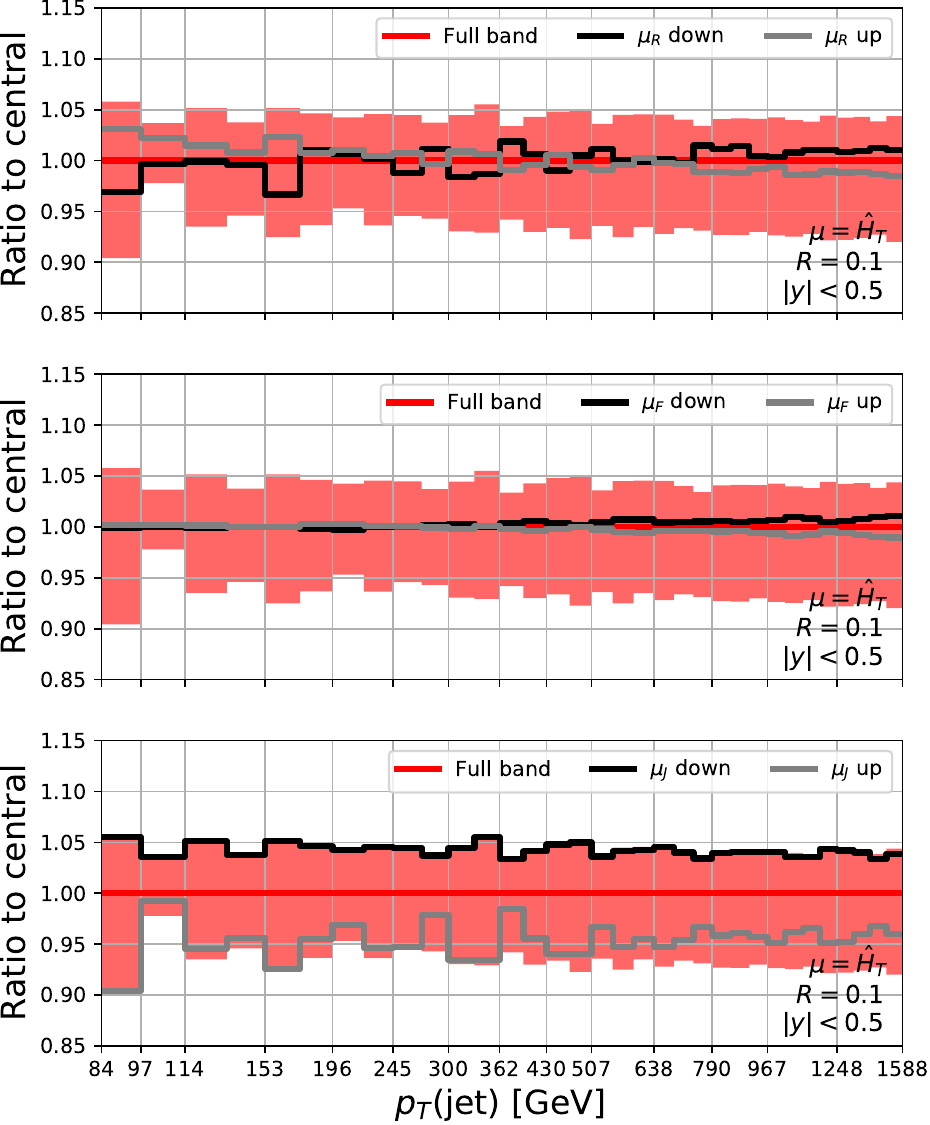}
\caption{A comparison of the impact of varying individual scales at NNLO+NNLL for a central scale of $\ptj$ (left) and $\hthat$ (right) for $R=0.1$.}\label{fig:ComparisonScalesR0.1}
\end{figure}

\begin{figure}
\includegraphics[width=0.49\textwidth]{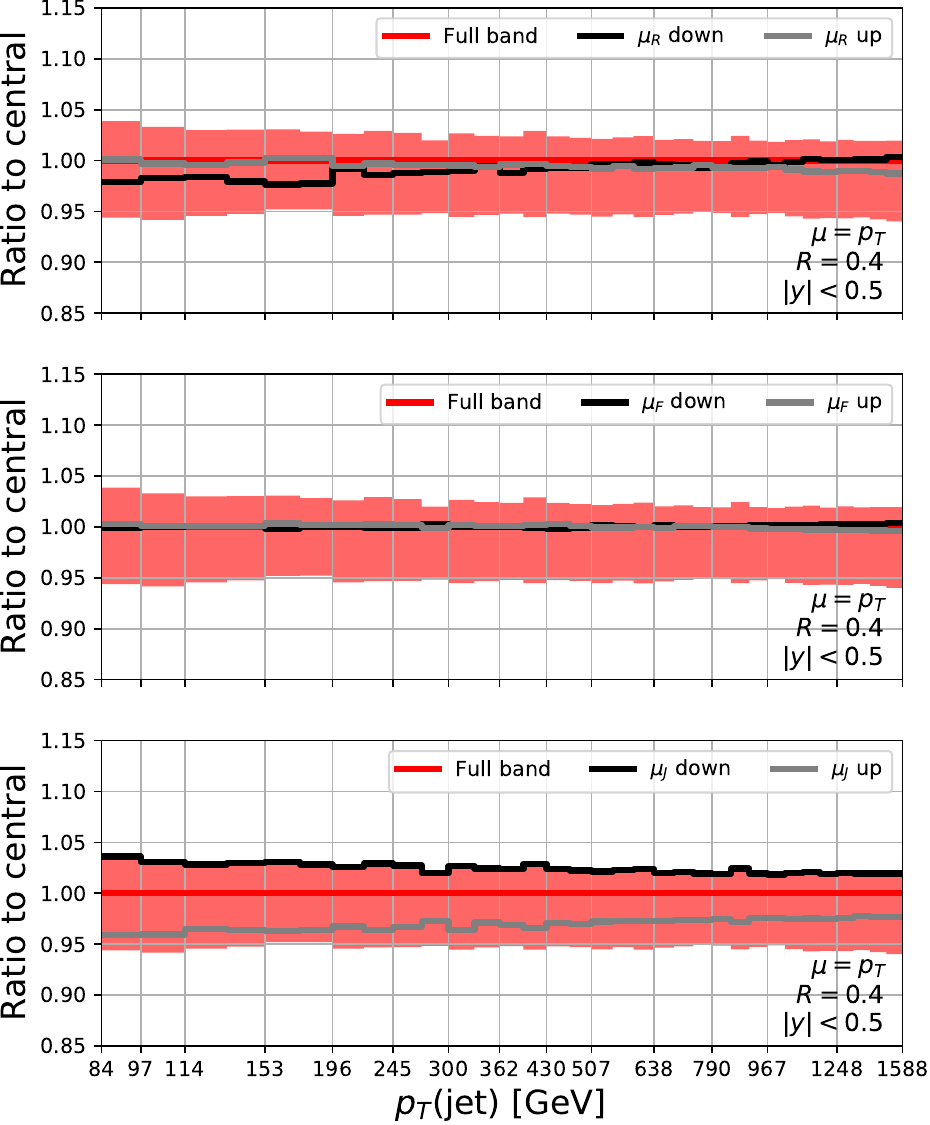}
\includegraphics[width=0.49\textwidth]{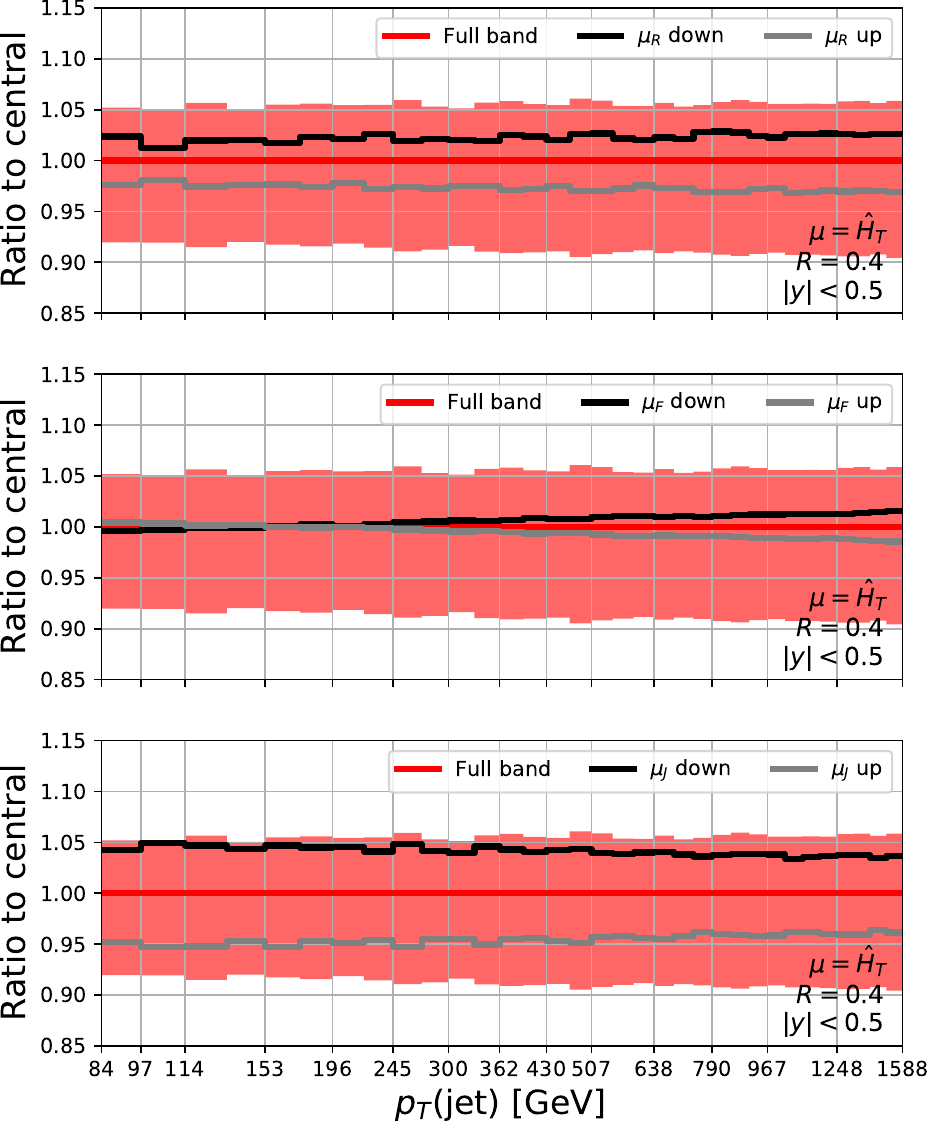}
\caption{As in Figure~\ref{fig:ComparisonScalesR0.1}, but for $R=0.4$.}\label{fig:ComparisonScalesR0.4}
\end{figure}

\begin{figure}
\includegraphics[width=0.49\textwidth]{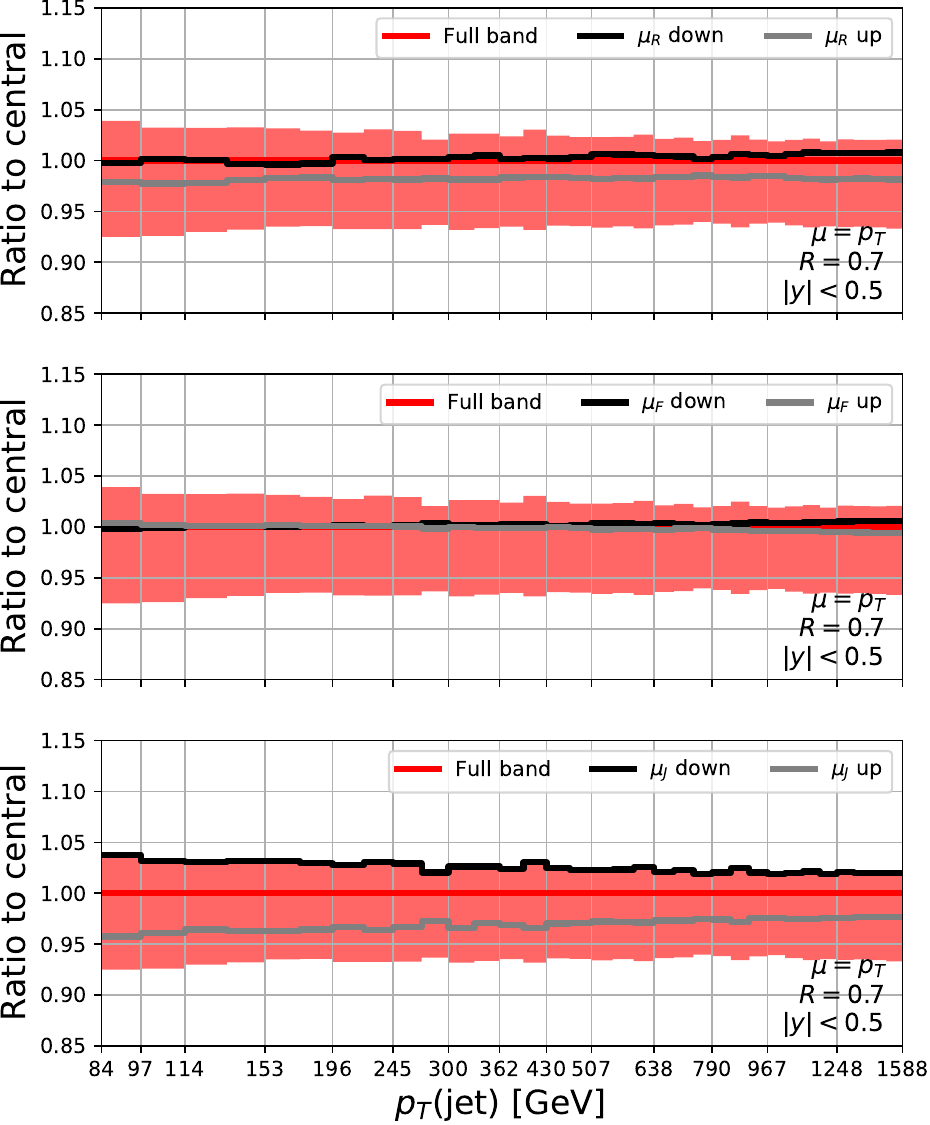}
\includegraphics[width=0.49\textwidth]{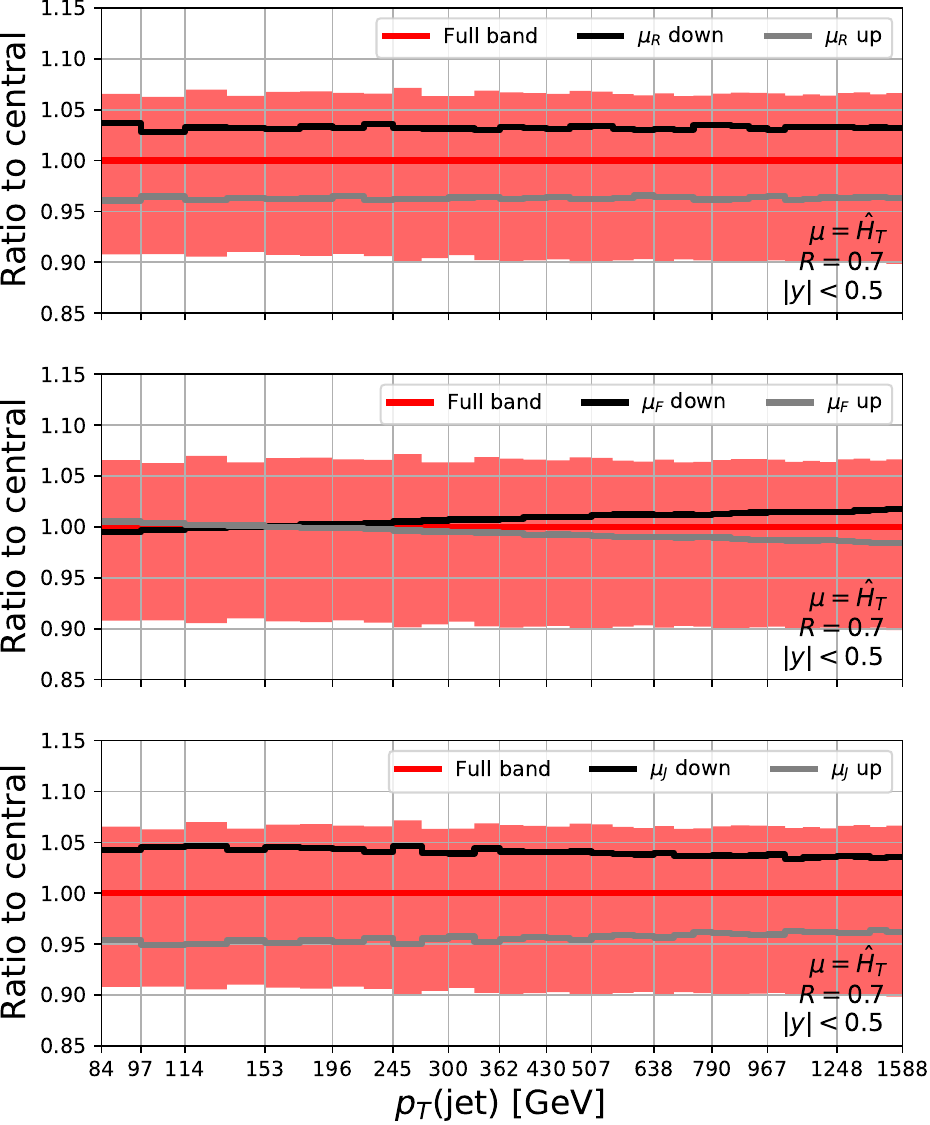}
\caption{As in Figure~\ref{fig:ComparisonScalesR0.1}, but for $R=0.7$.}\label{fig:ComparisonScalesR0.7}
\end{figure}

\begin{figure}
\includegraphics[width=0.49\textwidth]{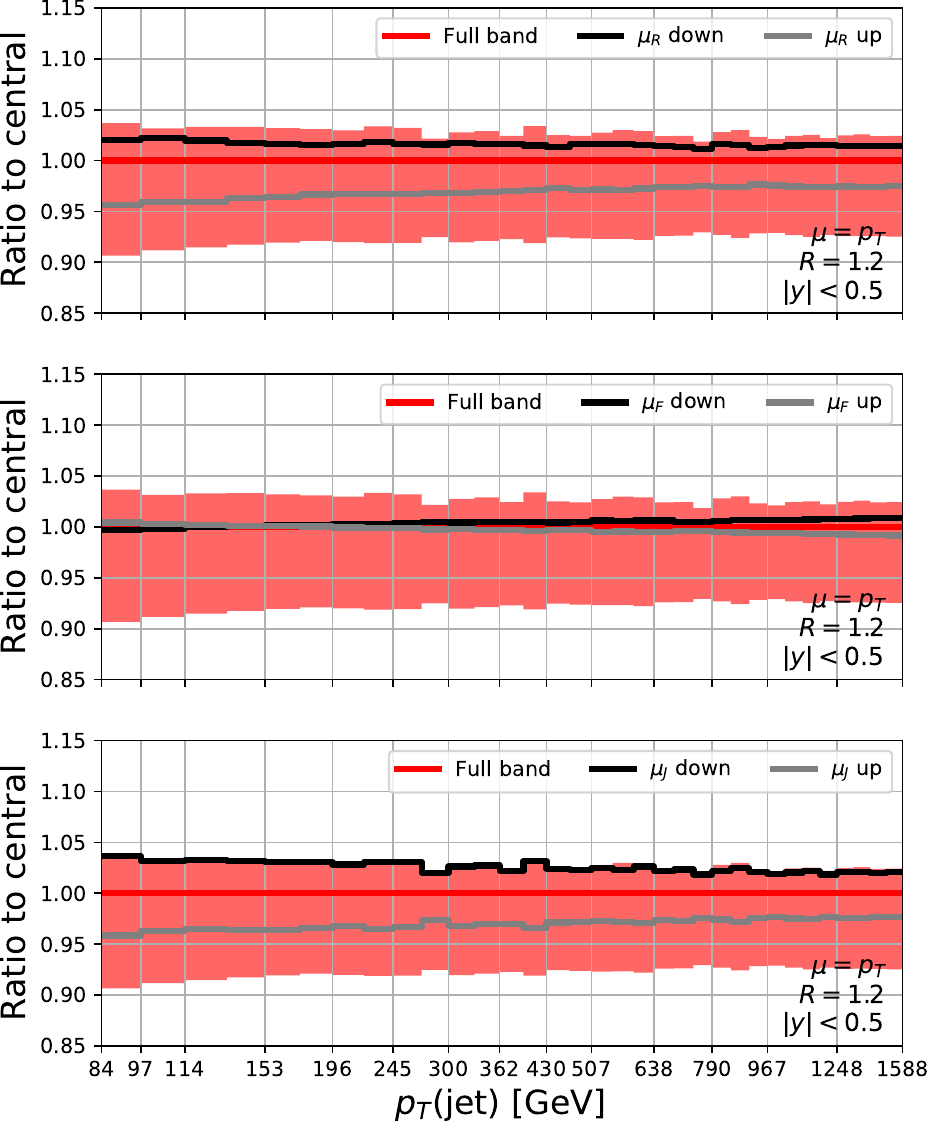}
\includegraphics[width=0.49\textwidth]{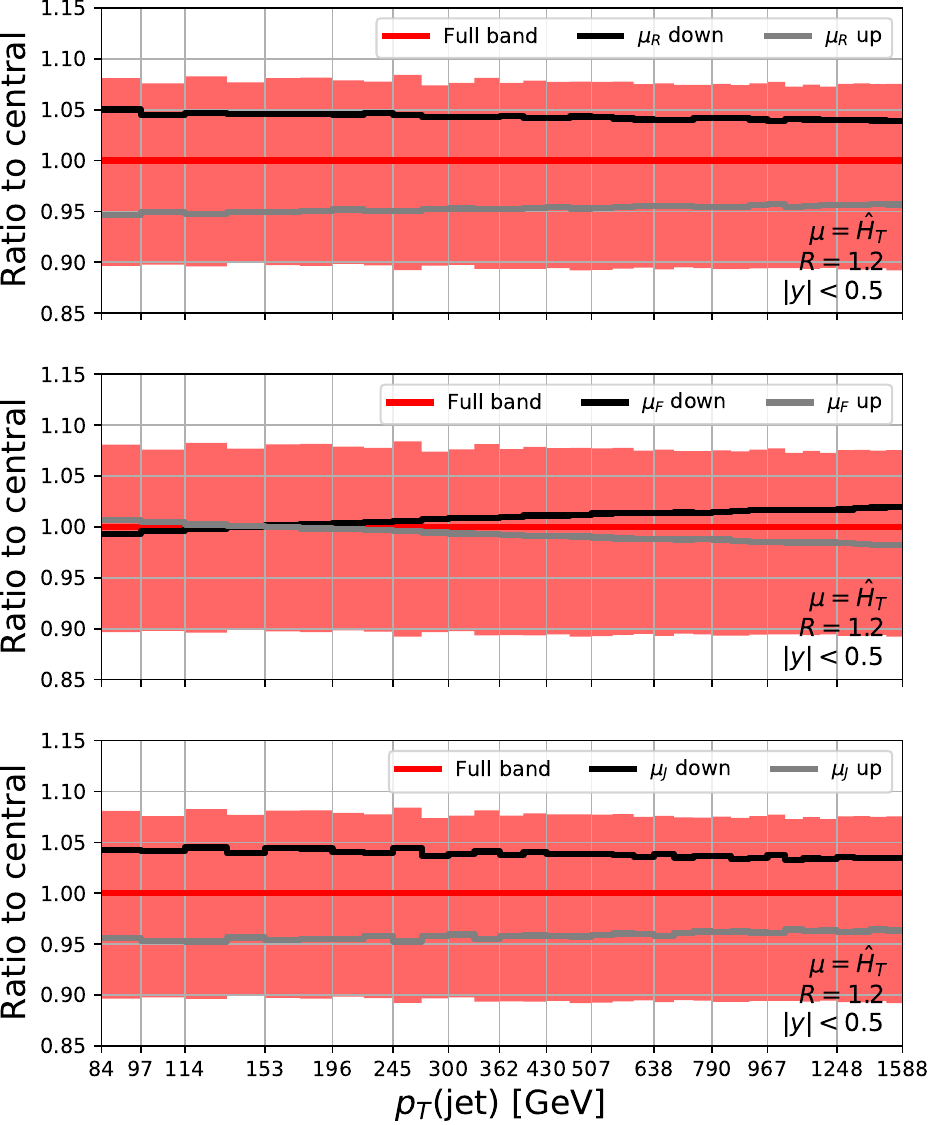}
\caption{As in Figure~\ref{fig:ComparisonScalesR0.1}, but for $R=1.2$.}\label{fig:ComparisonScalesR1.2}
\end{figure}

To provide more details on the scale variations of the NNLO+NNLL predictions for absolute transverse momentum spectra, we separate the scale variation into three variations: we keep all scales fixed except one, which we vary by a factor of $2$.
Compared to the full variations, we show this in Figure~\ref{fig:ComparisonScalesR0.1} to Figure~\ref{fig:ComparisonScalesR1.2}.

On the one hand, for $R=0.1$, we see that scale variation is dominated by the $\mur$ variations for the $\ptj$ scale but by the $\muj$ scale for the $\hthat$ scale choice.
On the other hand, for $R=0.4$ and $R=0.7$ jet cases, where fixed-order variations are very small, we see that the $\muj$ scale leads to the largest variation and that the $\mur$ variation is small.
Finally, for $R=1.2$, we see that $\mur$ and $\muj$ are similar in size.
The factorization scale ($\muf$) variation is not a dominant factor in any case.

\begin{figure}
\includegraphics[width=0.49\textwidth]{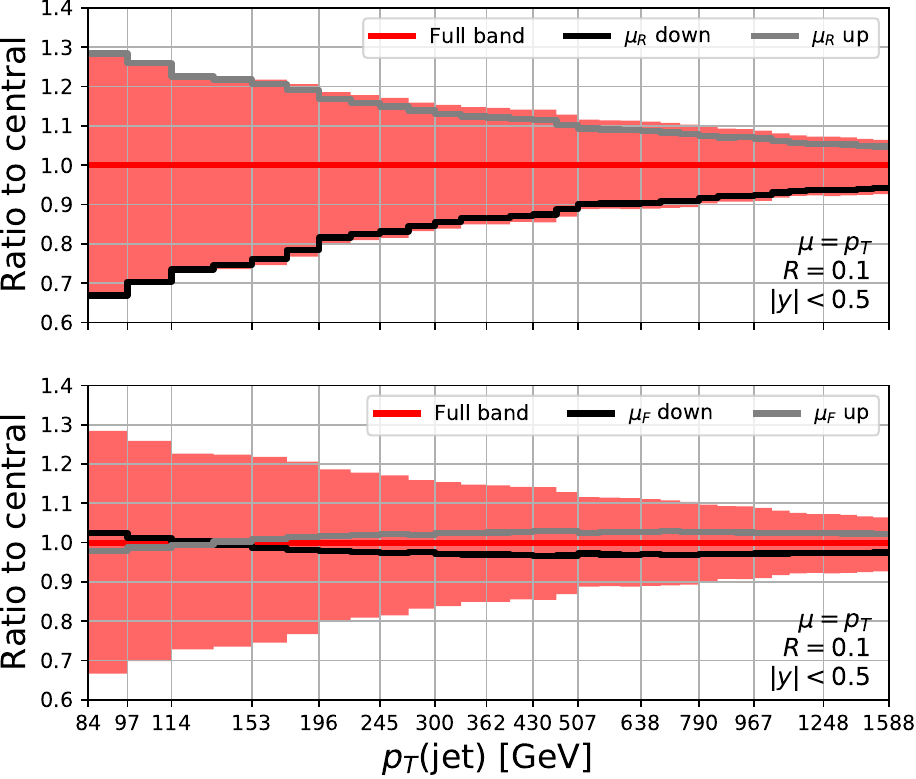}
\includegraphics[width=0.49\textwidth]{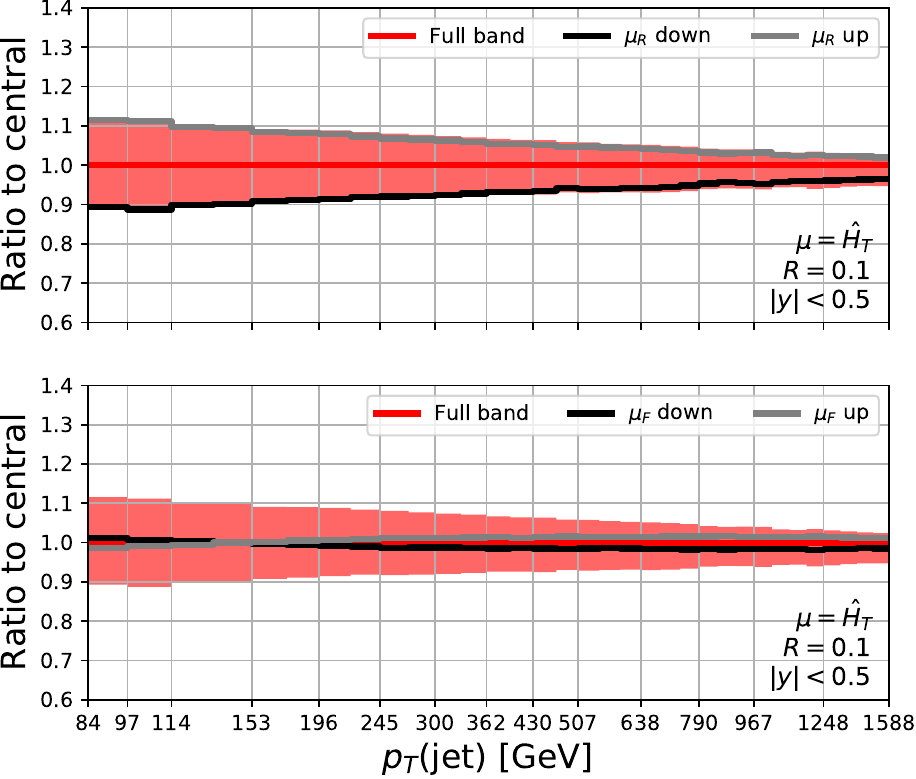}
\caption{A comparison of the impact of varying individual scales at NNLO for a central scale of $\ptj$ (left) and $\hthat$ (right) for $R=0.1$.}\label{fig:ComparisonScalesR0.1FO}
\end{figure}

\begin{figure}
\includegraphics[width=0.49\textwidth]{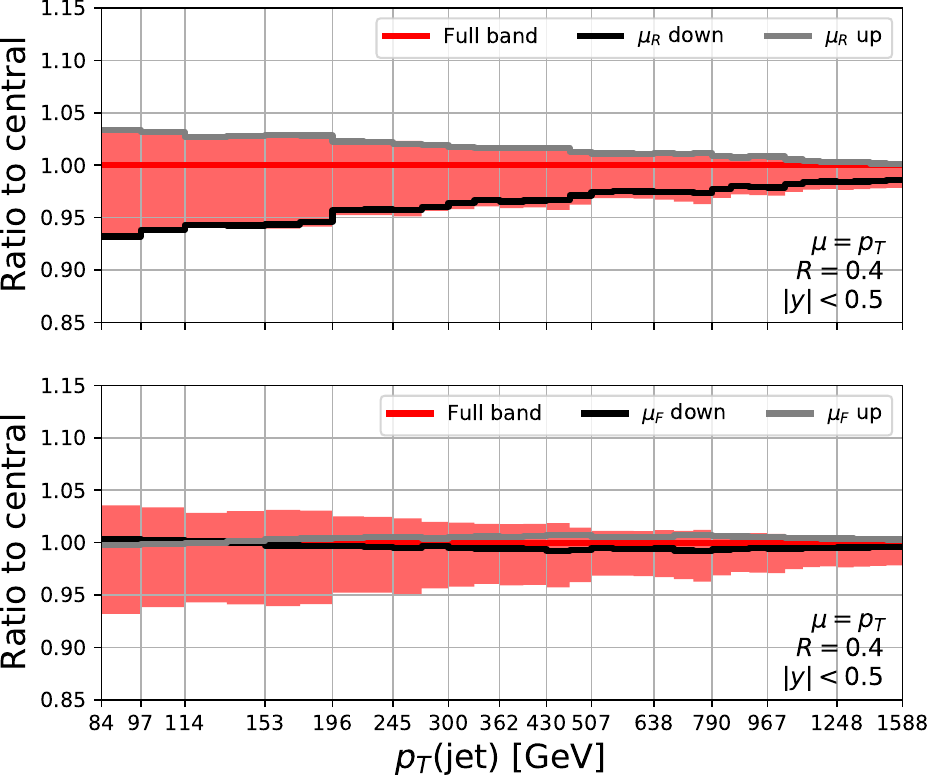}
\includegraphics[width=0.49\textwidth]{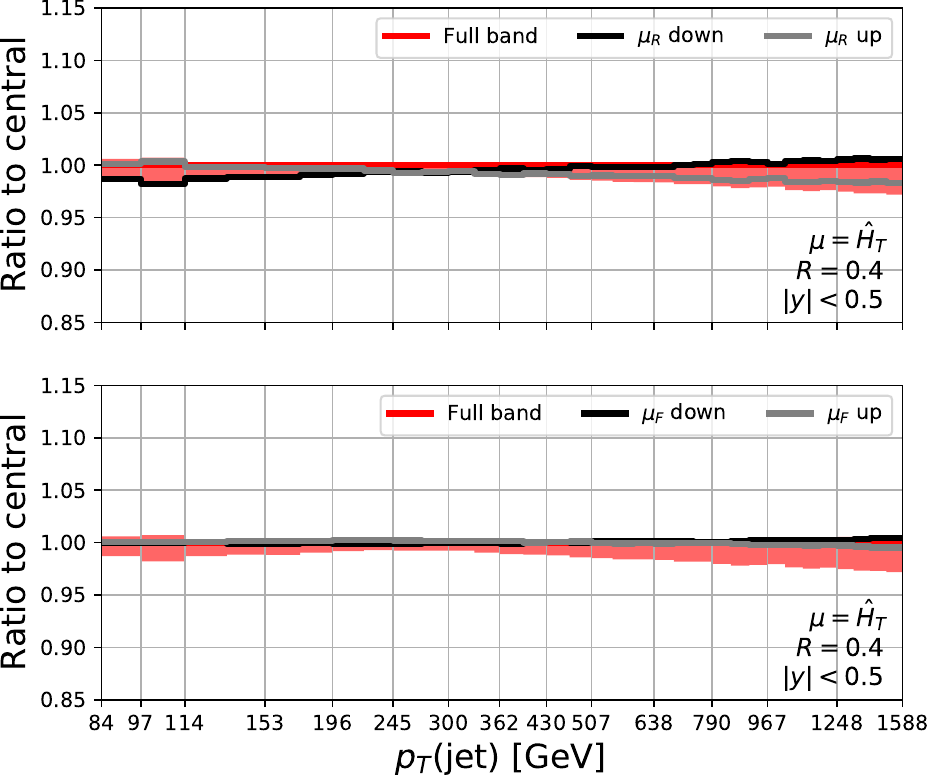}
\caption{As in Figure~\ref{fig:ComparisonScalesR0.1FO}, but for $R=0.4$.}\label{fig:ComparisonScalesR0.4FO}
\end{figure}

\begin{figure}
\includegraphics[width=0.49\textwidth]{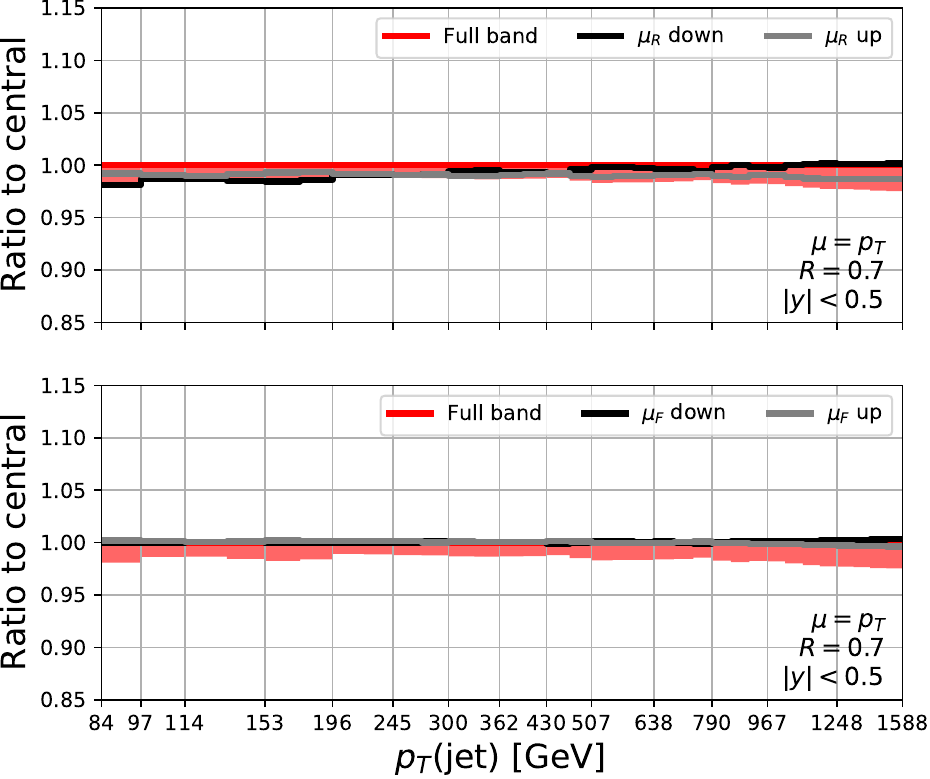}
\includegraphics[width=0.49\textwidth]{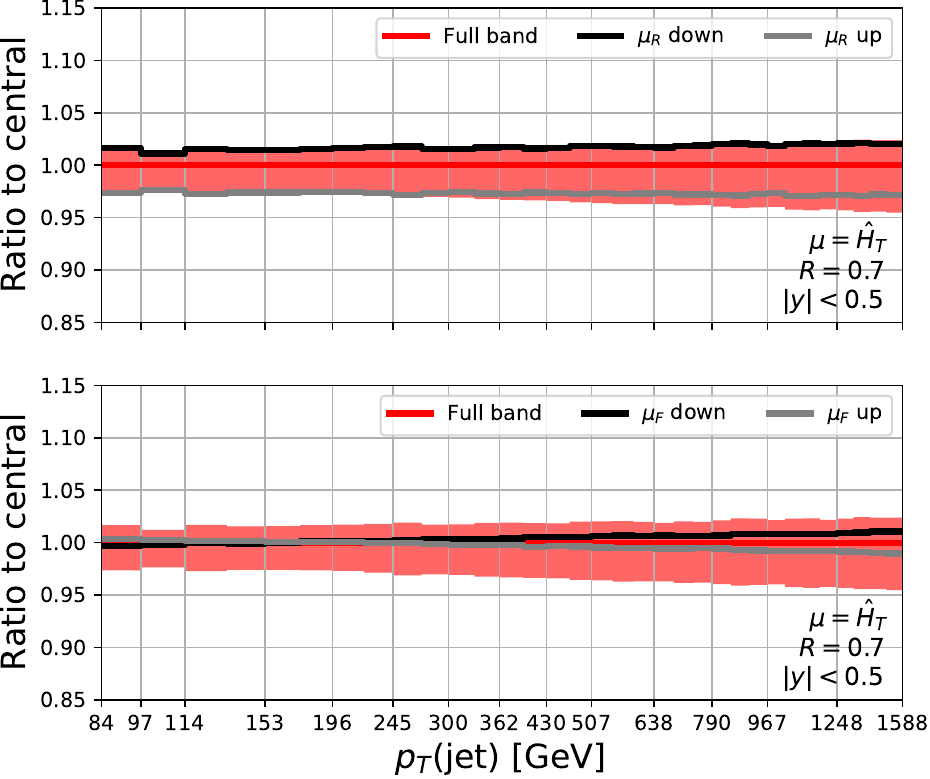}
\caption{As in Figure~\ref{fig:ComparisonScalesR0.1FO}, but for $R=0.7$.}\label{fig:ComparisonScalesR0.7FO}
\end{figure}

\begin{figure}
\includegraphics[width=0.49\textwidth]{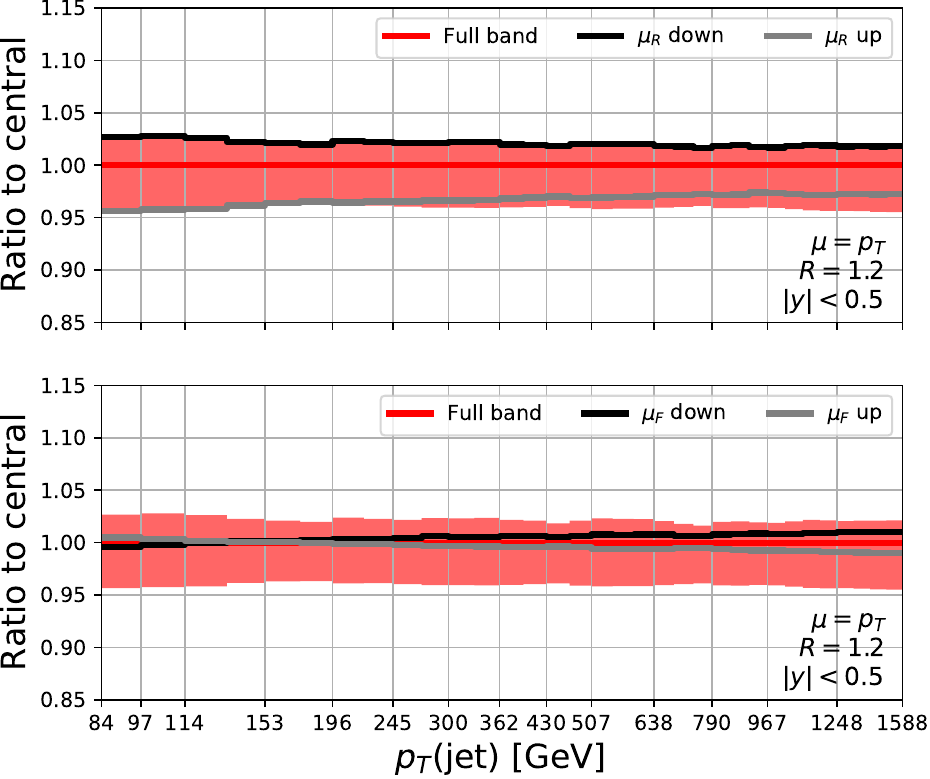}
\includegraphics[width=0.49\textwidth]{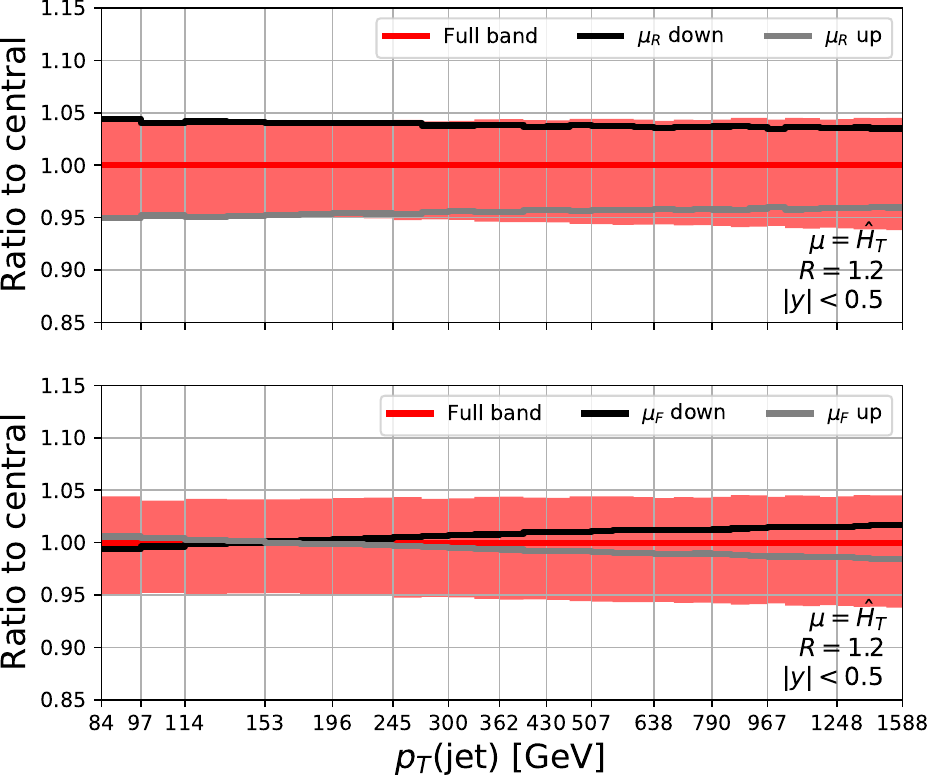}
\caption{As in Figure~\ref{fig:ComparisonScalesR0.1FO}, but for $R=1.2$.}\label{fig:ComparisonScalesR1.2FO}
\end{figure}

A breakdown into $\muf$ and $\mur$ variations for the absolute fixed-order predictions is provided in Figure~\ref{fig:ComparisonScalesR0.1FO} to Figure~\ref{fig:ComparisonScalesR1.2FO}, demonstrating that the $\mur$ scale variation is the largest component.
In comparison, for the resummed case, we can also see that the $\mur$ variation alone is different, most clearly seen when comparing the variations for $R=0.1$ in Figure~\ref{fig:ComparisonScalesR0.1} and Figure~\ref{fig:ComparisonScalesR0.1FO}.

\begin{figure}[t]
\includegraphics[width=0.49\textwidth]{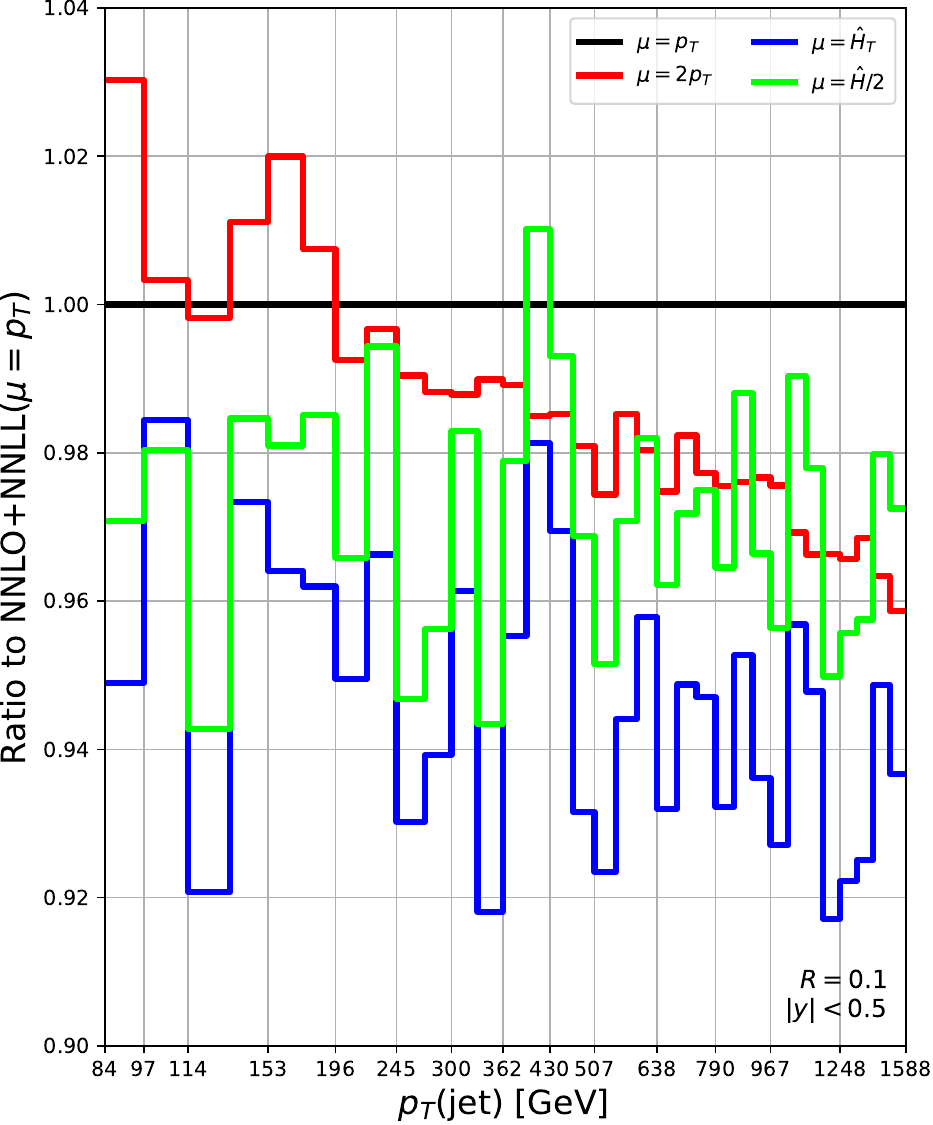}
\includegraphics[width=0.49\textwidth]{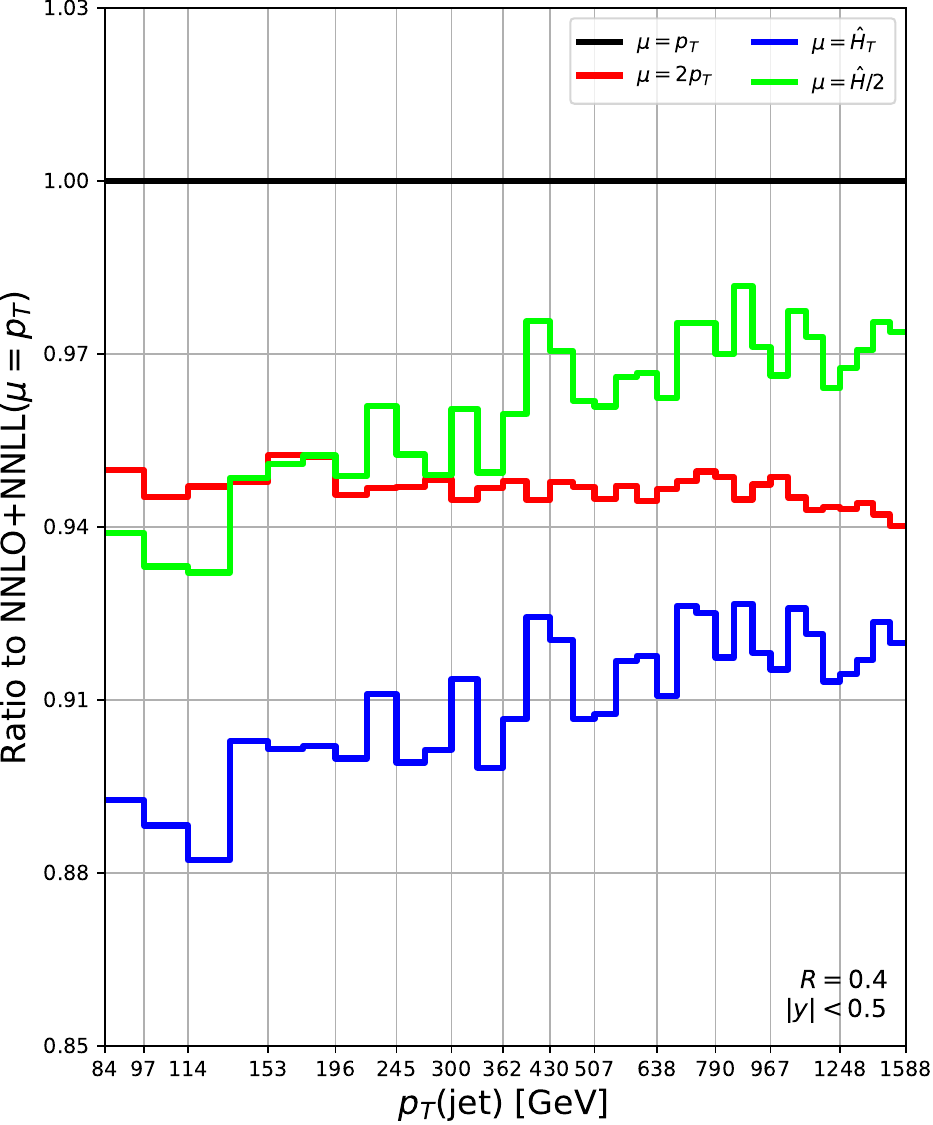}
\includegraphics[width=0.49\textwidth]{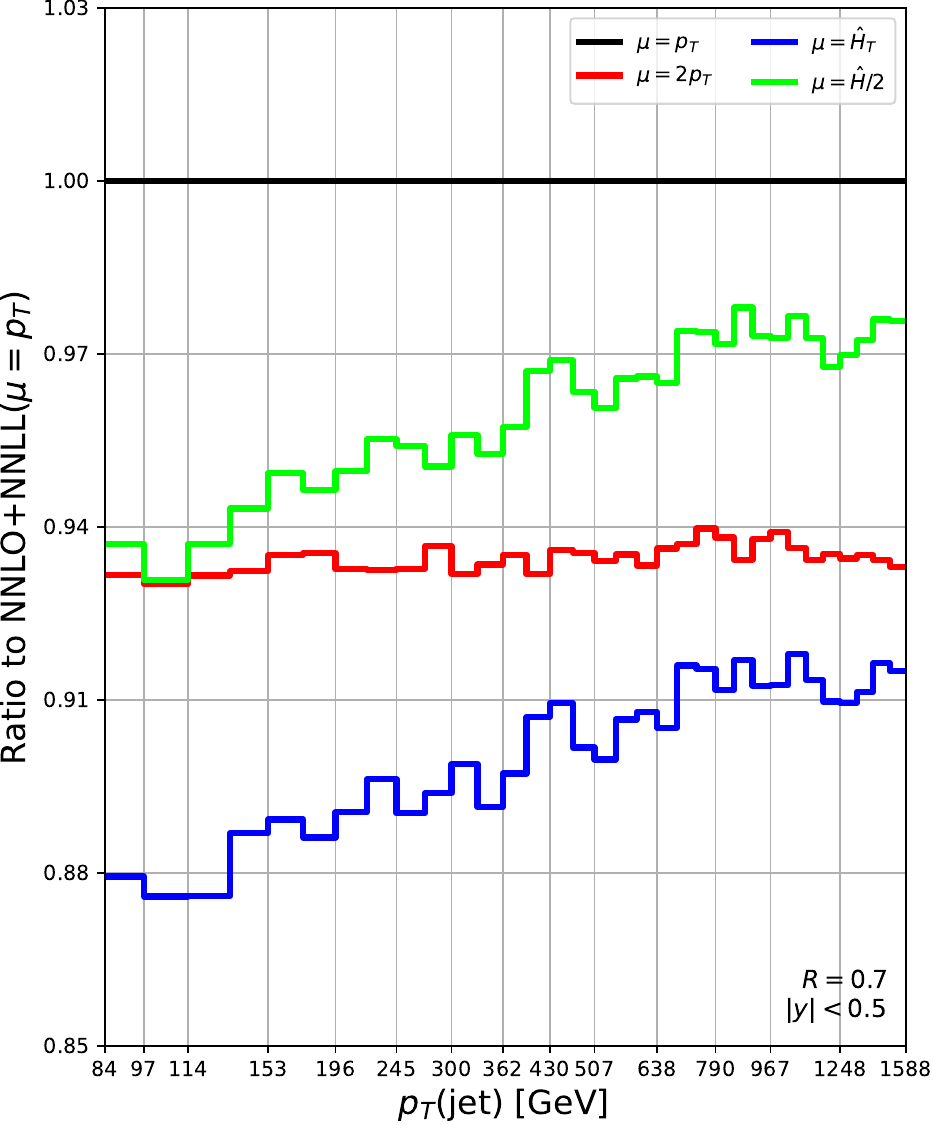}
\includegraphics[width=0.49\textwidth]{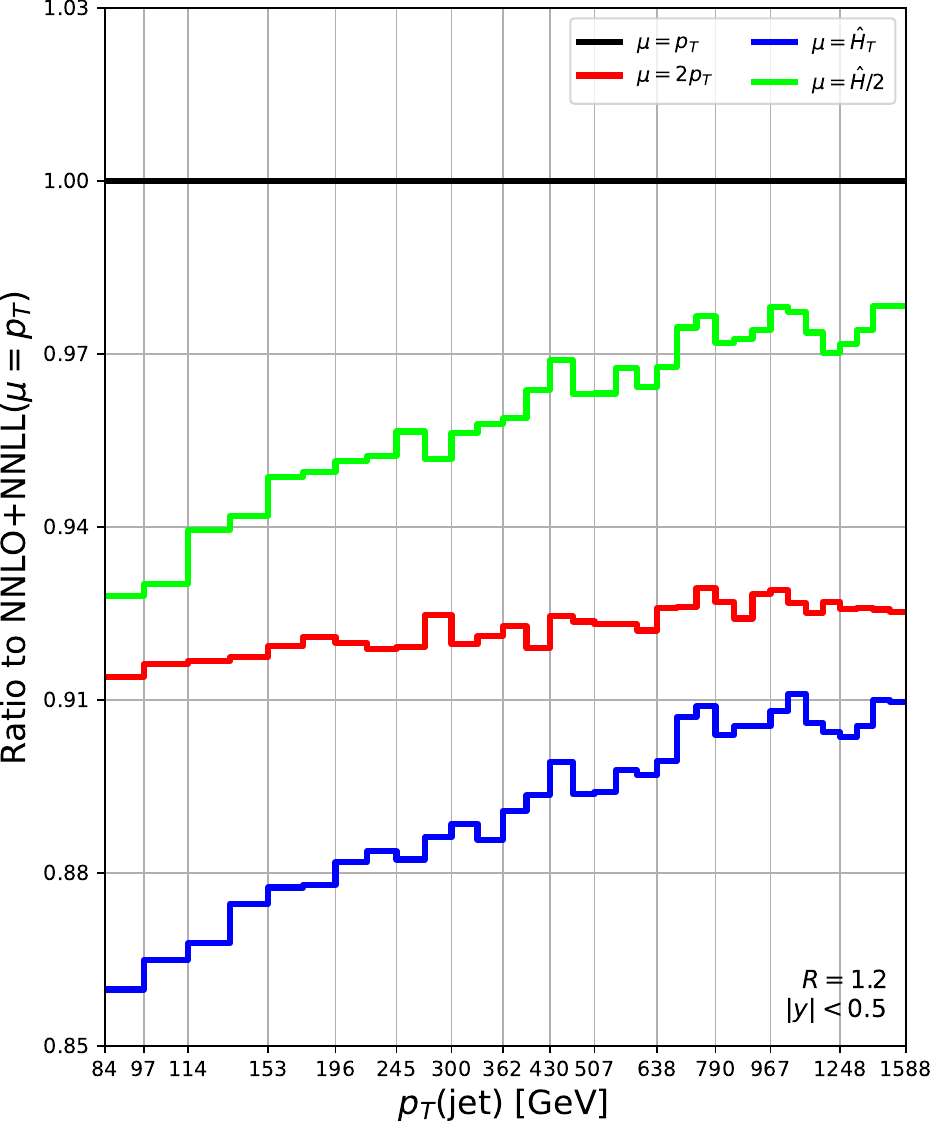}
\caption{A comparison of the NNLO+NNLL absolute spectrum for 4 different scale choices for $R=0.1$ (top left), $R=0.4$ (top right), $R=0.7$ (bottom left), $R=1.2$ (bottom right) for both scale choices.}\label{fig:ComparisonScales}
\end{figure}

In Figure~\ref{fig:ComparisonScales}, one can see the predictions for the scale choices $\ptj$, $2\ptj$, $\hthat$, and $\hthat/2$ for the resummed predictions.
Crucially, one observes that changing the prefactor in front of $\ptj$ reduces the cross section; as $R$ increases, the effect becomes stronger.
Changing from $\mu=\ptj$ to $\mu=2\ptj$ yields results closer to the $\mu=\hthat$ results, as expected (the difference now comes exclusively from real emission contributions). In particular, the prediction for $\mu=2\ptj$ appears to asymptotically approach the prediction for $\mu=\hthat$ at large transverse momentum, consistent with the expectation that dijet configurations dominate the high-$p_T$ regime.
Similarly, reducing the scale from $\hthat$ to $\hthat/2$ yields curves that are closer to those obtained using $\ptj$.

\begin{figure}
\includegraphics[width=0.49\textwidth]{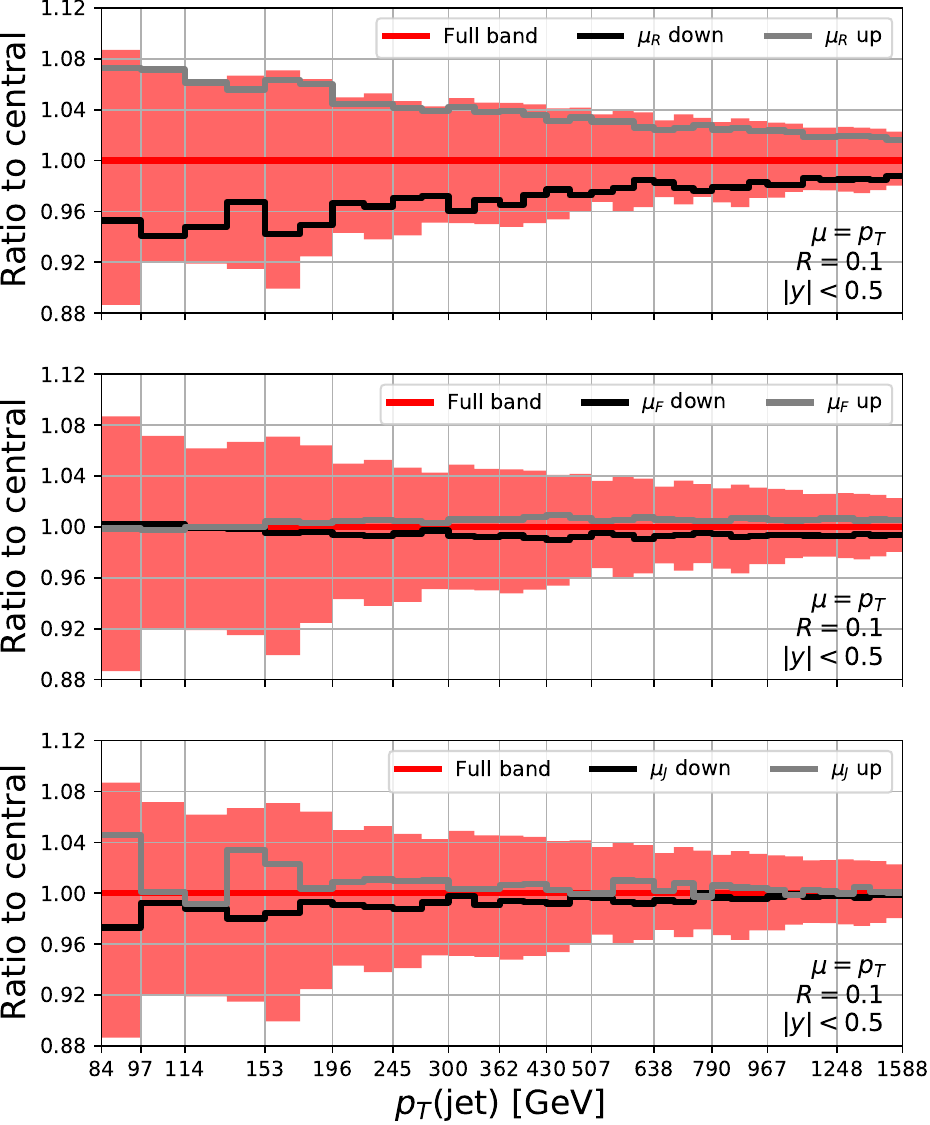}
\includegraphics[width=0.49\textwidth]{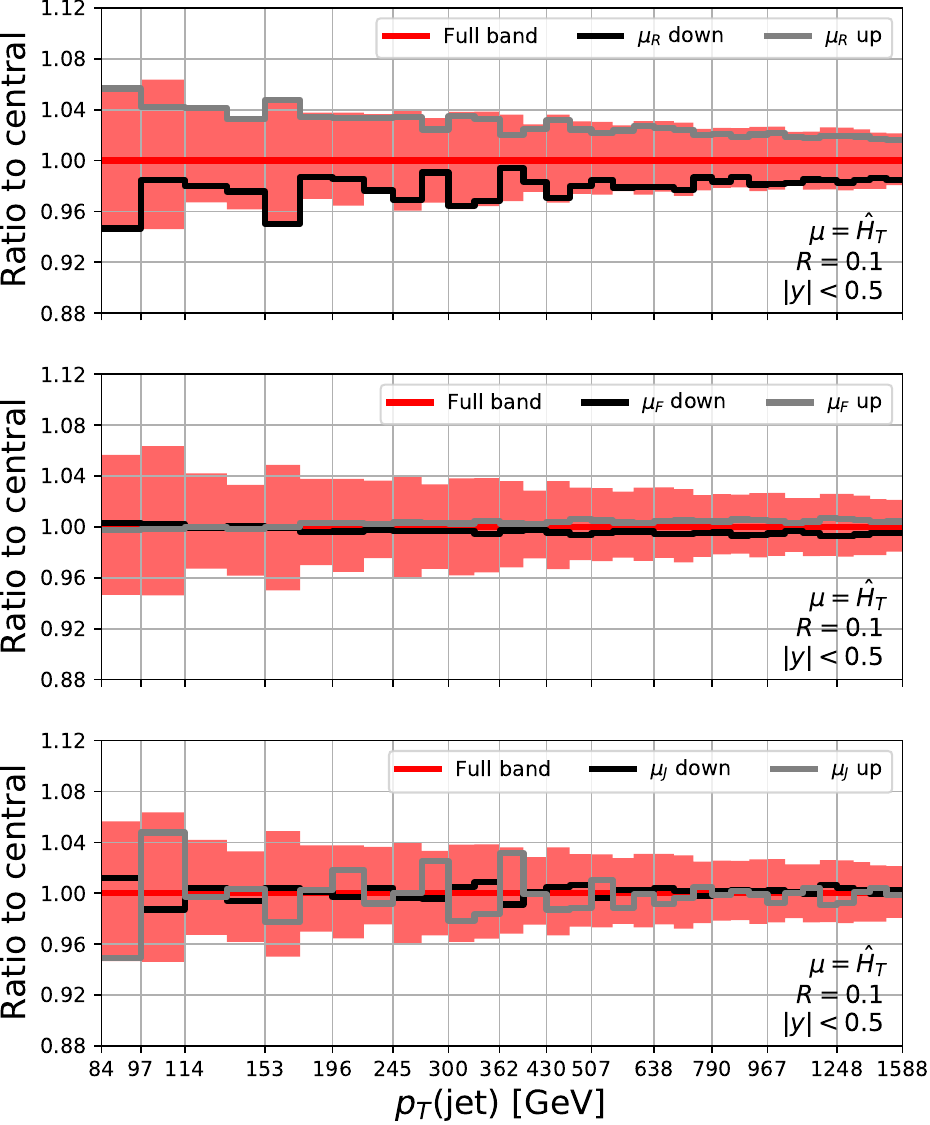}
\caption{A comparison of the impact of varying individual scales on the ratio between the $R=0.1$ and $R=0.4$ spectra at NNLO+NNLL for a central scale of $\ptj$ (left) and $\hthat$ (right).}\label{fig:ComparisonScalesR0.1Ratio}
\end{figure}

\begin{figure}
\includegraphics[width=0.49\textwidth]{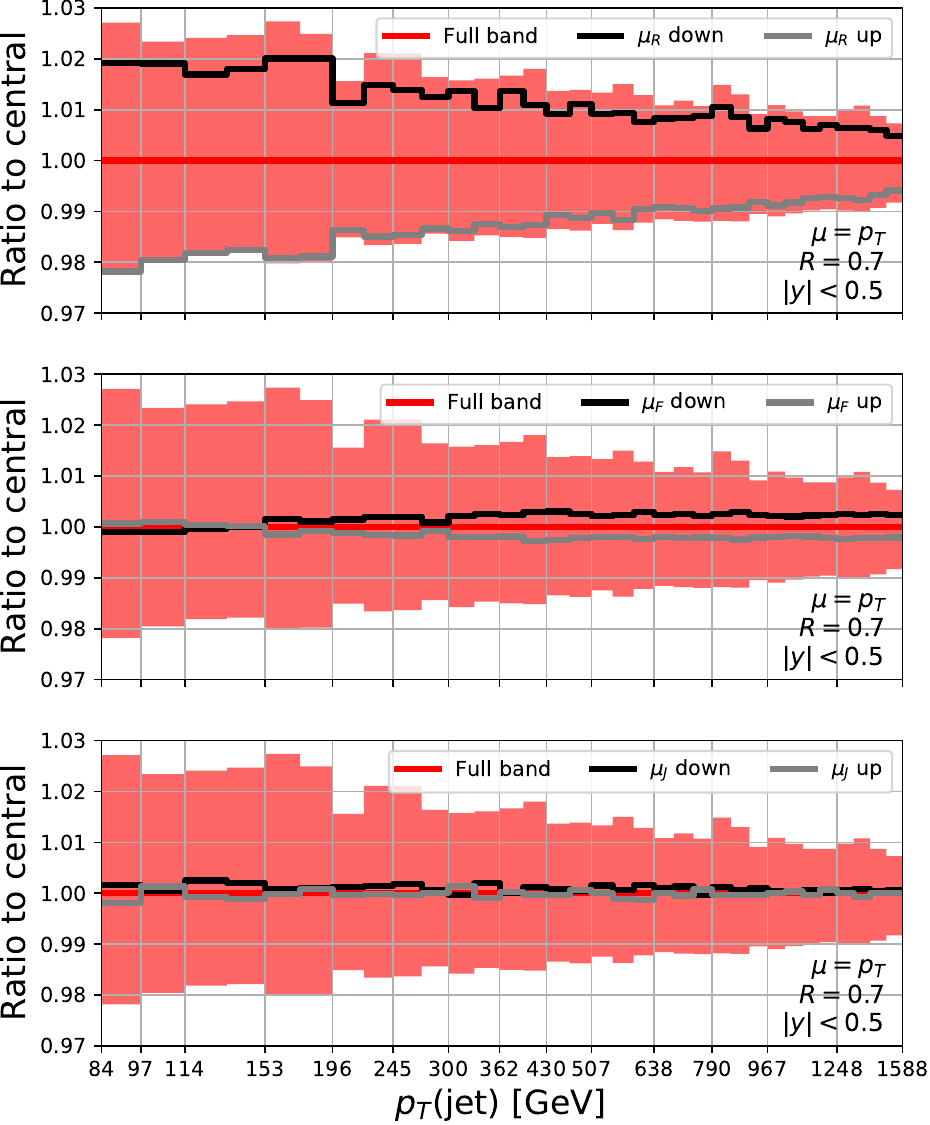}
\includegraphics[width=0.49\textwidth]{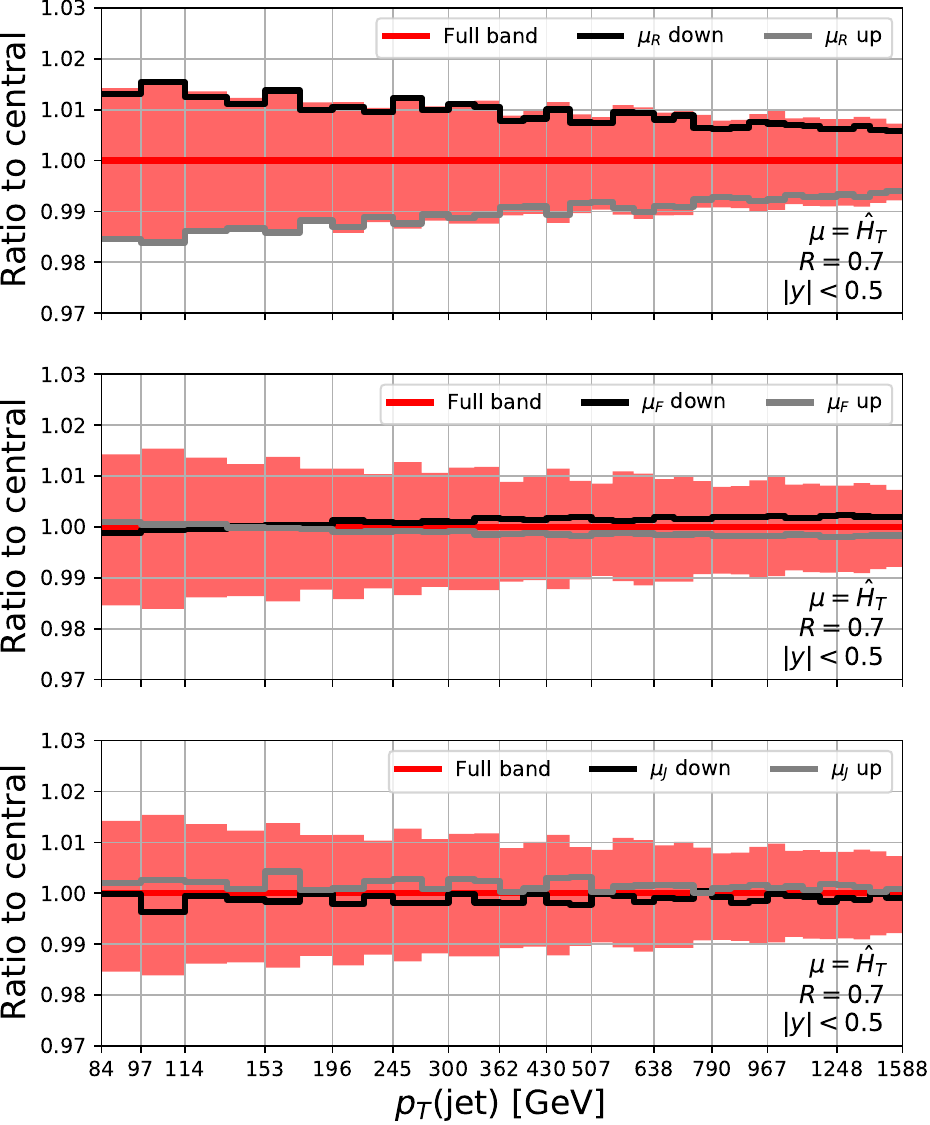}
\caption{As in Figure~\ref{fig:ComparisonScalesR0.1Ratio}, but for the ratio between $R=0.7$ and $R=0.4$.}\label{fig:ComparisonScalesR0.7Ratio}
\end{figure}

\begin{figure}
\includegraphics[width=0.49\textwidth]{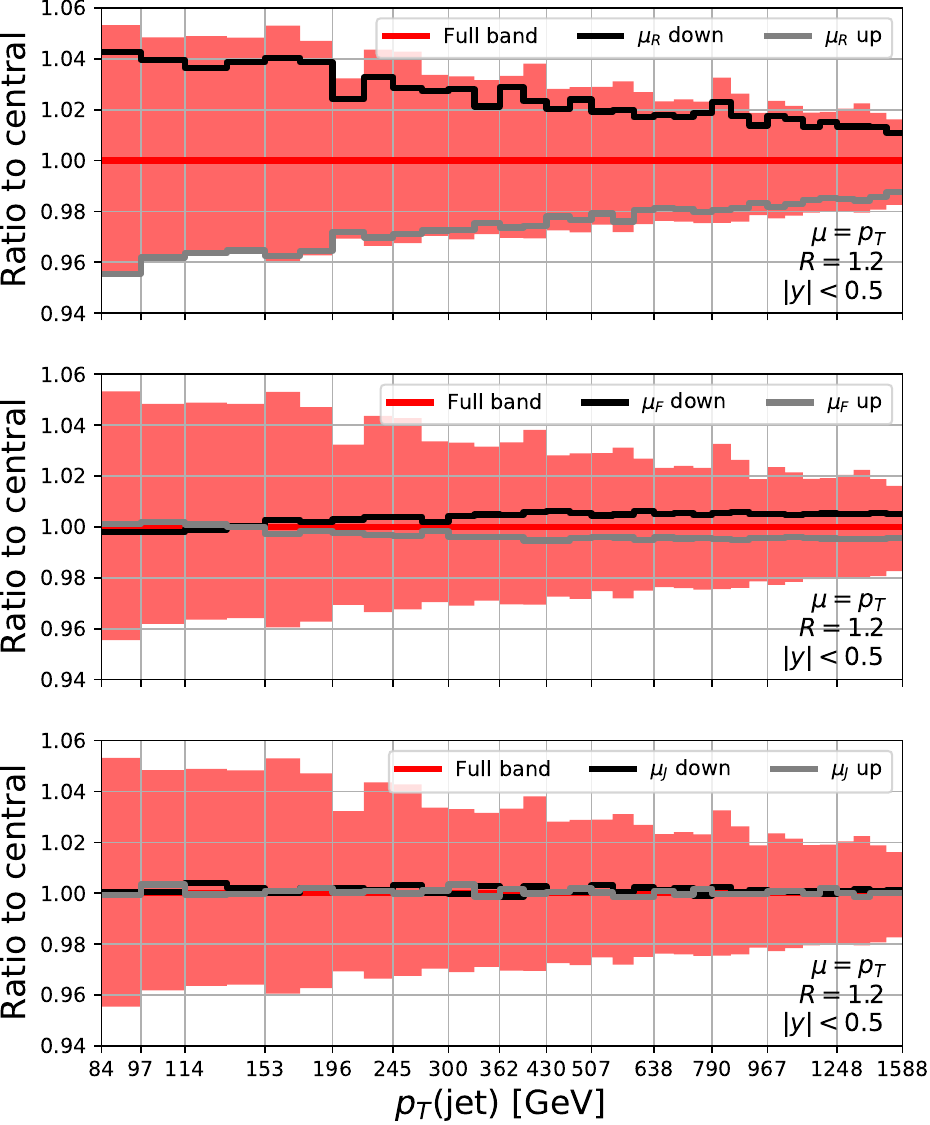}
\includegraphics[width=0.49\textwidth]{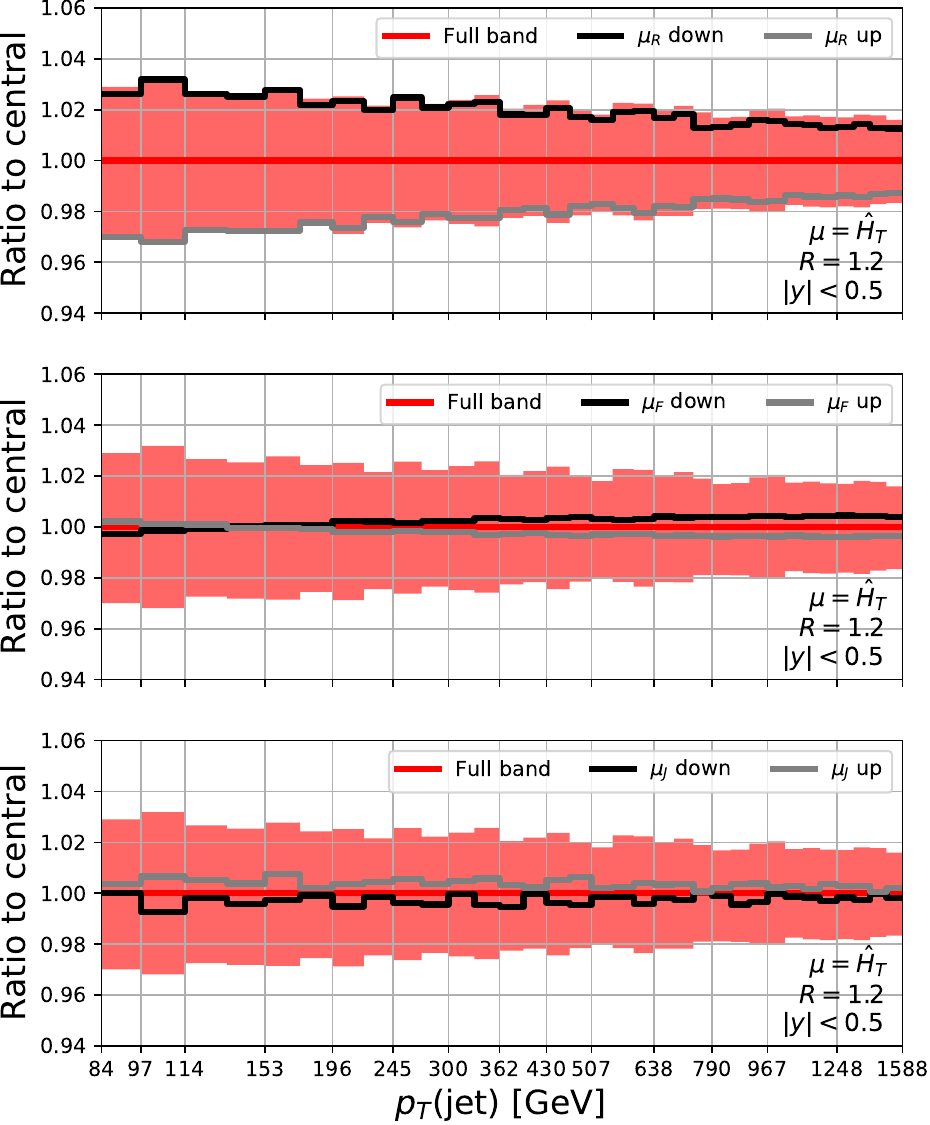}
\caption{As in Figure~\ref{fig:ComparisonScalesR0.1Ratio}, but for the ratio between $R=1.2$ and $R=0.4$.}\label{fig:ComparisonScalesR1.2Ratio}
\end{figure}

\begin{figure}
\includegraphics[width=0.49\textwidth]{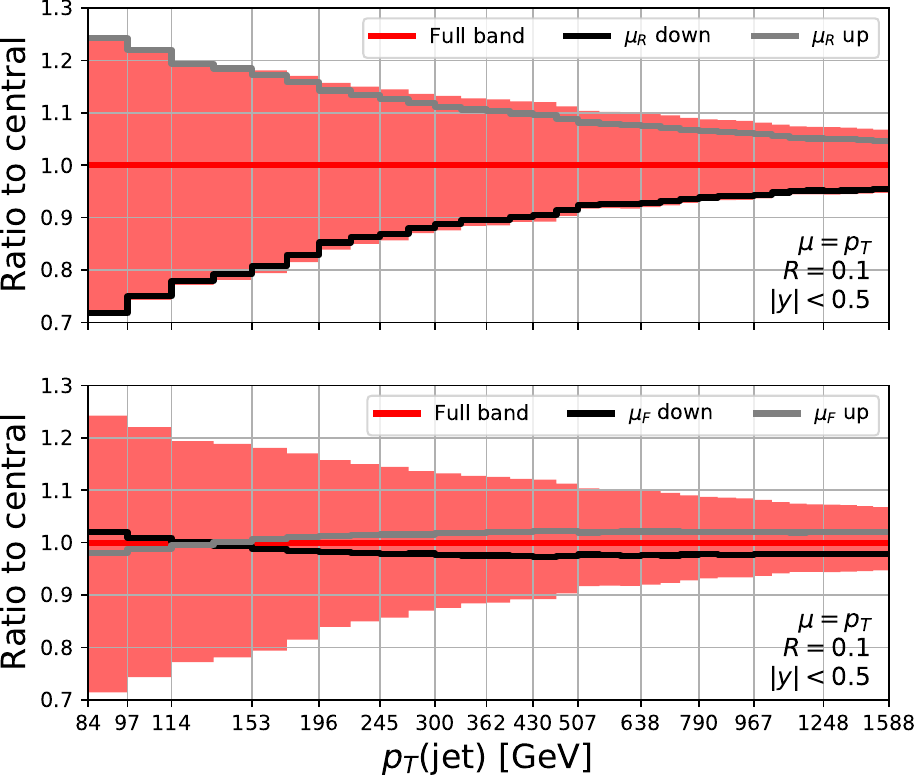}
\includegraphics[width=0.49\textwidth]{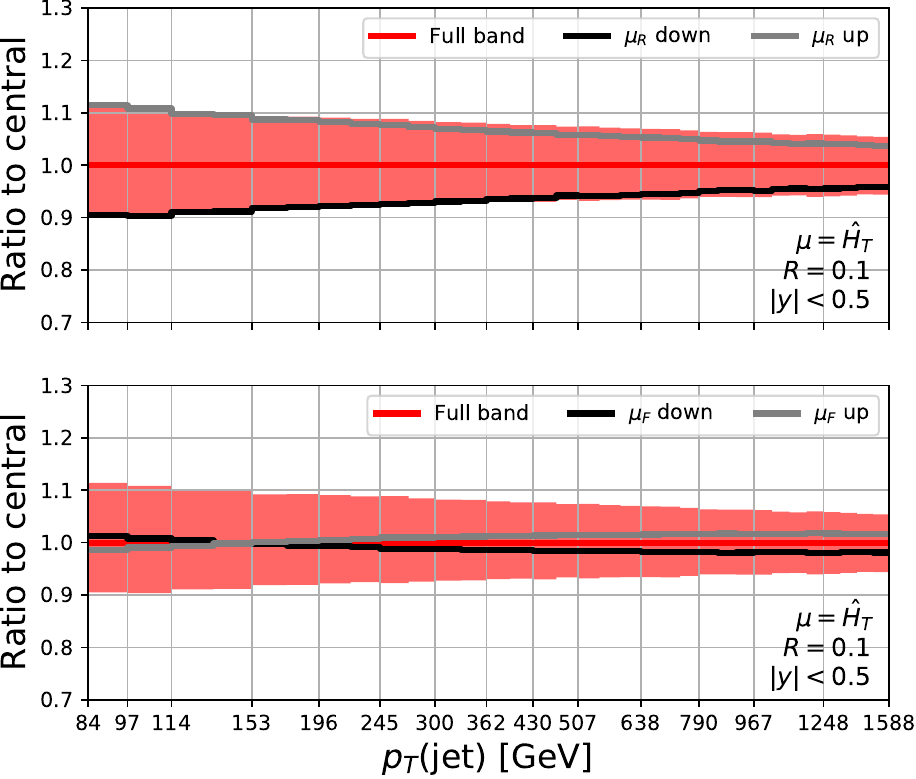}
\caption{A comparison of the impact of varying individual scales on the ratio between the $R=0.1$ and $R=0.4$ spectra at NNLO for a central scale of $\ptj$ (left) and $\hthat$ (right).}\label{fig:ComparisonScalesR0.1RatioFO}
\end{figure}

\begin{figure}
\includegraphics[width=0.49\textwidth]{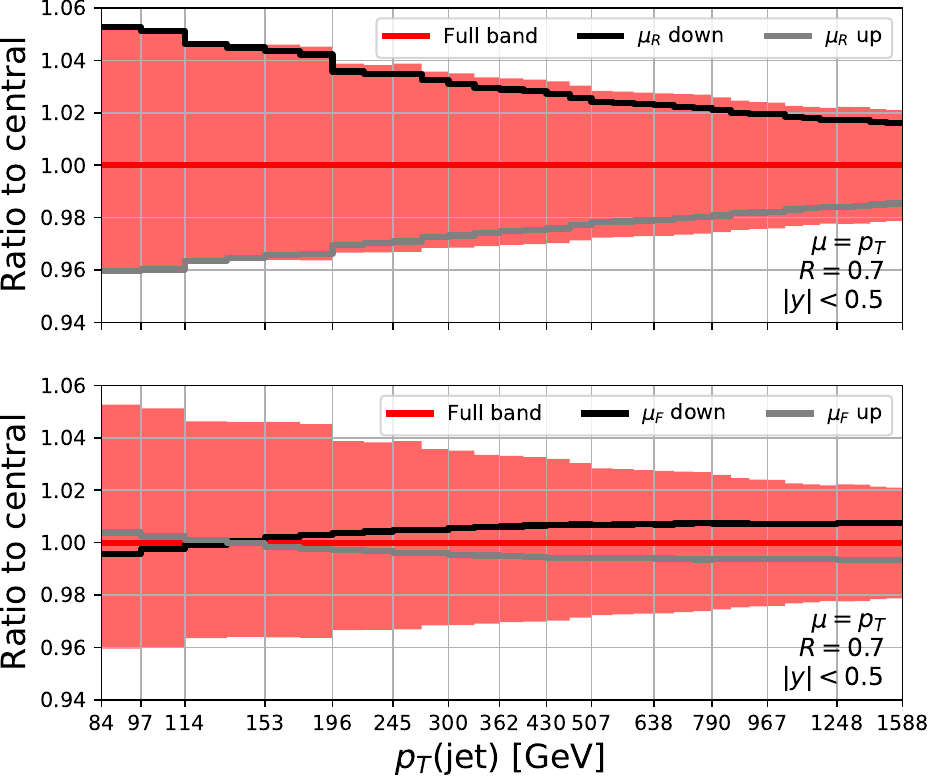}
\includegraphics[width=0.49\textwidth]{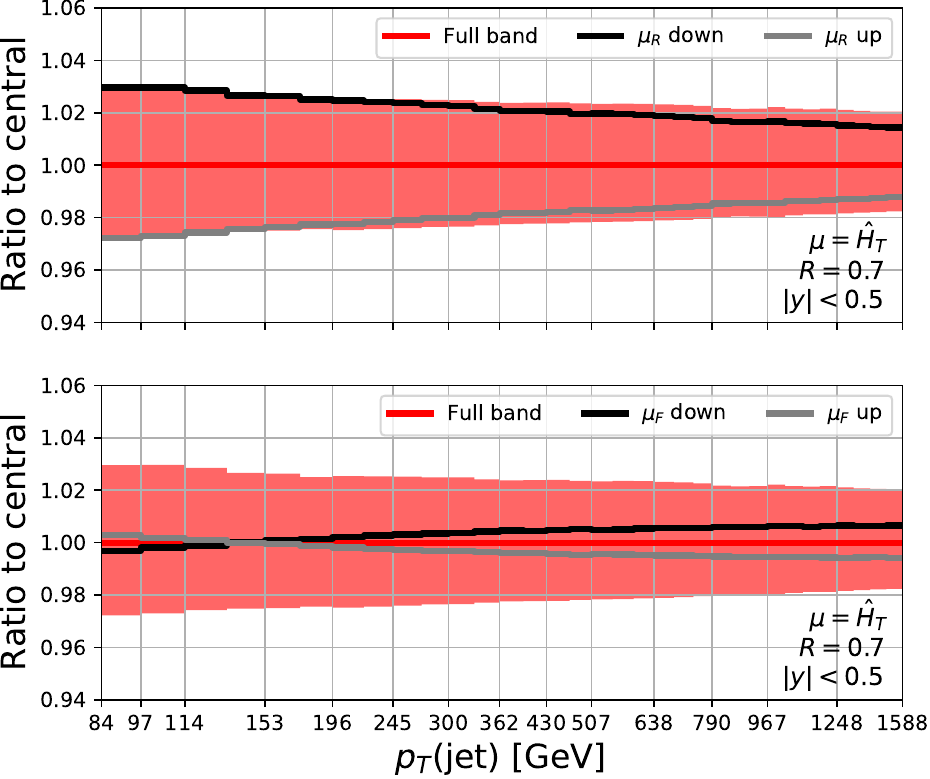}
\caption{As in Figure~\ref{fig:ComparisonScalesR0.1RatioFO}, but for the ratio between $R=0.7$ and $R=0.4$.}\label{fig:ComparisonScalesR0.7RatioFO}
\end{figure}

\begin{figure}
\includegraphics[width=0.49\textwidth]{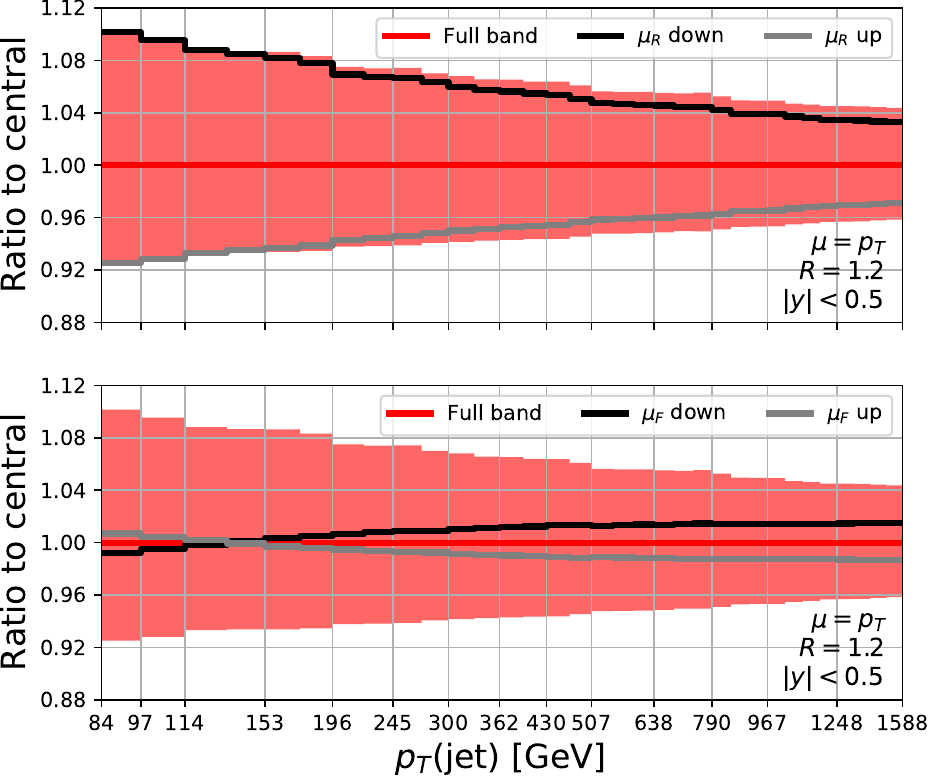}
\includegraphics[width=0.49\textwidth]{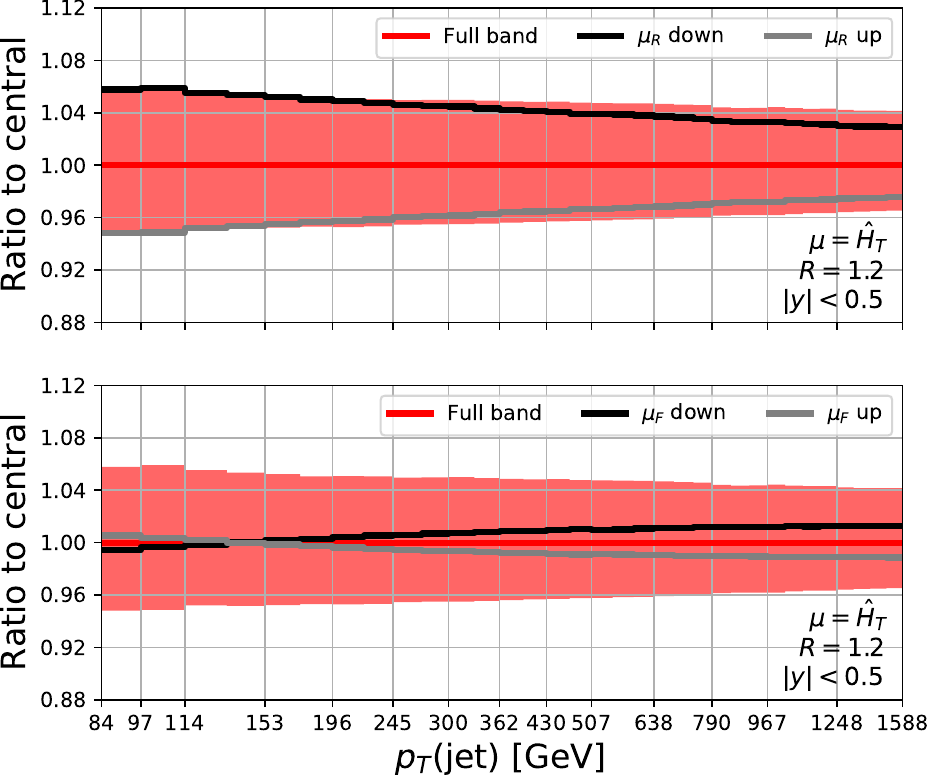}
\caption{As in Figure~\ref{fig:ComparisonScalesR0.1RatioFO}, but for the ratio between $R=1.2$ and $R=0.4$.}\label{fig:ComparisonScalesR1.2RatioFO}
\end{figure}

Finally, we provide the same breakdown for the ratio with respect to the $R=0.4$ radius in Figure~\ref{fig:ComparisonScalesR0.1Ratio} to Figure~\ref{fig:ComparisonScalesR1.2Ratio} for the NNLO+NNLL and in Figure~\ref{fig:ComparisonScalesR0.1RatioFO} to Figure~\ref{fig:ComparisonScalesR1.2RatioFO} for the NNLO predictions.
The most striking feature is that the $\muj$ dependence in the case of NNLO+NNLL predictions essentially vanishes regardless of the central scale choice, indicating a strong correlation of this variation across different jet radii.
For the resummed prediction, we also see a reduction of the $\mur$ variation sensitivity for small $R$ compared to the fixed-order case.


\clearpage

\bibliographystyle{JHEPmod}
\bibliography{lit}

@article{Huss:2025nlt,
    author = {Huss, Alexander and Huston, Joey and Jones, Stephen and Pellen, Mathieu and R{\"o}ntsch, Raoul},
    title = "{Les Houches 2023 -- Physics at TeV Colliders: Report on the Standard Model Precision Wishlist}",
    eprint = "2504.06689",
    archivePrefix = "arXiv",
    primaryClass = "hep-ph",
    reportNumber = "CERN-TH-2025-071, FR-PHENO-2025-005, IPPP/25/19, TIF-UNIMI-2025-9",
    doi = "10.21468/SciPostPhysCommRep.21",
    journal = "SciPost Phys. Comm. Rep.",
    pages = "21",
    year = "2026"
}

@article{Ablat:2024uvg,
    author = "Ablat, Alim and Dulat, Sayipjamal and Hou, Tie-Jiun and Huston, Joey and Nadolsky, Pavel and Sitiwaldi, Ibrahim and Xie, Keping and Yuan, C. -P.",
    collaboration = "CTEQ-TEA",
    title = "{Impact of LHC precision measurements of inclusive jet and dijet production on the CTEQ-TEA global PDF fit}",
    eprint = "2412.00350",
    archivePrefix = "arXiv",
    primaryClass = "hep-ph",
    doi = "10.1103/PhysRevD.111.036033",
    journal = "Phys. Rev. D",
    volume = "111",
    number = "3",
    pages = "036033",
    year = "2025"
}

@article{Hou:2019efy,
    author = "Hou, Tie-Jiun and others",
    title = "{New CTEQ global analysis of quantum chromodynamics with high-precision data from the LHC}",
    eprint = "1912.10053",
    archivePrefix = "arXiv",
    primaryClass = "hep-ph",
    reportNumber = "MSUHEP-19-025, PITT-PACC-1911, SMU-HEP-19-03",
    doi = "10.1103/PhysRevD.103.014013",
    journal = "Phys. Rev. D",
    volume = "103",
    number = "1",
    pages = "014013",
    year = "2021"
}

@article{Reyer:2019obz,
    author = {Reyer, Max and Sch{\"o}nherr, Marek and Schumann, Steffen},
    title = "{Full NLO corrections to 3-jet production and $\mathbf {R_{32}}$ at the LHC}",
    eprint = "1902.01763",
    archivePrefix = "arXiv",
    primaryClass = "hep-ph",
    reportNumber = "FR-PHENO-2018-016, CERN-TH-2018-275, IPPP/19/9, MCNET-19-03",
    doi = "10.1140/epjc/s10052-019-6815-3",
    journal = "Eur. Phys. J. C",
    volume = "79",
    number = "4",
    pages = "321",
    year = "2019"
}

@article{Alioli:2010xa,
    author = "Alioli, Simone and Hamilton, Keith and Nason, Paolo and Oleari, Carlo and Re, Emanuele",
    title = "{Jet pair production in POWHEG}",
    eprint = "1012.3380",
    archivePrefix = "arXiv",
    primaryClass = "hep-ph",
    reportNumber = "DESY-10-233, SFB-CPP-10-128, IPPP-10-100, DCPT-10-200, MCNET-10-23",
    doi = "10.1007/JHEP04(2011)081",
    journal = "JHEP",
    volume = "04",
    pages = "081",
    year = "2011"
}

@article{Gleisberg:2008ta,
    Archiveprefix = {arXiv},
    Author = {Gleisberg, T. and Hoeche, Stefan. and Krauss, F. and Schonherr, M. and Schumann, S. and Siegert, F. and Winter, J.},
    Doi = {10.1088/1126-6708/2009/02/007},
    Eprint = {0811.4622},
    Journal = {JHEP},
    Pages = {007},
    Primaryclass = {hep-ph},
    Reportnumber = {FERMILAB-PUB-08-477-T, SLAC-PUB-13420, ZU-TH-17-08, DCPT-08-138, IPPP-08-69, EDINBURGH-2008-30, MCNET-08-14},
    Title = {{Event generation with SHERPA 1.1}},
    Volume = {02},
    Year = {2009}}

@article{Czakon:2021ohs,
    author = "Czakon, Micha\l{} and Generet, Terry and Mitov, Alexander and Poncelet, Rene",
    title = "{B-hadron production in NNLO QCD: application to LHC t$ \overline{t} $ events with leptonic decays}",
    eprint = "2102.08267",
    archivePrefix = "arXiv",
    primaryClass = "hep-ph",
    reportNumber = "TTK-21-05, P3H-21-011, Cavendish-HEP-21/02",
    doi = "10.1007/JHEP10(2021)216",
    journal = "JHEP",
    volume = "10",
    pages = "216",
    year = "2021"
}

@article{Czakon:2019tmo,
    Archiveprefix = {arXiv},
    Author = {Czakon, Micha\l{} and van Hameren, Andreas and Mitov, Alexander and Poncelet, Rene},
    Doi = {10.1007/JHEP10(2019)262},
    Eprint = {1907.12911},
    Journal = {JHEP},
    Pages = {262},
    Primaryclass = {hep-ph},
    Title = {{Single-jet inclusive rates with exact color at $ \mathcal{O} $ ($ {\alpha}_s^4 $)}},
    Volume = {10},
    Year = {2019}}

@article{Liu:2017pbb,
    author = "Liu, Xiaohui and Moch, Sven-Olaf and Ringer, Felix",
    title = "{Threshold and jet radius joint resummation for single-inclusive jet production}",
    eprint = "1708.04641",
    archivePrefix = "arXiv",
    primaryClass = "hep-ph",
    reportNumber = "DESY-17-119",
    doi = "10.1103/PhysRevLett.119.212001",
    journal = "Phys. Rev. Lett.",
    volume = "119",
    number = "21",
    pages = "212001",
    year = "2017"
}

@article{Liu:2018ktv,
    Archiveprefix = {arXiv},
    Author = {Liu, Xiaohui and Moch, Sven-Olaf and Ringer, Felix},
    Doi = {10.1103/PhysRevD.97.056026},
    Eprint = {1801.07284},
    Journal = {Phys. Rev. D},
    Number = {5},
    Pages = {056026},
    Primaryclass = {hep-ph},
    Reportnumber = {DESY-18-007},
    Title = {{Phenomenology of single-inclusive jet production with jet radius and threshold resummation}},
    Volume = {97},
    Year = {2018}}

@article{Moch:2018hgy,
    author = "Moch, Sven-Olaf and Eren, Engin and Lipka, Katerina and Liu, Xiaohui and Ringer, Felix",
    title = "{Threshold and jet radius joint resummation for single-inclusive jet production}",
    eprint = "1808.04574",
    archivePrefix = "arXiv",
    primaryClass = "hep-ph",
    reportNumber = "DESY-18-135",
    doi = "10.22323/1.303.0002",
    journal = "PoS",
    volume = "LL2018",
    pages = "002",
    year = "2018"
}

@article{Kidonakis:1998bk,
    author = "Kidonakis, Nikolaos and Oderda, Gianluca and Sterman, George F.",
    title = "{Threshold resummation for dijet cross-sections}",
    eprint = "hep-ph/9801268",
    archivePrefix = "arXiv",
    reportNumber = "EDINBURGH-97-22, ITP-SB-97-78",
    doi = "10.1016/S0550-3213(98)00243-0",
    journal = "Nucl. Phys. B",
    volume = "525",
    pages = "299--332",
    year = "1998"
}

@article{Kidonakis:2000gi,
    author = "Kidonakis, Nikolaos and Owens, J. F.",
    title = "{Effects of higher order threshold corrections in high E(T) jet production}",
    eprint = "hep-ph/0007268",
    archivePrefix = "arXiv",
    reportNumber = "FSU-HEP-2000-0720",
    doi = "10.1103/PhysRevD.63.054019",
    journal = "Phys. Rev. D",
    volume = "63",
    pages = "054019",
    year = "2001"
}

@article{Dasgupta:2014yra,
    author = "Dasgupta, Mrinal and Dreyer, Fr\'ed\'eric and Salam, Gavin P. and Soyez, Gregory",
    title = "{Small-radius jets to all orders in QCD}",
    eprint = "1411.5182",
    archivePrefix = "arXiv",
    primaryClass = "hep-ph",
    reportNumber = "CERN-PH-TH-2014-222",
    doi = "10.1007/JHEP04(2015)039",
    journal = "JHEP",
    volume = "04",
    pages = "039",
    year = "2015"
}

@article{Kang:2016mcy,
    author = "Kang, Zhong-Bo and Ringer, Felix and Vitev, Ivan",
    title = "{The semi-inclusive jet function in SCET and small radius resummation for inclusive jet production}",
    eprint = "1606.06732",
    archivePrefix = "arXiv",
    primaryClass = "hep-ph",
    doi = "10.1007/JHEP10(2016)125",
    journal = "JHEP",
    volume = "10",
    pages = "125",
    year = "2016"
}

@article{Dai:2016hzf,
    author = "Dai, Lin and Kim, Chul and Leibovich, Adam K.",
    title = "{Fragmentation of a Jet with Small Radius}",
    eprint = "1606.07411",
    archivePrefix = "arXiv",
    primaryClass = "hep-ph",
    doi = "10.1103/PhysRevD.94.114023",
    journal = "Phys. Rev. D",
    volume = "94",
    number = "11",
    pages = "114023",
    year = "2016"
}

@article{Boughezal:2015dva,
    author = "Boughezal, Radja and Focke, Christfried and Liu, Xiaohui and Petriello, Frank",
    title = "{$W$-boson production in association with a jet at next-to-next-to-leading order in perturbative QCD}",
    eprint = "1504.02131",
    archivePrefix = "arXiv",
    primaryClass = "hep-ph",
    doi = "10.1103/PhysRevLett.115.062002",
    journal = "Phys. Rev. Lett.",
    volume = "115",
    number = "6",
    pages = "062002",
    year = "2015"
}

@article{Boughezal:2015dra,
    author = "Boughezal, Radja and Caola, Fabrizio and Melnikov, Kirill and Petriello, Frank and Schulze, Markus",
    title = "{Higgs boson production in association with a jet at next-to-next-to-leading order}",
    eprint = "1504.07922",
    archivePrefix = "arXiv",
    primaryClass = "hep-ph",
    reportNumber = "CERN-PH-TH-2015-056, TTP15-017",
    doi = "10.1103/PhysRevLett.115.082003",
    journal = "Phys. Rev. Lett.",
    volume = "115",
    number = "8",
    pages = "082003",
    year = "2015"
}

@article{Boughezal:2015aha,
    author = "Boughezal, Radja and Focke, Christfried and Giele, Walter and Liu, Xiaohui and Petriello, Frank",
    title = "{Higgs boson production in association with a jet at NNLO using jettiness subtraction}",
    eprint = "1505.03893",
    archivePrefix = "arXiv",
    primaryClass = "hep-ph",
    reportNumber = "FERMILAB-PUB-15-210-T",
    doi = "10.1016/j.physletb.2015.06.055",
    journal = "Phys. Lett. B",
    volume = "748",
    pages = "5--8",
    year = "2015"
}

@article{Chen:2014gva,
    author = "Chen, X. and Gehrmann, T. and Glover, E. W. N. and Jaquier, M.",
    title = "{Precise QCD predictions for the production of Higgs + jet final states}",
    eprint = "1408.5325",
    archivePrefix = "arXiv",
    primaryClass = "hep-ph",
    reportNumber = "IPPP-14-64, ZU-TH-27-14",
    doi = "10.1016/j.physletb.2014.11.021",
    journal = "Phys. Lett. B",
    volume = "740",
    pages = "147--150",
    year = "2015"
}

@article{Gehrmann-DeRidder:2015wbt,
    author = "Gehrmann-De Ridder, A. and Gehrmann, T. and Glover, E. W. N. and Huss, A. and Morgan, T. A.",
    title = "{Precise QCD predictions for the production of a Z boson in association with a hadronic jet}",
    eprint = "1507.02850",
    archivePrefix = "arXiv",
    primaryClass = "hep-ph",
    reportNumber = "IPPP-15-44, ZU-TH-23-15",
    doi = "10.1103/PhysRevLett.117.022001",
    journal = "Phys. Rev. Lett.",
    volume = "117",
    number = "2",
    pages = "022001",
    year = "2016"
}

@article{Boughezal:2015ded,
    author = "Boughezal, Radja and Campbell, John M. and Ellis, R. Keith and Focke, Christfried and Giele, Walter T. and Liu, Xiaohui and Petriello, Frank",
    title = "{Z-boson production in association with a jet at next-to-next-to-leading order in perturbative QCD}",
    eprint = "1512.01291",
    archivePrefix = "arXiv",
    primaryClass = "hep-ph",
    reportNumber = "FERMILAB-PUB-15-519-T, IPPP-15-79",
    doi = "10.1103/PhysRevLett.116.152001",
    journal = "Phys. Rev. Lett.",
    volume = "116",
    number = "15",
    pages = "152001",
    year = "2016"
}

@article{Currie:2016bfm,
    author = "Currie, J and Glover, E. W. N. and Pires, J",
    title = "{Next-to-Next-to Leading Order QCD Predictions for Single Jet Inclusive Production at the LHC}",
    eprint = "1611.01460",
    archivePrefix = "arXiv",
    primaryClass = "hep-ph",
    reportNumber = "IPPP-16-110, MPP-2016-322",
    doi = "10.1103/PhysRevLett.118.072002",
    journal = "Phys. Rev. Lett.",
    volume = "118",
    number = "7",
    pages = "072002",
    year = "2017"
}

@article{Currie:2017eqf,
    author = "Currie, James and Gehrmann-De Ridder, Aude and Gehrmann, Thomas and Glover, E. W. N. and Huss, Alexander and Pires, Joao",
    title = "{Precise predictions for dijet production at the LHC}",
    eprint = "1705.10271",
    archivePrefix = "arXiv",
    primaryClass = "hep-ph",
    reportNumber = "IPPP-17-45, ZU-TH-13-17, MPP-2017-107",
    doi = "10.1103/PhysRevLett.119.152001",
    journal = "Phys. Rev. Lett.",
    volume = "119",
    number = "15",
    pages = "152001",
    year = "2017"
}

@article{Campbell:2016lzl,
    author = "Campbell, John M. and Ellis, R. Keith and Williams, Ciaran",
    title = "{Direct Photon Production at Next-to\textendash{}Next-to-Leading Order}",
    eprint = "1612.04333",
    archivePrefix = "arXiv",
    primaryClass = "hep-ph",
    reportNumber = "IPPP-16-115, FERMILAB-PUB-16-585-T",
    doi = "10.1103/PhysRevLett.118.222001",
    journal = "Phys. Rev. Lett.",
    volume = "118",
    number = "22",
    pages = "222001",
    year = "2017",
    note = "[Erratum: Phys.Rev.Lett. 124, 259901 (2020)]"
}

@article{Nocera:2017zge,
    author = "Nocera, Emanuele Roberto and Ubiali, Maria",
    editor = "Klein, Uta",
    title = "{Constraining the gluon PDF at large x with LHC data}",
    eprint = "1709.09690",
    archivePrefix = "arXiv",
    primaryClass = "hep-ph",
    doi = "10.22323/1.297.0008",
    journal = "PoS",
    volume = "DIS2017",
    pages = "008",
    year = "2018"
}

@article{Harland-Lang:2017ytb,
    author = "Harland-Lang, L. A. and Martin, A. D. and Thorne, R. S.",
    title = "{The Impact of LHC Jet Data on the MMHT PDF Fit at NNLO}",
    eprint = "1711.05757",
    archivePrefix = "arXiv",
    primaryClass = "hep-ph",
    reportNumber = "IPPP-17-85",
    doi = "10.1140/epjc/s10052-018-5710-7",
    journal = "Eur. Phys. J. C",
    volume = "78",
    number = "3",
    pages = "248",
    year = "2018"
}

@article{Bellm:2019yyh,
    author = "Bellm, Johannes and others",
    title = "{Jet Cross Sections at the LHC and the Quest for Higher Precision}",
    eprint = "1903.12563",
    archivePrefix = "arXiv",
    primaryClass = "hep-ph",
    reportNumber = "FERMILAB-PUB-19-100-T, CERN-TH-2019-038, SLAC-PUB-17411, LAPTH-019/19, CFTP/19-006, ZU-TH 16/19, LU-TP-19-10, IPPP/19/22, UWTHPH-2019-9, MCnet-19-06",
    doi = "10.1140/epjc/s10052-019-7574-x",
    journal = "Eur. Phys. J. C",
    volume = "80",
    number = "2",
    pages = "93",
    year = "2020"
}

@article{AbdulKhalek:2020jut,
    author = "Abdul Khalek, Rabah and others",
    title = "{Phenomenology of NNLO jet production at the LHC and its impact on parton distributions}",
    eprint = "2005.11327",
    archivePrefix = "arXiv",
    primaryClass = "hep-ph",
    reportNumber = "CERN-TH-2020-073, IPPP/20/17, Nikhef/2020-001, TIF-UNIMI-2020-3, ZU-TH 15/20, TIF-UNIMI-2020-3,
  ZU-TH 15/20",
    doi = "10.1140/epjc/s10052-020-8328-5",
    journal = "Eur. Phys. J. C",
    volume = "80",
    number = "8",
    pages = "797",
    year = "2020"
}

@article{Currie:2018xkj,
    author = "Currie, James and Gehrmann-De Ridder, Aude and Gehrmann, Thomas and Glover, E. W. Nigel and Huss, Alexander and Pires, Jo\~ao",
    title = "{Infrared sensitivity of single jet inclusive production at hadron colliders}",
    eprint = "1807.03692",
    archivePrefix = "arXiv",
    primaryClass = "hep-ph",
    reportNumber = "IPPP-18-38, ZU-TH-24-18, CFTP-18-010, IPPP/18/38, ZU-TH 24/18, CFTP/18-010, CERN-TH-2018-159",
    doi = "10.1007/JHEP10(2018)155",
    journal = "JHEP",
    volume = "10",
    pages = "155",
    year = "2018"
}

@article{Czakon:2024tjr,
    author = "Czakon, Micha{\l} and Generet, Terry and Mitov, Alexander and Poncelet, Rene",
    title = "{Open B-Hadron Production at Hadron Colliders in QCD at Next-to-Next-to-Leading-Order and Next-to-Next-to-Leading-Logarithmic Accuracy}",
    eprint = "2411.09684",
    archivePrefix = "arXiv",
    primaryClass = "hep-ph",
    reportNumber = "Cavendish-HEP-24/01, IFJPAN-IV-2024-13, P3H-24-087, TTK-24-48",
    doi = "10.1103/b6pf-rj4h",
    journal = "Phys. Rev. Lett.",
    volume = "135",
    number = "16",
    pages = "161903",
    year = "2025"
}

@article{Badger:2023mgf,
    author = "Badger, Simon and Czakon, Micha\l{} and Hartanto, Heribertus Bayu and Moodie, Ryan and Peraro, Tiziano and Poncelet, Rene and Zoia, Simone",
    title = "{Isolated photon production in association with a jet pair through next-to-next-to-leading order in QCD}",
    eprint = "2304.06682",
    archivePrefix = "arXiv",
    primaryClass = "hep-ph",
    reportNumber = "Cavendish-HEP-23/02, P3H-23-022, TTK-23-09",
    doi = "10.1007/JHEP10(2023)071",
    journal = "JHEP",
    volume = "10",
    pages = "071",
    year = "2023"
}

@article{Alvarez:2023fhi,
    author = "Alvarez, Manuel and Cantero, Josu and Czakon, Michal and Llorente, Javier and Mitov, Alexander and Poncelet, Rene",
    title = "{NNLO QCD corrections to event shapes at the LHC}",
    eprint = "2301.01086",
    archivePrefix = "arXiv",
    primaryClass = "hep-ph",
    reportNumber = "Cavendish-HEP-22/11, P3H-22-129, TTK-22-49",
    doi = "10.1007/JHEP03(2023)129",
    journal = "JHEP",
    volume = "03",
    pages = "129",
    year = "2023"
}

@article{Lim:2024nsk,
    author = "Lim, Matthew A. and Poncelet, Rene",
    title = "{Robust estimates of theoretical uncertainties at fixed-order in perturbation theory}",
    eprint = "2412.14910",
    archivePrefix = "arXiv",
    primaryClass = "hep-ph",
    reportNumber = "IFJPAN-IV-2024-15",
    doi = "10.1103/7g5k-4y3v",
    journal = "Phys. Rev. D",
    volume = "112",
    number = "11",
    pages = "L111901",
    year = "2025"
}

@article{Czakon:2021mjy,
    author = "Czakon, Michal and Mitov, Alexander and Poncelet, Rene",
    title = "{Next-to-Next-to-Leading Order Study of Three-Jet Production at the LHC}",
    eprint = "2106.05331",
    archivePrefix = "arXiv",
    primaryClass = "hep-ph",
    reportNumber = "Cavendish-HEP-21/09, P3H-21-043, TTK-21-20",
    doi = "10.1103/PhysRevLett.127.152001",
    journal = "Phys. Rev. Lett.",
    volume = "127",
    number = "15",
    pages = "152001",
    year = "2021",
    note = "[Erratum: Phys.Rev.Lett. 129, 119901 (2022)]"
}

@article{Chawdhry:2021hkp,
    author = "Chawdhry, Herschel A. and Czakon, Michal and Mitov, Alexander and Poncelet, Rene",
    title = "{NNLO QCD corrections to diphoton production with an additional jet at the LHC}",
    eprint = "2105.06940",
    archivePrefix = "arXiv",
    primaryClass = "hep-ph",
    reportNumber = "Cavendish-HEP-21/07, OUTP-21-13P, P3H-21-032",
    doi = "10.1007/JHEP09(2021)093",
    journal = "JHEP",
    volume = "09",
    pages = "093",
    year = "2021"
}

@article{Chen:2022tpk,
    author = "Chen, X. and Gehrmann, T. and Glover, E. W. N. and Huss, A. and Mo, J.",
    title = "{NNLO QCD corrections in full colour for jet production observables at the LHC}",
    eprint = "2204.10173",
    archivePrefix = "arXiv",
    primaryClass = "hep-ph",
    reportNumber = "ZU-TH 11/22, KA-TP-07-2022, IPPP/22/20, P3H-22-037, CERN-TH-2022-067",
    doi = "10.1007/JHEP09(2022)025",
    journal = "JHEP",
    volume = "09",
    pages = "025",
    year = "2022"
}

@article{CMS:2020caw,
    author = "Sirunyan, Albert M and others",
    collaboration = "CMS",
    title = "{Dependence of inclusive jet production on the anti-k$_{T}$ distance parameter in pp collisions at $ \sqrt{\mathrm{s}} $ = 13 TeV}",
    eprint = "2005.05159",
    archivePrefix = "arXiv",
    primaryClass = "hep-ex",
    reportNumber = "CMS-SMP-19-003, CERN-EP-2020-040",
    doi = "10.1007/JHEP12(2020)082",
    journal = "JHEP",
    volume = "12",
    pages = "082",
    year = "2020"
}

@article{Czakon:2014oma,
    author = "Czakon, M. and Heymes, D.",
    title = "{Four-dimensional formulation of the sector-improved residue subtraction scheme}",
    eprint = "1408.2500",
    archivePrefix = "arXiv",
    primaryClass = "hep-ph",
    reportNumber = "TTK-14-16",
    doi = "10.1016/j.nuclphysb.2014.11.006",
    journal = "Nucl. Phys. B",
    volume = "890",
    pages = "152--227",
    year = "2014"
}

@article{Czakon:2010td,
    author = "Czakon, M.",
    title = "{A novel subtraction scheme for double-real radiation at NNLO}",
    eprint = "1005.0274",
    archivePrefix = "arXiv",
    primaryClass = "hep-ph",
    doi = "10.1016/j.physletb.2010.08.036",
    journal = "Phys. Lett. B",
    volume = "693",
    pages = "259--268",
    year = "2010"
}

@article{Dasgupta:2016bnd,
    Archiveprefix = {arXiv},
    Author = {Dasgupta, Mrinal and Dreyer, Fr\'ed\'eric A. and Salam, Gavin P. and Soyez, Gregory},
    Doi = {10.1007/JHEP06(2016)057},
    Eprint = {1602.01110},
    Journal = {JHEP},
    Pages = {057},
    Primaryclass = {hep-ph},
    Reportnumber = {CERN-TH-2016-020},
    Title = {{Inclusive jet spectrum for small-radius jets}},
    Volume = {06},
    Year = {2016}}

@article{Cacciari:2011ze,
    author = "Cacciari, Matteo and Houdeau, Nicolas",
    title = "{Meaningful characterisation of perturbative theoretical uncertainties}",
    eprint = "1105.5152",
    archivePrefix = "arXiv",
    primaryClass = "hep-ph",
    doi = "10.1007/JHEP09(2011)039",
    journal = "JHEP",
    volume = "09",
    pages = "039",
    year = "2011"
}

@article{Bonvini:2020xeo,
    author = "Bonvini, Marco",
    title = "{Probabilistic definition of the perturbative theoretical uncertainty from missing higher orders}",
    eprint = "2006.16293",
    archivePrefix = "arXiv",
    primaryClass = "hep-ph",
    doi = "10.1140/epjc/s10052-020-08545-z",
    journal = "Eur. Phys. J. C",
    volume = "80",
    number = "10",
    pages = "989",
    year = "2020"
}

@article{Duhr:2021mfd,
    author = "Duhr, Claude and Huss, Alexander and Mazeliauskas, Aleksas and Szafron, Robert",
    title = "{An analysis of Bayesian estimates for missing higher orders in perturbative calculations}",
    eprint = "2106.04585",
    archivePrefix = "arXiv",
    primaryClass = "hep-ph",
    reportNumber = "CERN-TH-2021-058",
    doi = "10.1007/JHEP09(2021)122",
    journal = "JHEP",
    volume = "09",
    pages = "122",
    year = "2021"
}

@article{Ghosh:2022lrf,
    author = "Ghosh, Aishik and Nachman, Benjamin and Plehn, Tilman and Shire, Lily and Tait, Tim M. P. and Whiteson, Daniel",
    title = "{Statistical patterns of theory uncertainties}",
    eprint = "2210.15167",
    archivePrefix = "arXiv",
    primaryClass = "hep-ph",
    doi = "10.21468/SciPostPhysCore.6.2.045",
    journal = "SciPost Phys. Core",
    volume = "6",
    pages = "045",
    year = "2023"
}

@article{Tackmann:2024kci,
    author = "Tackmann, Frank J.",
    title = "{Beyond scale variations: perturbative theory uncertainties from nuisance parameters}",
    eprint = "2411.18606",
    archivePrefix = "arXiv",
    primaryClass = "hep-ph",
    reportNumber = "DESY-19-021",
    doi = "10.1007/JHEP08(2025)098",
    journal = "JHEP",
    volume = "08",
    pages = "098",
    year = "2025"
}

@article{Lee:2024icn,
    author = "Lee, Kyle and Moult, Ian and Zhang, Xiaoyuan",
    title = "{Revisiting single inclusive jet production: timelike factorization and reciprocity}",
    eprint = "2409.19045",
    archivePrefix = "arXiv",
    primaryClass = "hep-ph",
    reportNumber = "MIT-CTP 5766",
    doi = "10.1007/JHEP05(2025)129",
    journal = "JHEP",
    volume = "05",
    pages = "129",
    year = "2025"
}

@article{Czakon:2025yti,
    author = "Czakon, Micha{\l} and Generet, Terry and Mitov, Alexander and Poncelet, Rene",
    title = "{Identified Hadron Production at Hadron Colliders in Next-to-Next-to-Leading-Order QCD}",
    eprint = "2503.11489",
    archivePrefix = "arXiv",
    primaryClass = "hep-ph",
    reportNumber = "Cavendish-HEP-25/01, IFJPAN-IV-2025-6, P3H-25-016, TTK-25-09",
    doi = "10.1103/mhkl-vt96",
    journal = "Phys. Rev. Lett.",
    volume = "135",
    number = "17",
    pages = "17",
    year = "2025"
}

@article{NNPDF:2017mvq,
    author = "Ball, Richard D. and others",
    collaboration = "NNPDF",
    title = "{Parton distributions from high-precision collider data}",
    eprint = "1706.00428",
    archivePrefix = "arXiv",
    primaryClass = "hep-ph",
    reportNumber = "EDINBURGH-2017-08, NIKHEF-2017-006, OUTP-17-04P, TIF-UNIMI-2017-3, CAVENDISH-HEP-17-06, CERN-TH-2017-077, Edinburgh 2017/08,
  Nikhef/2017-006, OUTP-17-04P,TIF-UNIMI-2017-3",
    doi = "10.1140/epjc/s10052-017-5199-5",
    journal = "Eur. Phys. J. C",
    volume = "77",
    number = "10",
    pages = "663",
    year = "2017"
}

@article{Bailey:2020ooq,
    author = "Bailey, S. and Cridge, T. and Harland-Lang, L. A. and Martin, A. D. and Thorne, R. S.",
    title = "{Parton distributions from LHC, HERA, Tevatron and fixed target data: MSHT20 PDFs}",
    eprint = "2012.04684",
    archivePrefix = "arXiv",
    primaryClass = "hep-ph",
    reportNumber = "IPPP/20/58",
    doi = "10.1140/epjc/s10052-021-09057-0",
    journal = "Eur. Phys. J. C",
    volume = "81",
    number = "4",
    pages = "341",
    year = "2021"
}

@article{NNPDF:2021njg,
    author = "Ball, Richard D. and others",
    collaboration = "NNPDF",
    title = "{The path to proton structure at 1{\%} accuracy}",
    eprint = "2109.02653",
    archivePrefix = "arXiv",
    primaryClass = "hep-ph",
    reportNumber = "Edinburgh 2021/12, Nikhef-2021-013, TIF-UNIMI-2021-11",
    doi = "10.1140/epjc/s10052-022-10328-7",
    journal = "Eur. Phys. J. C",
    volume = "82",
    number = "5",
    pages = "428",
    year = "2022"
}

@article{H1:2021xxi,
    author = "Abt, I. and others",
    collaboration = "H1, ZEUS",
    title = "{Impact of jet-production data on the next-to-next-to-leading-order determination of HERAPDF2.0 parton distributions}",
    eprint = "2112.01120",
    archivePrefix = "arXiv",
    primaryClass = "hep-ex",
    reportNumber = "DESY-21-206",
    doi = "10.1140/epjc/s10052-022-10083-9",
    journal = "Eur. Phys. J. C",
    volume = "82",
    number = "3",
    pages = "243",
    year = "2022"
}

@article{Generet:2025vth,
    author = "Generet, Terry and Lee, Kyle and Moult, Ian and Poncelet, Rene and Zhang, Xiaoyuan",
    title = "{Small radius inclusive jet production at the LHC through NNLO+NNLL}",
    eprint = "2503.21866",
    archivePrefix = "arXiv",
    primaryClass = "hep-ph",
    reportNumber = "Cavendish-HEP-25/02, MIT-CTP 5857, IFJPAN-IV-2025-7",
    doi = "10.1007/JHEP08(2025)015",
    journal = "JHEP",
    volume = "08",
    pages = "015",
    year = "2025"
}

@article{Andersen:2023kuj,
    author = "Andersen, Jeppe R. and Duclou{\'e}, Bertrand and Elrick, Conor and Hassan, Hitham and Maier, Andreas and Nail, Graeme and Paltrinieri, J{\'e}r{\'e}my and Papaefstathiou, Andreas and Smillie, Jennifer M.",
    title = "{HEJ 2.2: W boson pairs and Higgs boson plus jet production at high energies}",
    eprint = "2303.15778",
    archivePrefix = "arXiv",
    primaryClass = "hep-ph",
    reportNumber = "IPPP/23/18, DCPT/23/36, DESY-23-038",
    doi = "10.21468/SciPostPhysCodeb.21",
    journal = "SciPost Phys. Codeb.",
    pages = "21",
    year = "2023"
}

@article{Andersen:2009nu,
    author = "Andersen, Jeppe R. and Smillie, Jennifer M.",
    title = "{Constructing All-Order Corrections to Multi-Jet Rates}",
    eprint = "0908.2786",
    archivePrefix = "arXiv",
    primaryClass = "hep-ph",
    reportNumber = "CERN-PH-TH-2009-153, DCPT-09-122, IPPP-09-61",
    doi = "10.1007/JHEP01(2010)039",
    journal = "JHEP",
    volume = "01",
    pages = "039",
    year = "2010"
}

@article{Andersen:2009he,
    author = "Andersen, Jeppe R. and Smillie, Jennifer M.",
    title = "{The Factorisation of the t-channel Pole in Quark-Gluon Scattering}",
    eprint = "0910.5113",
    archivePrefix = "arXiv",
    primaryClass = "hep-ph",
    reportNumber = "CERN-PH-TH-2009-202",
    doi = "10.1103/PhysRevD.81.114021",
    journal = "Phys. Rev. D",
    volume = "81",
    pages = "114021",
    year = "2010"
}

@article{Andersen:2011hs,
    author = "Andersen, Jeppe R. and Smillie, Jennifer M.",
    title = "{Multiple Jets at the LHC with High Energy Jets}",
    eprint = "1101.5394",
    archivePrefix = "arXiv",
    primaryClass = "hep-ph",
    reportNumber = "CP3-ORIGINS-2011-02, EDINBURGH-2011-03",
    doi = "10.1007/JHEP06(2011)010",
    journal = "JHEP",
    volume = "06",
    pages = "010",
    year = "2011"
}

\end{document}